\documentclass [aps,twocolumn,showpacs,reprint,pre,12pt]{revtex4-1}
\usepackage{graphicx}
\usepackage{textcomp}
\usepackage{psfrag}
\usepackage{float}
\usepackage{amsmath}
\usepackage{mathtools}
\usepackage{xcolor}
\usepackage{amssymb}
\usepackage{romannum}
\usepackage{afterpage}
\usepackage{url}
\usepackage{hyperref}
\usepackage{caption}
\usepackage{subcaption}
\usepackage[us,12hr]{datetime}
\usepackage[normalem]{ulem}
\begin{document}

\title{Structure and dynamics of finite three-dimensional Yukawa clusters in complex plasmas : Newtonian versus Langevin Dynamics}
\author{Hirakjyoti Sarma}
\email{hirakphy2019@gmail.com}
\affiliation{Department of Physics, Tezpur University, Napaam, Tezpur- 784028, India}

\author {Nilakshi Das}
\email{ndas@tezu.ernet.in}
\affiliation{Department of Physics, Tezpur University, Napaam, Tezpur- 784028, India}
\date{\today ; \currenttime}

\begin{abstract}
	The structure and dynamics of a harmonically confined three dimensional finite dust cluster are investigated via both Langevin Dynamics (LD) and frictionless Molecular Dynamics (fMD) simulation. The static structure of the system is analyzed through the Radial Distribution Function, Center-two-particle correlation function(C2P) and angular correlation function. The intra-shell angular correlation, Radial Distribution Function and C2P remains largely unaffected by the dynamics employed. However, the inter-shell angular correlation exhibits sharp peaks at irregular angular intervals in fMD which are not seen in LD indicating strongly correlated motion of the two spherical shells in the cluster in fMD. The single particle dynamics of the cluster is characterized by the Van - Hove self autocorrelation function and Mean Squared Displacement (MSD). Notably, the Van - Hove autocorrelation function in fMD simulations exhibits narrower width and higher peak height as compared to the LD simulations, suggesting greater particle mobility in LD. Trajectory analysis reveals a rotational motion of the particles about a common axis in fMD which disappears with progressively increasing friction coefficient. We show that the disappearance of rotational motion with the introduction of neutral friction in the dynamics is due to the much faster relaxation of the interparticle distance as well as interparticle angular separation in LD as compared to fMD.    
\end{abstract}


\maketitle

	\onecolumngrid
\section{Introduction}
Clusters formed by trapping charged particles in space are achieved in different physical systems for example ions in Paul or Penning traps \cite{dubin1999trapped}, charged dust particles in plasmas \cite{bonitz2010complex}, electrons in quantum dots \cite{ashoori1996electrons} etc. Ions can be routinely trapped in the electromagnetic traps for several days in thermal equilibrium. Quantum dots have made it possible to perform atomic physics experiments in regimes that were previously inaccessible in real atoms. By studying these systems, valuable insights may be obtained about the microscopic dynamics of charged particles in finite systems.  

In these systems the confinement leads to some unique collective and single particle behavior which is normally not observed in bulk homogeneous systems. A characteristic property of harmonically trapped finite charged particle clusters is the occurence of concentric spherical shells in 3D and concentric circular rings in 2D \cite{ludwig2005structure}. 

Particle dynamics in an interacting many body system can be studied by using both Hamiltonian as well as Langevin Dynamics. The phase space evolution in a Hamiltonian Dynamical system is given by Liouville equation whereas the corresponding evolution in Langevin dynamical systems is governed by Fokker-Planck equation \cite{zwanzig2001nonequilibrium}. Static, dynamic and thermodynamic properties of finite trapped system of particles can be studied by using either Newtonian or Langevin Dynamics method. For example, the formation of dynamical states by harmonically confined charged dust particles in two dimensions was investigated via frictionless MD simulation \cite{maity2020dynamical}. They observed different modes of the cluster depending upon the total number of particles and the screening strengths, for example intershell rotation, radial oscillation etc. It was observed in the Langevin Dynamics simulation of a one dimensional chain of trapped ions that in the absence of the stochastic force the competition of the ions with thermal motion leads to a transition from perfect one dimensional chain to a zig-zag chain arrangement. Stochastic force was found to play an important role in nucleation as well as in defect formation \cite{wen2019interacting}. 

Rotation of dust clusters in complex plasma was widely investigated in the past by many researchers. Traditionally the rotation of dust clusters in complex plasma experiments in the presence of a magnetic field is attributed to the drag force exerted by the ions due to $\mathbf{E}\times \mathbf{B}$ drift on the dust particles  \cite{sato2001dynamics,konopka2000rigid}. However, it was observed by Carstensen et al. \cite{carstensen2009effect} that the neutral gas may be set into rotation by ion-neutral collisions and the neutral gas in turn advects the dust particles causing rotation of the cluster. Cluster rotation was also induced by the torque applied by an electron beam by Ticos et al. \cite{ticocs2021rotation}. Rotation is also observed in clusters without any external forcing agency. For example, the nomal mode analysis of 2D clusters by solving the dynamical matrix and finding the eigenvalues and eigenvectors reveals that the cluster contains rotational mode in addition to the other modes such as, center-of-mass mode, breathing mode, vortex-antivortex mode etc \cite{melzer2003mode, melzer2019physics}. 

The neutral component in a complex plasma system provides damping to the dust particle motion thereby ``cooling down" the system of dust particles \cite{fortov2009complex}. In a dusty plasma system the presence of ion flow gives rise to attractive wake potential among the dust particles. Besides dust-neutral collision, the ions also collide with the neutrals and it was found that in the strong collisional limit the attractive wake potential vanishes \cite{bezbaruah2016collisional}. 

The motivation for the present work, is to understand the effects of dust-neutral collision in the static and dynamic properties of a harmonically confined charged dust cluster in complex plasma.  In bulk charged-stabilized colloidal systems the effect of solvent friction on structural relaxation was studied by L\"{o}wen et al \cite{lowen1991brownian} by comparing MD and BD simulations. They observed qualitatively different structural relaxation in the two different dynamical evolutions. In the present work, we study the effect of dust-neutral collision on the structure and dynamics of  finite charged particle clusters that can be formed in dusty plasmas. Because of spatial confinement these clusters lack translational invariance. In a bulk homogeneous system, translational invariance allows for the presence of shear or compressional wave modes extending throughout the system. On the other hand, a harmonically confined finite charged particle cluster doesn't support pure shear or compressional modes due to the presence of a boundary \cite{melzer2003mode}. These so called Yukawa clusters  contain multiple time scales. The interplay among these time scales was recently shown to give rise to various interesting collective and single particle dynamical phenomena in two-dimensional finite dust clusters via Langevin Dynamics simulation \cite{sarma2024phonon}. Although the finite charged particle clusters have been studied by using frictionless MD simulation as well as Langevin Dynamics simulation by many researchers in the past \cite{totsuji2002competition,ludwig2005structure, bonitz2006structural, ludwig2012ion}, a comparative study between Newtonian and Langevin Dynamics has not been done yet. 

The remainder of the paper is organized as follows : section \Romannum{2} describes the model adopted. Section \Romannum{3} describes the simulation technique and the quantities used to understand the static and dynamic properties of the Yukawa cluster. The results are presented in section \Romannum{4} followed by the concluding remarks in section \Romannum{5}. 
\section{The Model}
We consider a system of $N$ interacting dust particles each having mass $m$ and charge $q_d$ in three dimensions immersed in  plasma obeying quasineutrality. The interaction among the particles is assumed to be governed by Yukawa potential energy,
\begin{equation}
	V_{Y}(r_{ij})=\frac{q_d^2}{4\pi\epsilon_0 r_{ij}}\exp{(-r_{ij}/\lambda_d)},
	\label{intPot}
\end{equation}
where, $r_{ij}$ is interparticle distance between the $i$th and $j$th particle and $\lambda_d$ is the Debye length for the dust particles. The particles are assumed to be harmonically confined i.e, each dust particle experiences a force due to a global harmonic oscillator potential,
\begin{equation}
	V_{C}(r) = \frac{1}{2}m\omega_0^2 r^2,
	\label{confPot}
\end{equation}
where, $\omega_{0}$ is the harmonic confinement strength and $r$ is the distance from the center of the simulation box. The harmonic confining potential was shown to mimic the superposition of gravity, ion drag, electrostatic and thermophoretic force acting on a dust particle \cite{arp2005confinement}. 

The equation of motion of the $i^{th}$ particle considering Langevin Dynamics can thus be written as :
\begin{equation}
	m\ddot{\mathbf{r}}_i =-\mathbf{\nabla}\sum_{j\neq i}^{N} V_Y(r_{ij})
	-m\omega_{0}^{2}\mathbf{r}_i -\nu m \dot{\mathbf{r}_i} + \mathbf{A}_i(t),
	\label{EOM_F}
\end{equation}
where, $r_{ij}=|\textbf{r}_i - \textbf{r}_j|$ is the distance between the $ith$ and $jth$ particles, $\nu$ is the dust-neutral collision frequency and $\textbf{A}_i(t)$ denotes the random force acting on the $i^{th}$ dust particle due to collision with the neutrals \cite{feng2014superdiffusion}. The random force term is assumed to obey the following relation according to the fluctuation - dissipation theorem  \cite{allen2017computer}, 
\begin{equation}
	\big<A_{i \alpha}(0)A_{i \beta}(t)\big> = 2m\nu k_B T_d \delta( t) \delta_{\alpha \beta},\;\; \alpha, \beta \in \{x,y,z\}
\end{equation}  
where, $T_d$ denotes dust kinetic temperature and $\delta(t)$ is delta function. 

The random force term $\mathbf{A}_i(t)$ in the equation of motion (Eq. \ref{EOM_F}) is not a differentiable function.  That's why, the differential increment of the random force term ($dA_i$) is proportional to $\Delta t^{1/2}$ in contrast to the other terms present in the force equation which are proportional to $\Delta t$, $\Delta t$ being the value of a time step in the simulation \cite{allen2017computer}.

In numerical simulations, the random force term on the $i$th particle ($\mathbf{A}_i(t)$) can be represented as \cite{kumar2018spiral}, 
\begin{equation}
	\mathbf{A}_i(t) = \sqrt{2m \nu k_B T_d / \Delta t} \; \mathbf{G}_i,
\end{equation}
where, $\mathbf{G}_i$ is a Gaussian white noise.
Equation \ref{EOM_F} can be expressed in dimensionless form as follows :
\begin{equation}
	\ddot{\mathbf{r}^{\prime}}_{i} =\Gamma \kappa \sum_{j \neq i}^{N} \frac{\Big[1 + r_{ij}^{\prime}\Big]}{r_{ij}^{\prime\;3}}\exp{(-r_{ij}^{\prime})} \mathbf{r}_{ij}^{\prime} + \omega_{0}^{\prime\; 2} \mathbf{r}_i^{\prime} - \nu^{\prime} \dot{\mathbf{r}}_i^{\prime} + \sqrt{2\nu^{\prime}/\Delta t^{\prime}} \mathbf{G}_i,	
\end{equation} 
where, $\mathbf{r}^{\prime}=\mathbf{r} / \lambda_d$, $\omega_{0}^{\prime} = t_0 \omega_{0}$, $\nu^{\prime}=t_0 \nu$ and $\Delta t^{\prime}=\Delta t / t_0.$ The scale factor for time is defined as, $t_0 = \sqrt{\frac{m\lambda_{d}^2}{k_B T_d}}.$ It is clear from above that the present system is characterized by four dimensionless parameters, $\Gamma$, $\kappa$, $\omega_{0}^{\prime}$ and $\nu^{\prime}$. $\Gamma$ and $\kappa$ respectively denote the Coulomb coupling parameter and screening parameter which are defined as, $\Gamma= \frac{q_d^2}{4\pi \epsilon_0 r_{av} k_B T_d}$ and $\kappa=\frac{r_{av}}{\lambda_{d}}$, where $r_{av}$ is the mean interparticle distance of the system of dust particles. 
\section{Numerical approach}
The case of a harmonically confined three-dimensional cluster consisting of $N$ particles is investigated by considering two different dynamical evolution methods. We use frictionless Molecular Dynamics (fMD) on one hand to study both the static and dynamic structural properites and then the same quantities are evaluated by incorporating the effect of dust-neutral collision in to the picture via Langevin Dynamics (LD). The friction coefficient is set to zero in Equation \ref{EOM_F} to derive the Newtonian equation of motion, which is then solved numerically in fMD simulations. We consider the $N$ dust particles to be point sized each having a mass ($m$) and charge ($q_d$) equal to $6.99\times10^{-13}\;kg$ and $-1.78\times10^{-16}\;C$ respectively and initially distribute them randomly inside a cubical simulation box with random initial velocities whose edge length ($L$) is determined by the number of particles ($N$) and assigned initial number density ($n_d$) which is fixed as $10^{11}\;m^{-3}$. The value of confinement stregth is fixed as $\omega_{0}=50\;Hz$, unless explicitly specified otherwise. In case of fMD we integrate the equations of motion of the particles by using a second order velocity verlet algorithm. To simulate the system of particles at a desired temperature, we use a velocity rescaling procedure which is known as the Berendsen thermostat \cite{berendsen1984molecular,morishita2000fluctuation}. In the case of LD runs we integrate the equations of motion, by using the BAOAB algorithm which was shown to perform well in the limits of both low and high friction \cite{allen2017computer, leimkuhler2013robust}. 

In the present work we use different quantities to probe the static structure of the system of particles. First of all, the Radial Distribution Function (RDF) is evaluated at different values of parameters defining the system. The RDF is defined as, 
\begin{equation}
	g(r) = \frac{1}{n_d N}\Big<\sum_{i=1}^{N} \sum_{j\neq i}^{N} \delta(r-r_{ij})\Big>,
	\label{RDF-definition}
\end{equation}
where, $n_d=N/L^3$ denotes the number density of the dust particles. $g(r)$ is proportional to the probability of finding a dust particle from another particle at a distance $r\pm \Delta r$, where $\Delta r$ is a differential increment in the distance.  The RDF as it is evaluated here, cannot distinguish between radial and angular correlation. To distinguish between the radial and angular correlations present in the cluster a two-particle correlation function known as the center-two-particle correlation function (C2P) has been used here which was developed by Thomsen et al.\cite{thomsen2015resolving}. This is a two particle correlation function of radial positions with respect to the center of the cluster and the angle between the radial coordinates. It is defined as the ratio of the correlated two particle
density to the uncorrelated or ideal two particle denstity,

\begin{equation}
	f(r_{1,}r_{2,}\theta)=\frac{\rho^{corr}_2(r_{1,}r_{2,}\theta)}{\rho^{id}_2(r_{1,}r_{2,}\theta)},
\end{equation}

where,

\begin{equation}
	\rho^{id}_2(r_{1,}r_{2,}\theta)=8r_{1}^{2}r_{2}^{2}\sin(\theta)\rho(r_{1})\rho(r_{2}).
\end{equation}

In the above $\rho^{corr}_2(r_{1,}r_{2,}\theta)$ is sampled from simulations
and $\rho(r)$ is the single particle radial densitiy. $\rho^{id}_2(r_{1,}r_{2,}\theta)$
is the ideal pair density of the same cluster if it is filled homogeneously
with shells about the trap center. The intra-shell angular correlation is then obtained from the C2P by integrating out the radial coordinates $r_1$ and $r_2$ over the range of a shell of the cluster.

To understand the dynamic structure of the cluster we use the self part of Van Hove autocorrelation function which is defined as \cite{haile1992molecular}, 
\begin{equation}
	G_s(r,t) = \frac{1}{N}\Big<\sum_{i=1}^{N}\delta(r-|\mathbf{r}_i(0)-\mathbf{r}_i(t)|)\Big>.
	\label{vanHoveDefinition}
\end{equation}

This function is proportional to the probability of finding a particle at distance $r$ at time $t$ given that the particle was at the origin at $t=0$. In the context of our simulation, the angle bracket $\Big<..\Big>$ denotes average over a sufficiently large number of time origins. While calculating $G_s(r,t)$ the bin width is chosen as $\Delta r = 0.001\;\lambda_d.$

A simulation run lasts for $5.5\times10^{6}$ time steps and the value of a time step is chosen as $4\times10^{-4}\; t_0$. The equilibration period lasts for $5\times10^{5}$ time steps and the data for calculating the static and dynamic quantities are recorded for the next $5\times10^{6}$ time steps. 

%%%%%%%%%%
\section{Results and Discussion}
\begin{figure}
	\centering
	\includegraphics[width=0.7\linewidth]{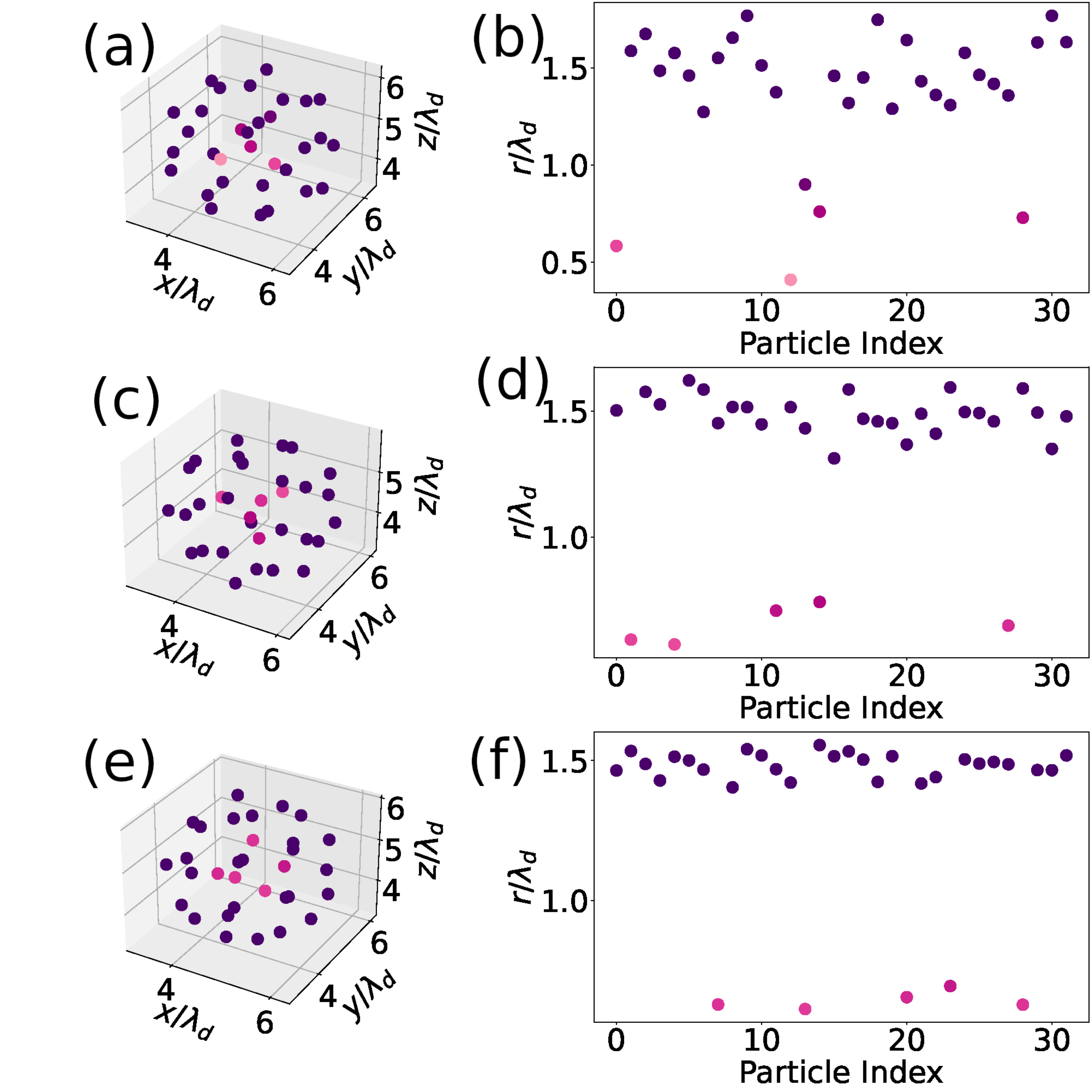}
	\caption{(Color online) Snapshots of particle positions along with their radial distances from the center of the cluster having $N=32$ particles at different values of Coulomb coupling parameter $\Gamma$ obtained using fMD simulation. In the figure, $\Gamma=5.14$ for (a) and (b), $\Gamma=51.43$ for (c) and (d) and $\Gamma=1543.14$ for (e) and (f). For all the cases $\kappa=1.8$.}
	\label{particle-snapshots}
\end{figure}

The particles in the harmonically confined 3D cluster arrange themselves into nested spherical shells. A few snapshots of the particle positions obtained from fMD simulations at different values of Coulomb coupling parameter for a cluster having $N=32$ particles are shown along with the radial distances of the particles from the center of the harmonic trap in Fig. \ref{particle-snapshots}. It is seen that the thickness of the spherical shells decreases with increase in Coupling strength.
\subsection{Static structural properties}
To get an idea of the static structure of the cluster we obtain RDF, C2P and angular correlation functions and plotted them at different values of coupling and screening parameters. We obtain these properties using both frictionless Molecular Dynamics and Langevin Dynamics simulation. It is expected that the static properties doesn't change with the dynamics considered.
\subsubsection{Frictionless Molecular Dynamics}
%\subsubsection{Changes with respect to $\Gamma$}

\begin{figure}[htbp]
	\centering
	\includegraphics[width=0.3\linewidth]{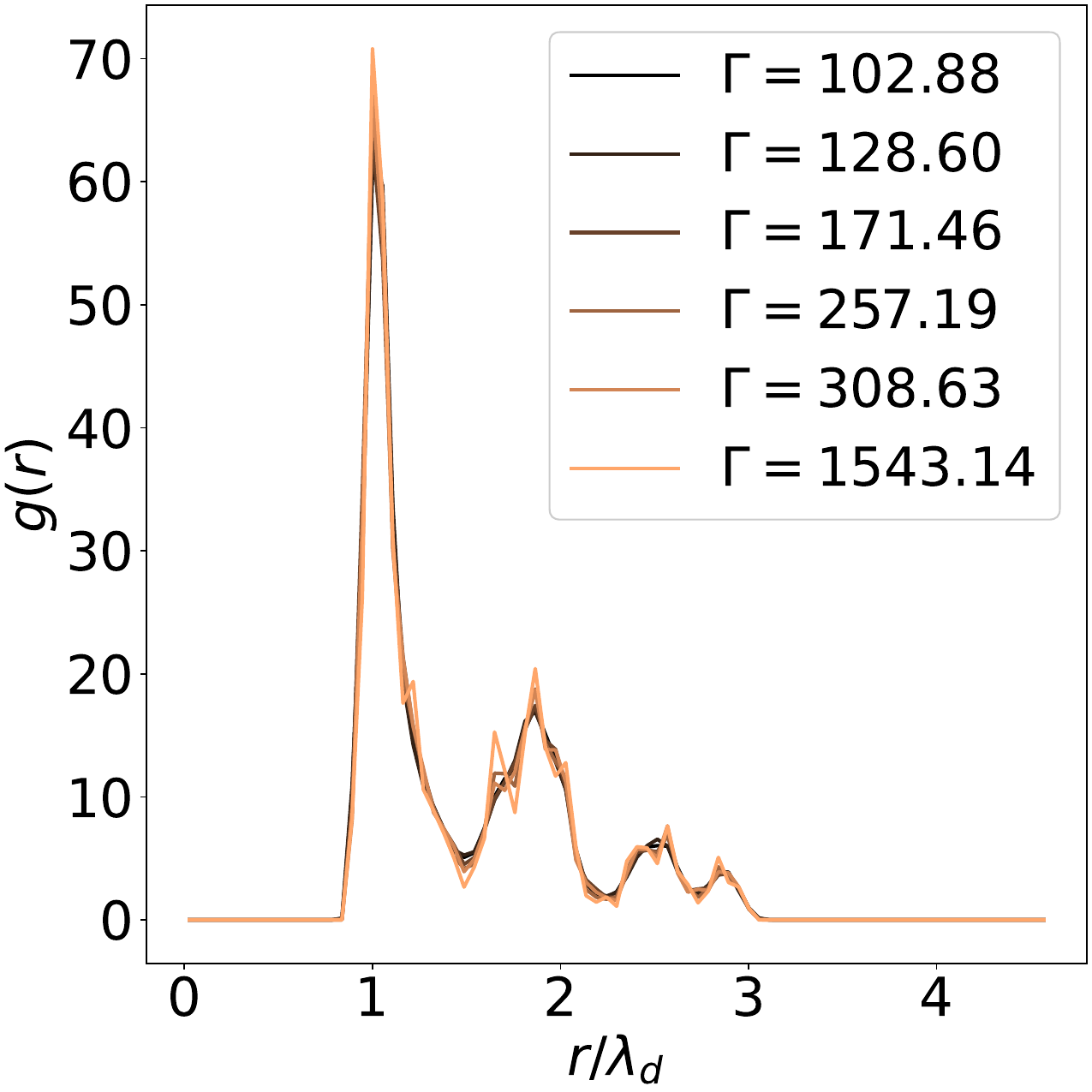}
	\caption{(Color online) Radial distribution function of particles at different values of coupling strengths at fixed values of screening parameter $\kappa=1.8$  for a cluster having $N=32$ particles.}
	\label{rdf-wg}
\end{figure}

\begin{figure}[htbp]
	\centering
	\includegraphics[width=0.3\linewidth]{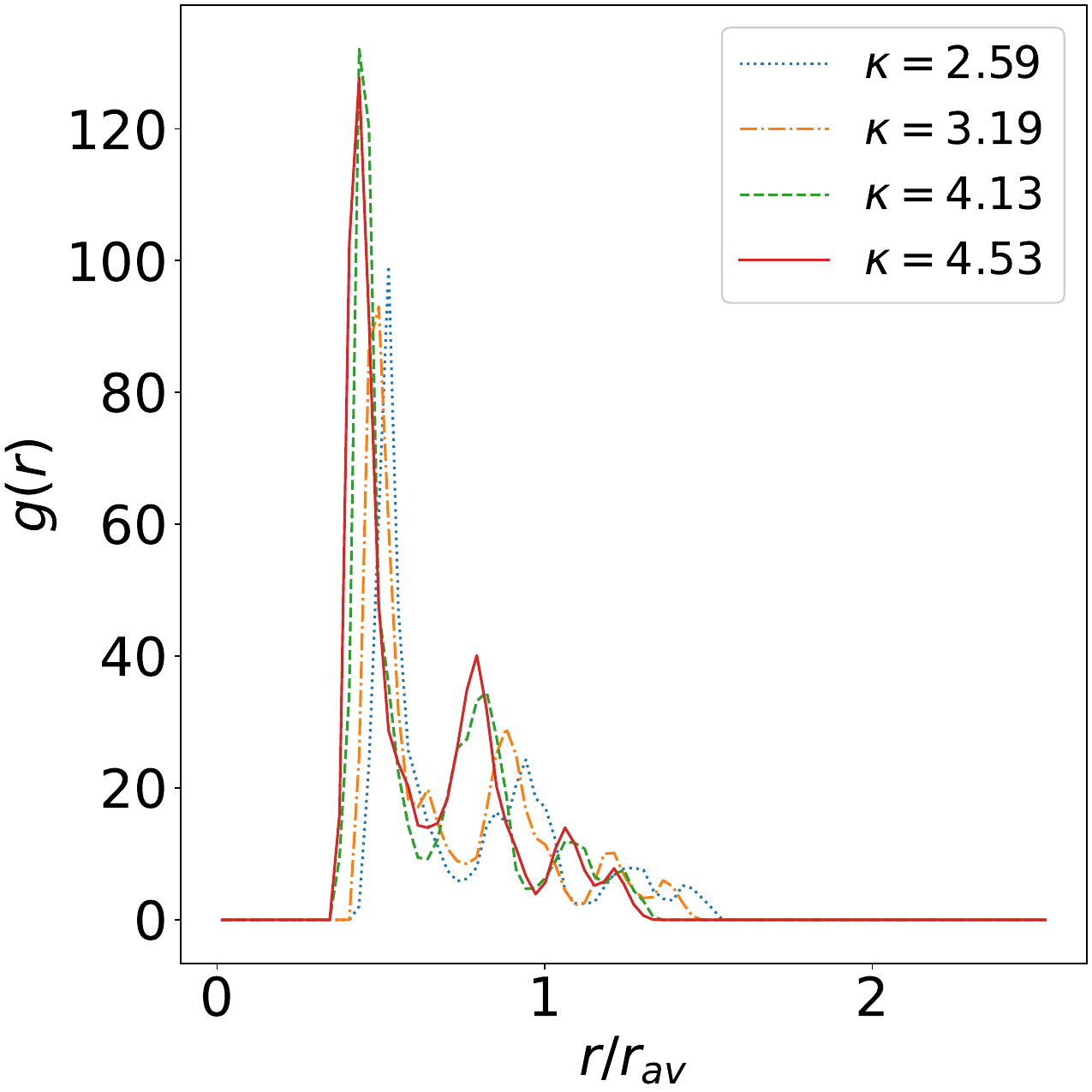}
	\caption{(Color online) Radial distribution functions at different values of screening parameter for a fixed value of coupling strength $\Gamma=257.19.$ for a cluster of $N=32$ particles.}
	\label{RDF-wk}
\end{figure}
In order to understand the effect of variation of coupling parameter on the static structure of the dust cluster we have plotted the Radial Distribution Function. 
Fig. \ref{rdf-wg} shows the Radial Distribution Functions evaluated at different values of coupling parameter keeping the screening parameter and total number of particles fixed. The RDFs suggest that the system remains in a liquid like state upto a certain value of coupling parameter $\Gamma$ and then on further increasing the coupling parameter the system exhibits order as in a partially crystallized state.

\begin{figure}[htbp]
	\begin{subfigure}{0.45\columnwidth}
		\includegraphics[width=0.5\linewidth]{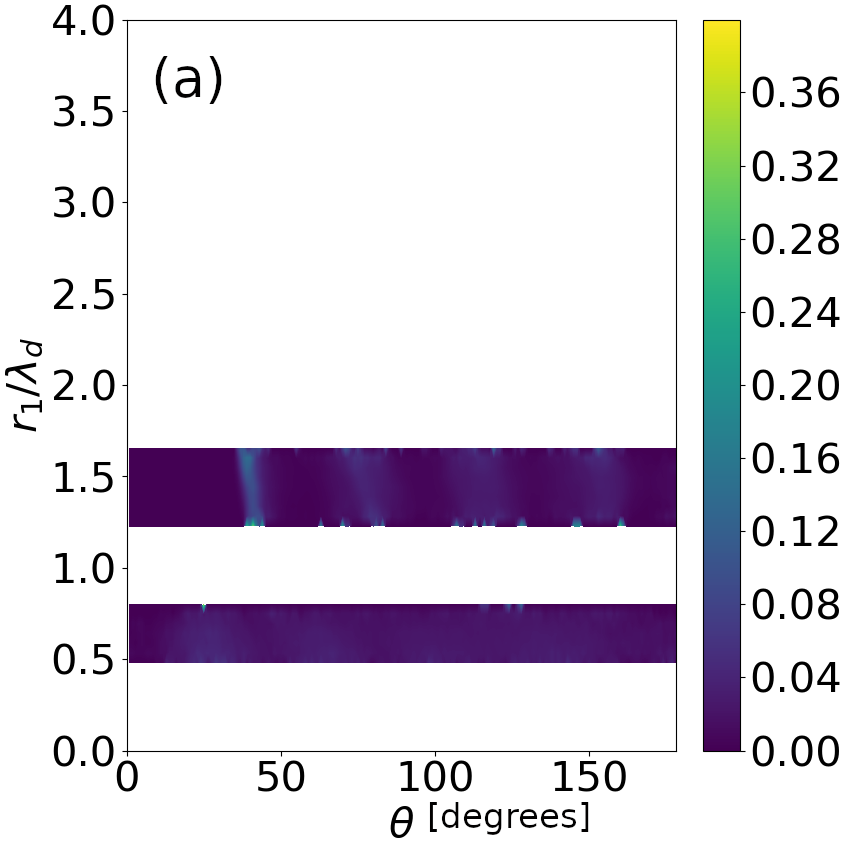}
		
	\end{subfigure}
	\begin{subfigure}{0.45\columnwidth}
		\includegraphics[width=0.5\linewidth]{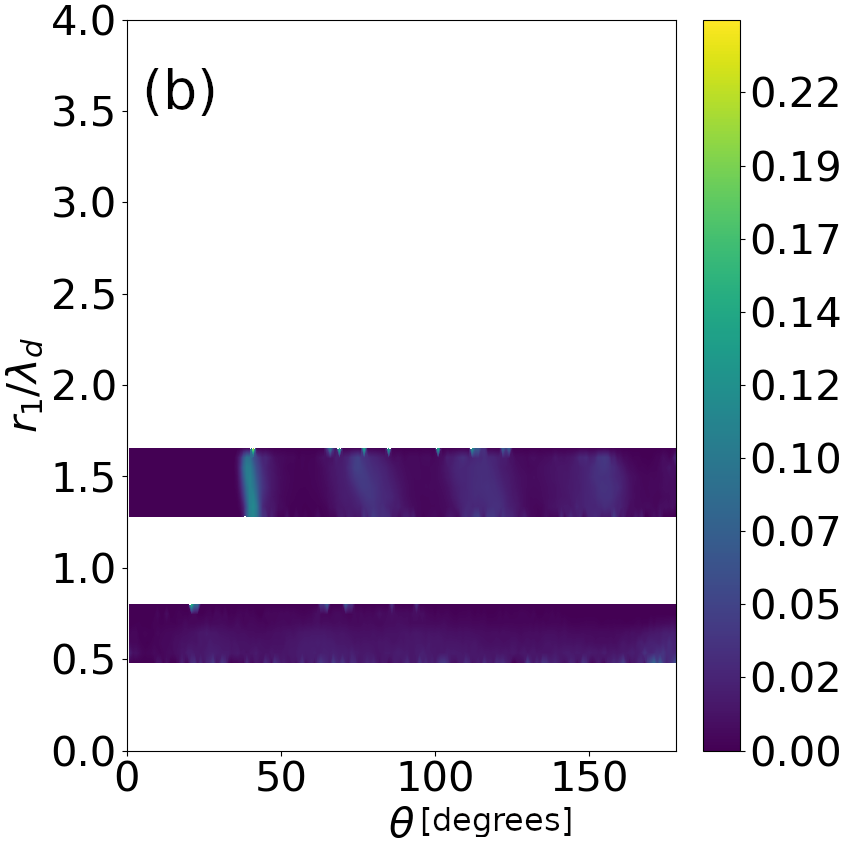}
		
	\end{subfigure}
	\begin{subfigure}{0.45\columnwidth}
		\includegraphics[width=0.5\linewidth]{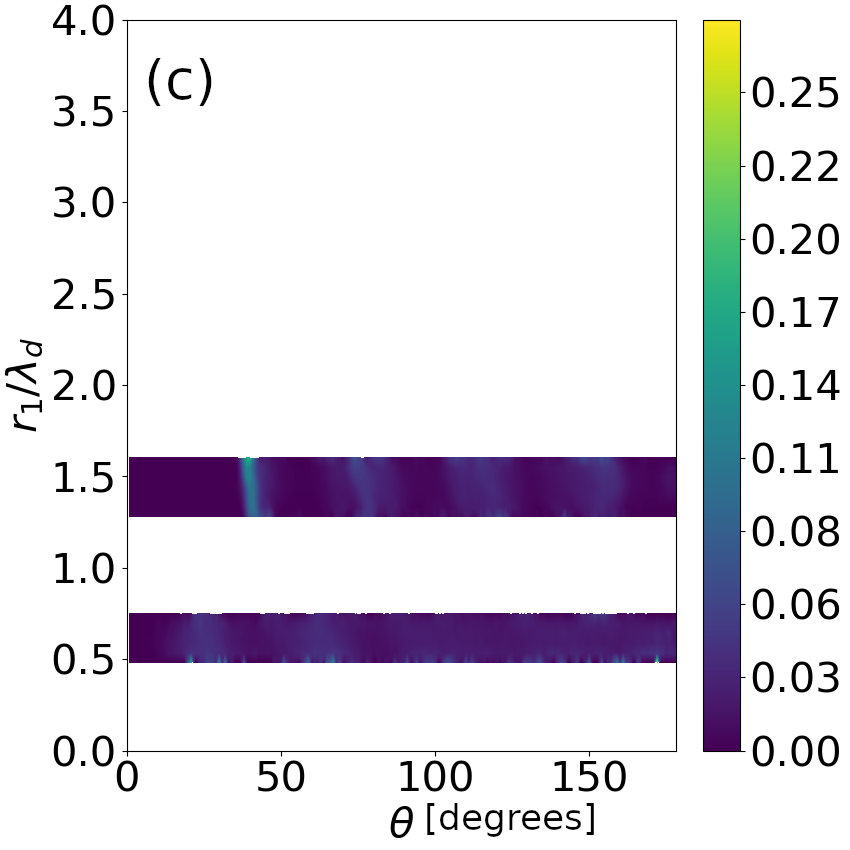}
		
	\end{subfigure}
	\begin{subfigure}{0.45\columnwidth}
		\includegraphics[width=0.5\linewidth]{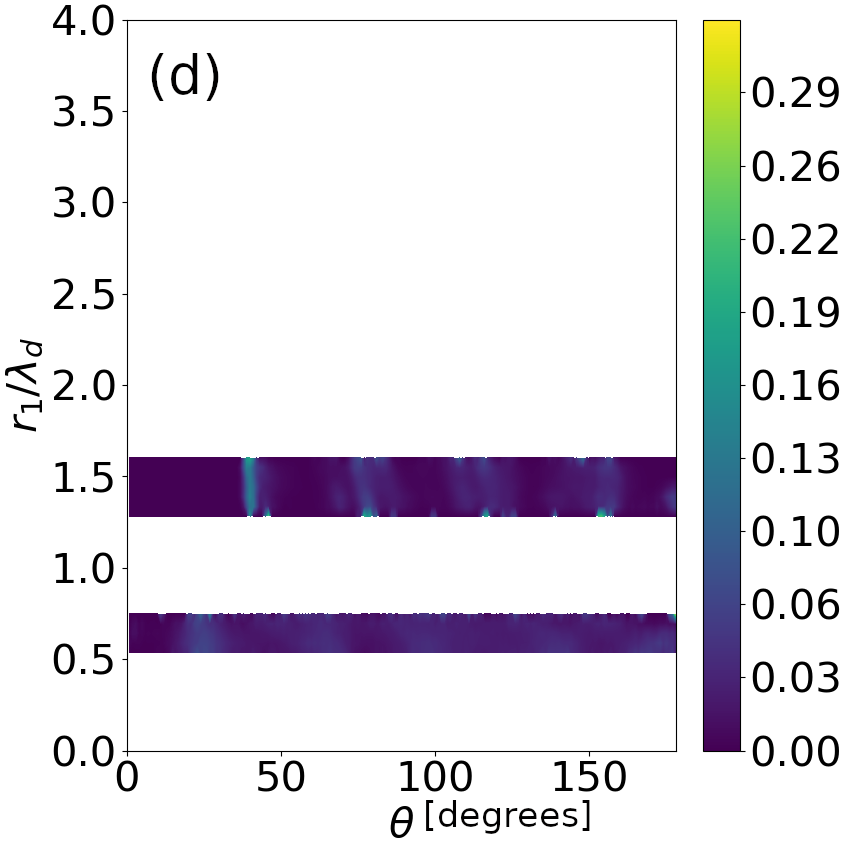}
	\end{subfigure}	
	\caption{(Color online) C2P at different values of coupling strengths (a) $\Gamma=102.88$, (b) $\Gamma=128.60$, (c) $\Gamma=171.46$, (d) $\Gamma=257.19$ by keeping screening parameter fixed as $\kappa=1.8$ for a cluster having $N=32$ particles.}
	\label{c2pWG_withoutFric}
\end{figure}

\begin{figure}[htbp]
	\centering
	\includegraphics[width=0.3\linewidth]{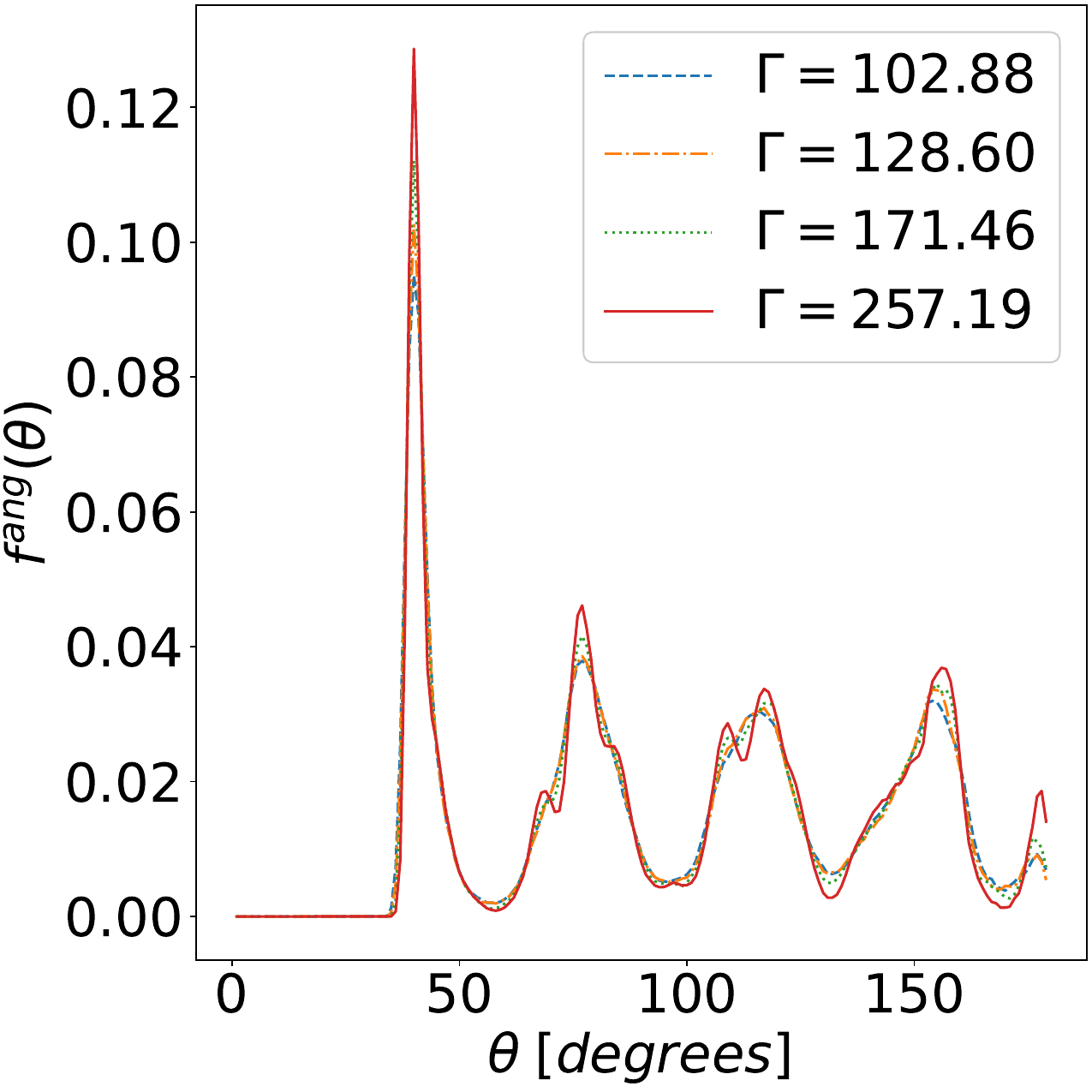}
	\caption{(Color online) Intra-shell angular correlation function at different values of coupling strengths at $\kappa=1.8$ for a cluster of $N=32$ particles. }
	\label{AngCorrWg_withoutFric}
\end{figure}

\begin{figure}[htbp]
	\begin{subfigure}{0.45\columnwidth}
		\includegraphics[width=0.5\linewidth]{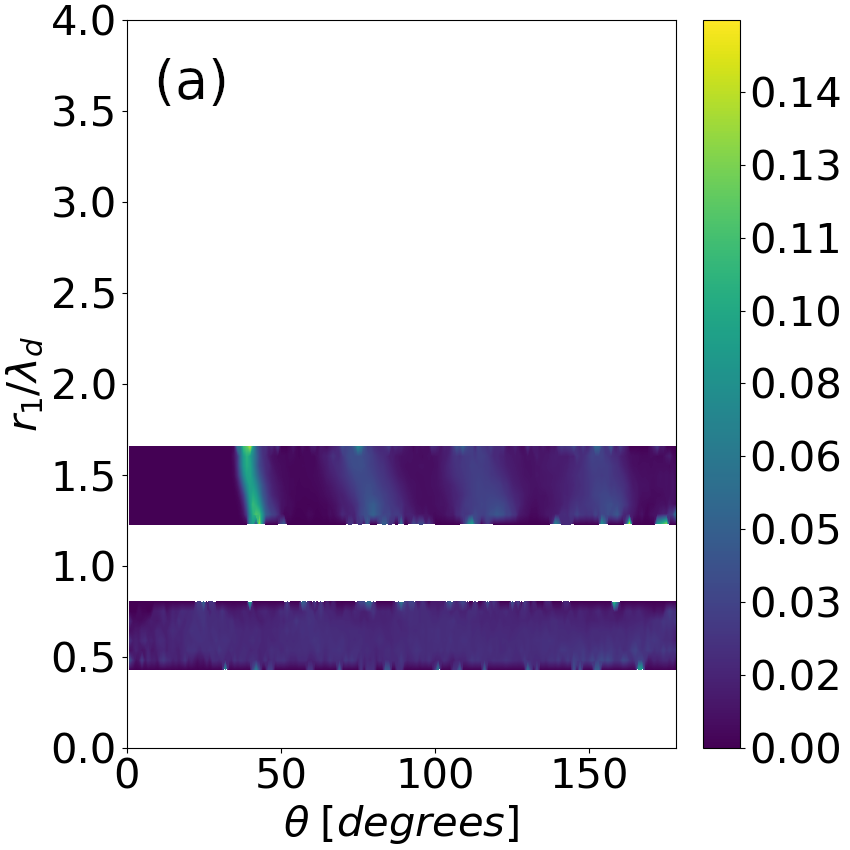}
		
	\end{subfigure}
	\begin{subfigure}{0.45\columnwidth}
		\includegraphics[width=0.5\linewidth]{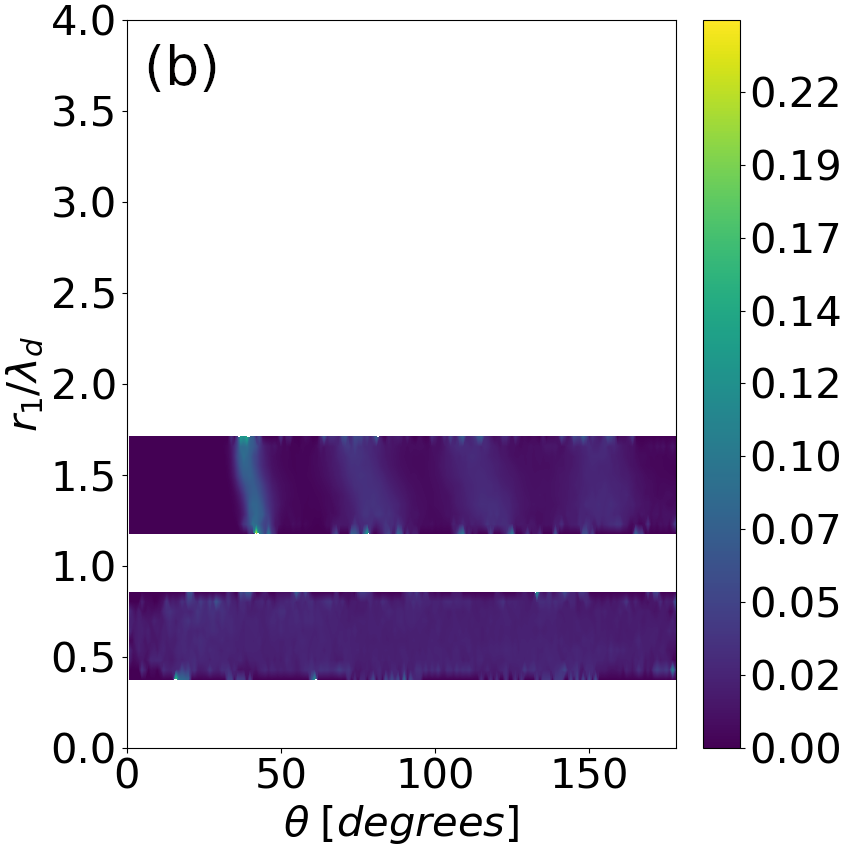}
		
	\end{subfigure}
	\begin{subfigure}{0.45\columnwidth}
		\includegraphics[width=0.5\linewidth]{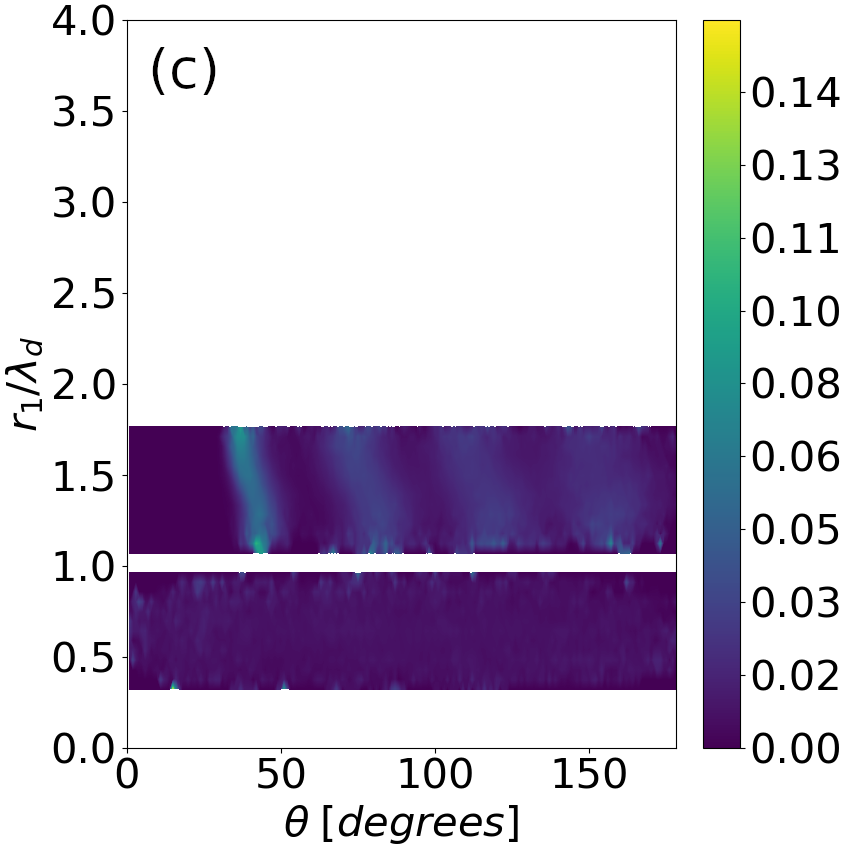}
		
	\end{subfigure}
	\begin{subfigure}{0.45\columnwidth}
		\includegraphics[width=0.5\linewidth]{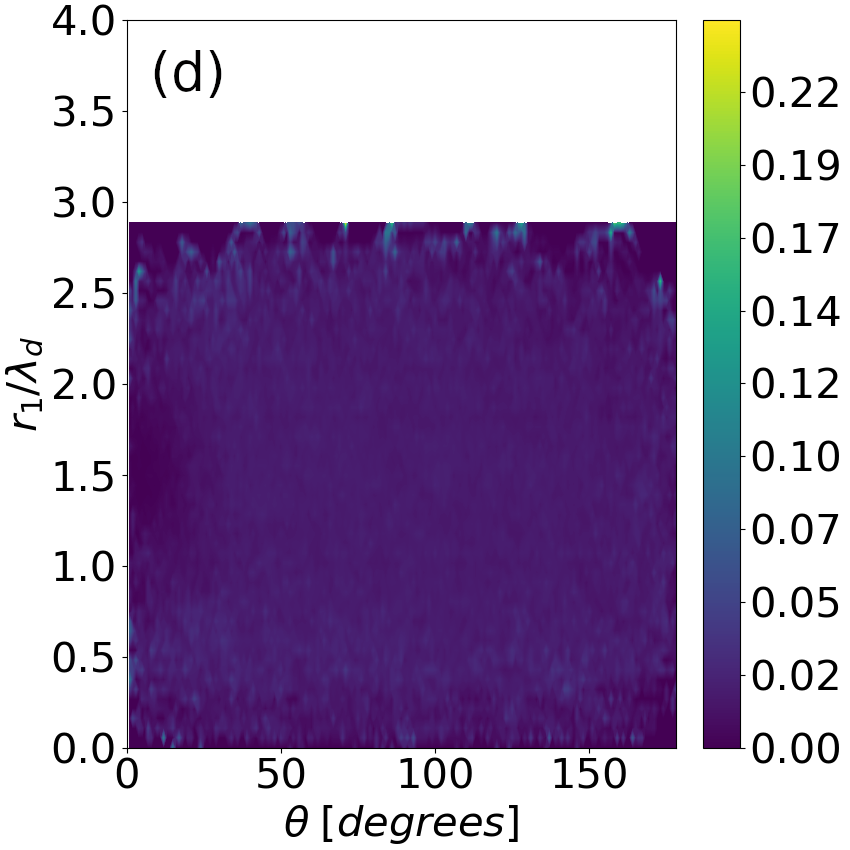}
	\end{subfigure}	
	\caption{(Color online) C2P at different values of coupling strengths (a) $\Gamma=77.16$, (b) $\Gamma=38.16$, (c) $\Gamma=19.29$, (d) $\Gamma=0.86$ for $\kappa=1.8.$ for a cluster having $N=32$ particles.}
	\label{c2pWG_withoutFric2}
\end{figure}

One of the important parameters of a Yukawa system is the screening parameter. The radial distribution functions at different values of screening parameter $\kappa$ by keeping $\Gamma=257.19$ and total number of particles $N=32$ are shown in Fig. \ref{RDF-wk}. The first peak of the RDFs shifts towards left with increase in $\kappa$. This is due to the fact that with increase in the screening parameter, the Yukawa repulsion among the particles reduces and the screening independent  attractive harmonic potential brings the particles to closer interparticle separation.
In Fig. \ref{c2pWG_withoutFric} C2P is plotted by integrating out the second radial coordinate $r_{2}$ from $f(r_{1,}r_{2,}\theta)$. The integration is done over the range of the outer shell. Fig. \ref{c2pWG_withoutFric} shows the C2P plots for different coupling strengths at fixed value of screening parameter $\kappa$ and total number of particles. The two strips seen in the Fig. \ref{c2pWG_withoutFric}
corresponds to the particles located in two distinct shells. The upper strip represents intra-shell correlation among the particles of outer shell whereas, the lower strip represents inter-shell correlation among
the particles of inner and outer shell. The dark shade in blue at angular separation around $40^{\circ}$ and radial distance 
$\sim 1.5$ (in units of $\lambda_{d}$) indicates a strong correlation.
Relatively weaker correlation is maintained by the particles at around angles $80^{\circ}$, $120^{\circ}$ and $160^{\circ}$. This result is in agreement with the result of the intra-shell angular correlation of Fig. \ref{AngCorrWg_withoutFric}. The angular correlation is obtained from the integrated C2P by integrating out the radial coordinate $r_1$ over the range of the outer shell. From the angular correlation results it is seen that, at $\Gamma=257.19$ substructure appears in the second and third peaks which indicates a structural transition in the cluster. A still weaker inter-shell correlation is seen in the lower strip. As the temperature is increased or $\Gamma$ is lowered, merging of the two shells is seen at around $\Gamma=19.29$ accompanied by the exchange of the particles in the two shells as shown in Fig. \ref{c2pWG_withoutFric2}. 

\begin{figure}[htbp]
	\begin{subfigure}{0.45\columnwidth}
		\includegraphics[width=0.5\linewidth]{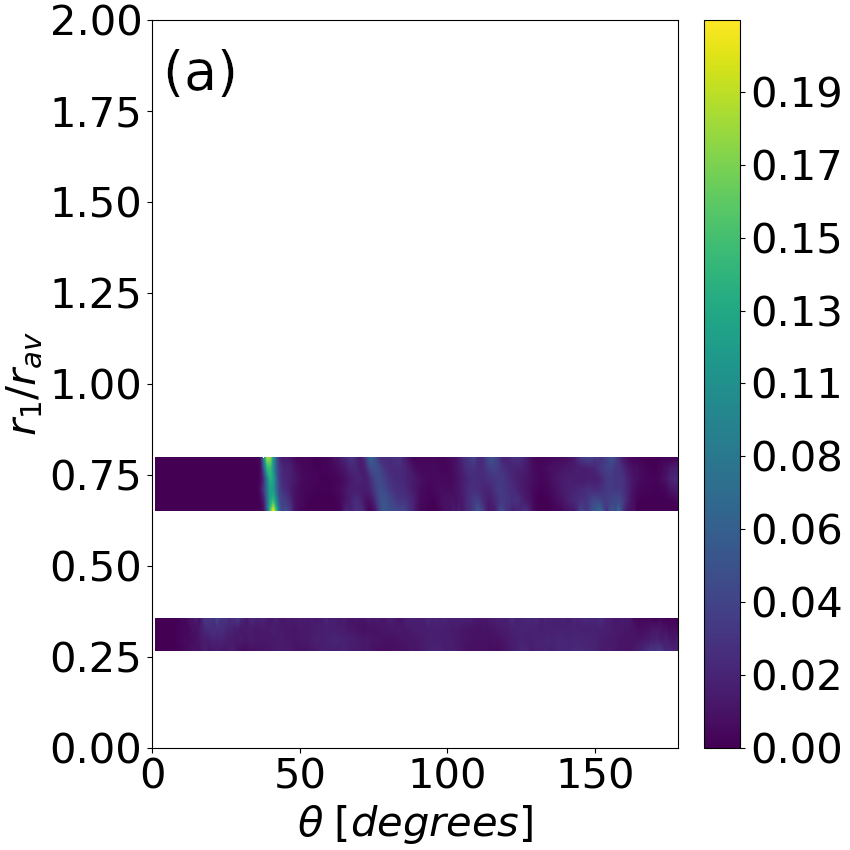}
		
	\end{subfigure}
	\begin{subfigure}{0.45\columnwidth}
		\includegraphics[width=0.5\linewidth]{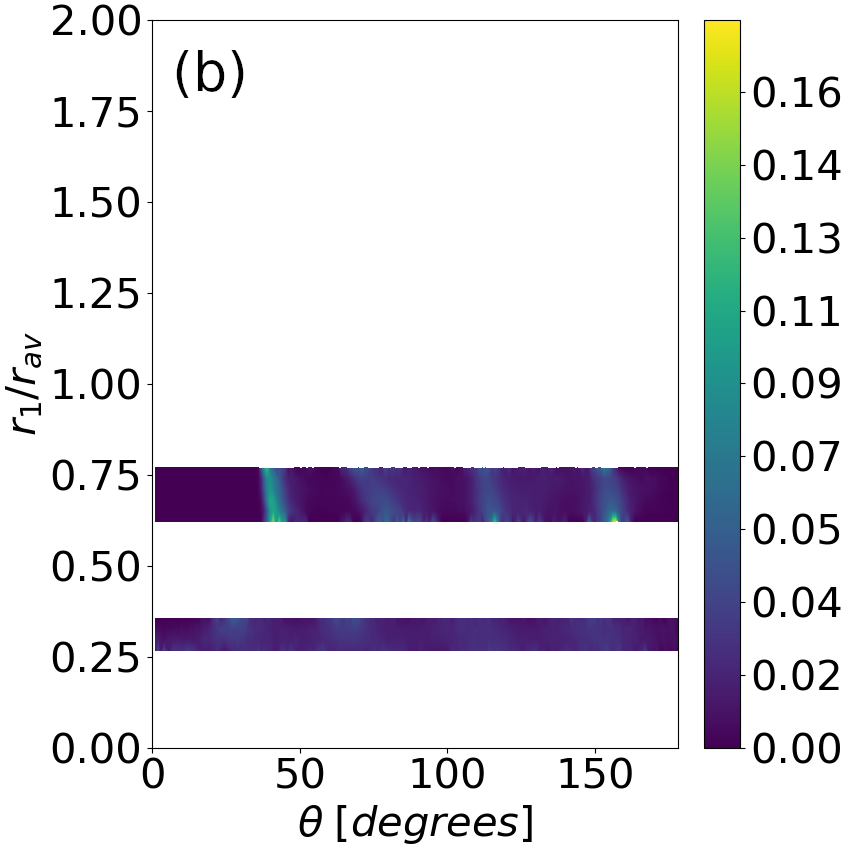}
		
	\end{subfigure}
	\begin{subfigure}{0.45\columnwidth}
		\includegraphics[width=0.5\linewidth]{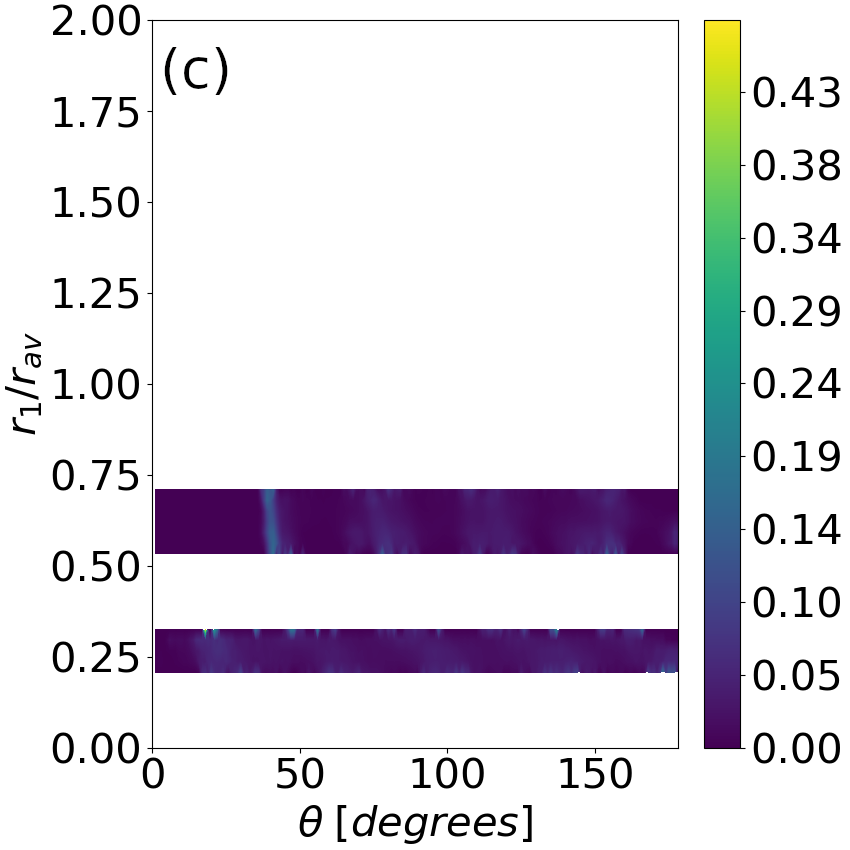}
		
	\end{subfigure}
	%\hspace{0.7cm}
	\begin{subfigure}{0.45\columnwidth}
		\includegraphics[width=0.5\linewidth]{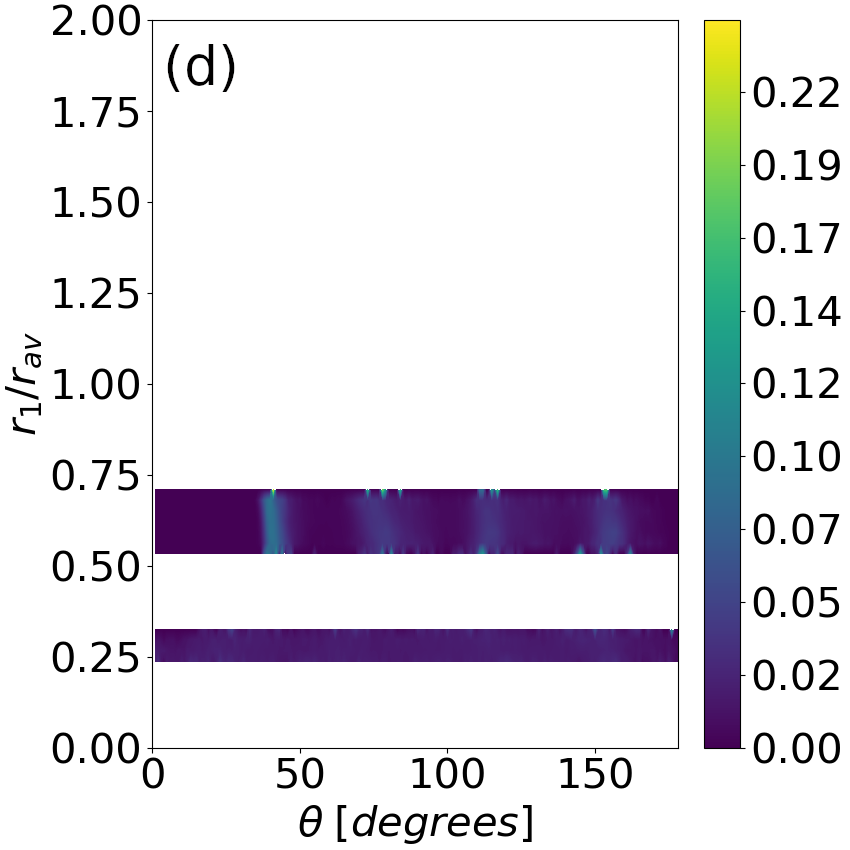}
	\end{subfigure}	
	\caption{(Color online) C2P at different values of screening parameters (a) $\kappa=2.59$, (b) $\kappa=3.19$, (c) $\kappa=4.13$, (d) $\kappa=4.53$ for $\Gamma=257.19$ and $N=32.$}
	\label{c2p_wK_withoutFric}
\end{figure}

\begin{figure}[htbp]
	\centering
	\includegraphics[width=0.3\linewidth]{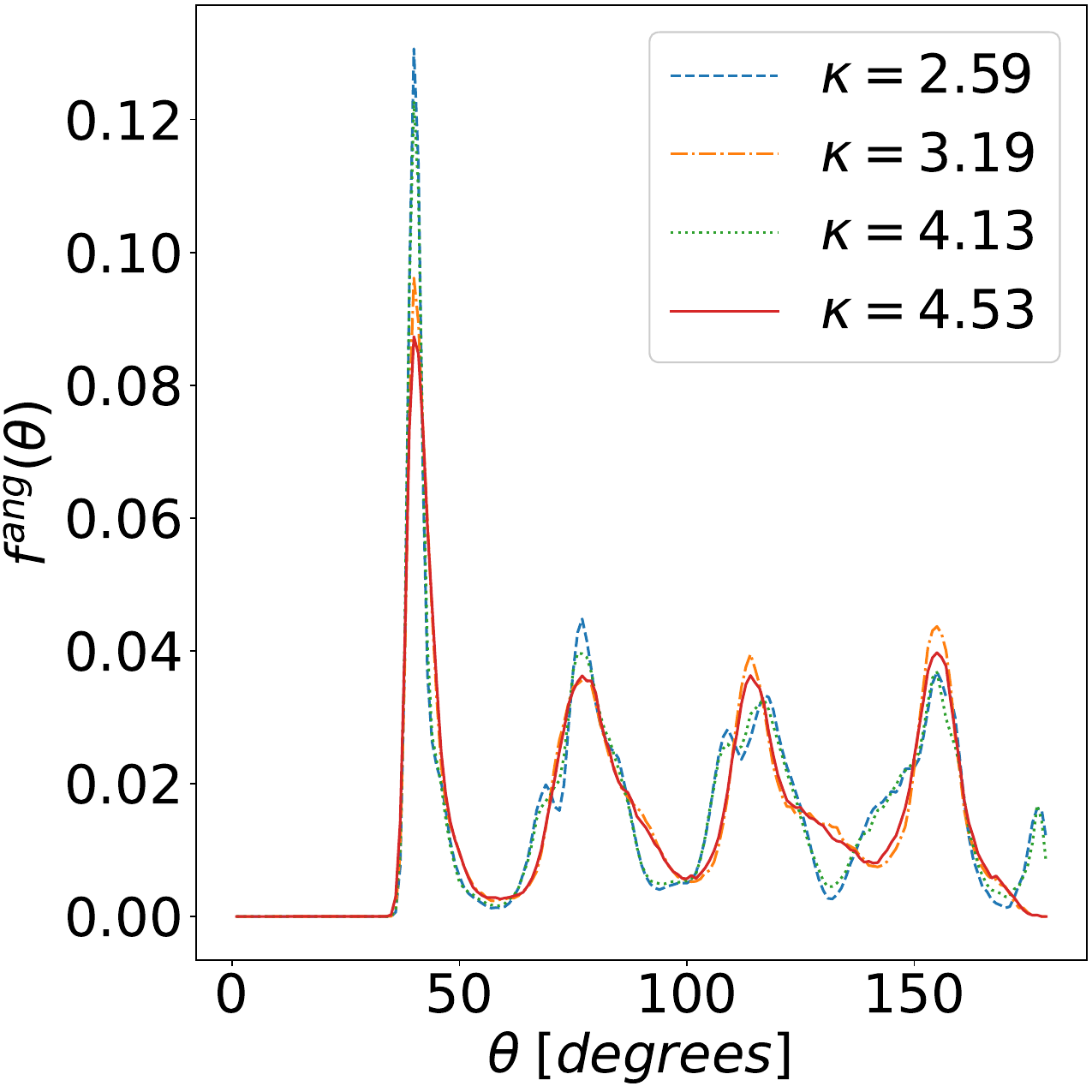}
	\caption{(Color online) Intra-shell angular correlation function at different values of screening parameter at a fixed value of coupling strength $\Gamma=257.19$ for a cluster with $N=32.$} 
	\label{Angcorr-wK-withoutFric}
\end{figure}

The center-two-particle correlation functions at different values of screening parameter and the corresponding intra-shell angular correlation functions are shown in Fig. \ref{c2p_wK_withoutFric} and \ref{Angcorr-wK-withoutFric} respectively. The C2P shows both intra-shell correlation among the outer shell particles (the outer strip ) and inter-shell correlation among the particles from the outer and inner shells (the inner strip). It is to be noted that the distances in the C2P plots of Fig. \ref{c2p_wK_withoutFric} is scaled by $r_{av}$ instead of $\lambda_d$. This has been done to enable valid comparison between the figures as $\kappa$ is changed by changing $\lambda_d$. It is seen from the figures that with increase in screening, both the strips appear at smaller distances from the trap center indicating a decrease in the radii of the cluster shells. Bonitz et al. observed a reduction of the shell roughness on increasing screening parameter and fixed dust kinetic temperature which was found to be significant for the inner shell \cite{bonitz2006structural}. From the angular correlation plots in Fig. \ref{Angcorr-wK-withoutFric} it is seen that, the 2nd and 4th minima in the angular correlation plots tend to shift towards large angles for $\kappa=3.19$ and $\kappa=4.53$. 

%\subsubsection{Changes with respect to $\kappa$}
\subsubsection{Langevin Dynamics}

\begin{figure}[htbp]
	\begin{subfigure}{0.45\columnwidth}
		\includegraphics[width=0.5\linewidth]{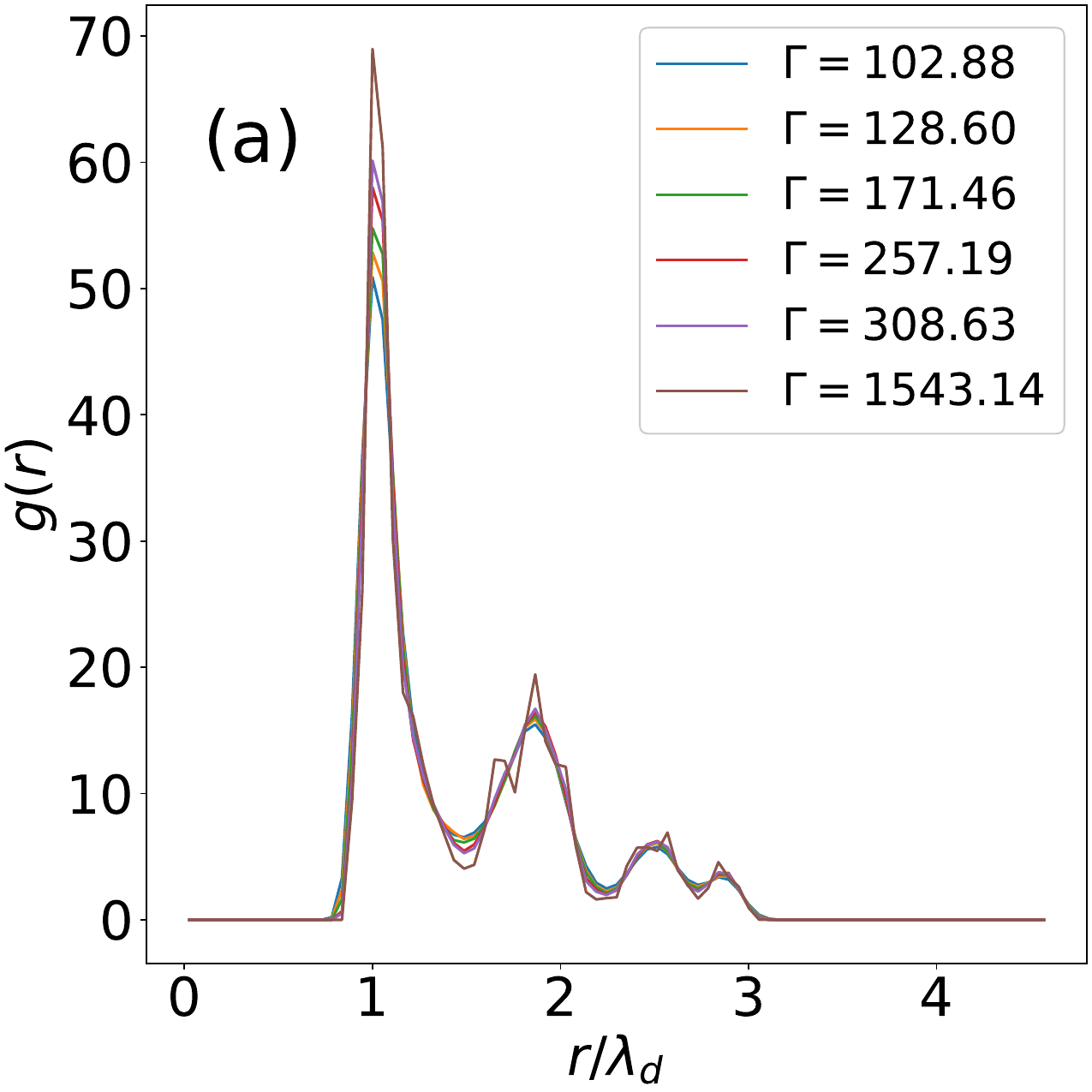}
		
	\end{subfigure}
	\begin{subfigure}{0.45\columnwidth}
		\includegraphics[width=0.5\linewidth]{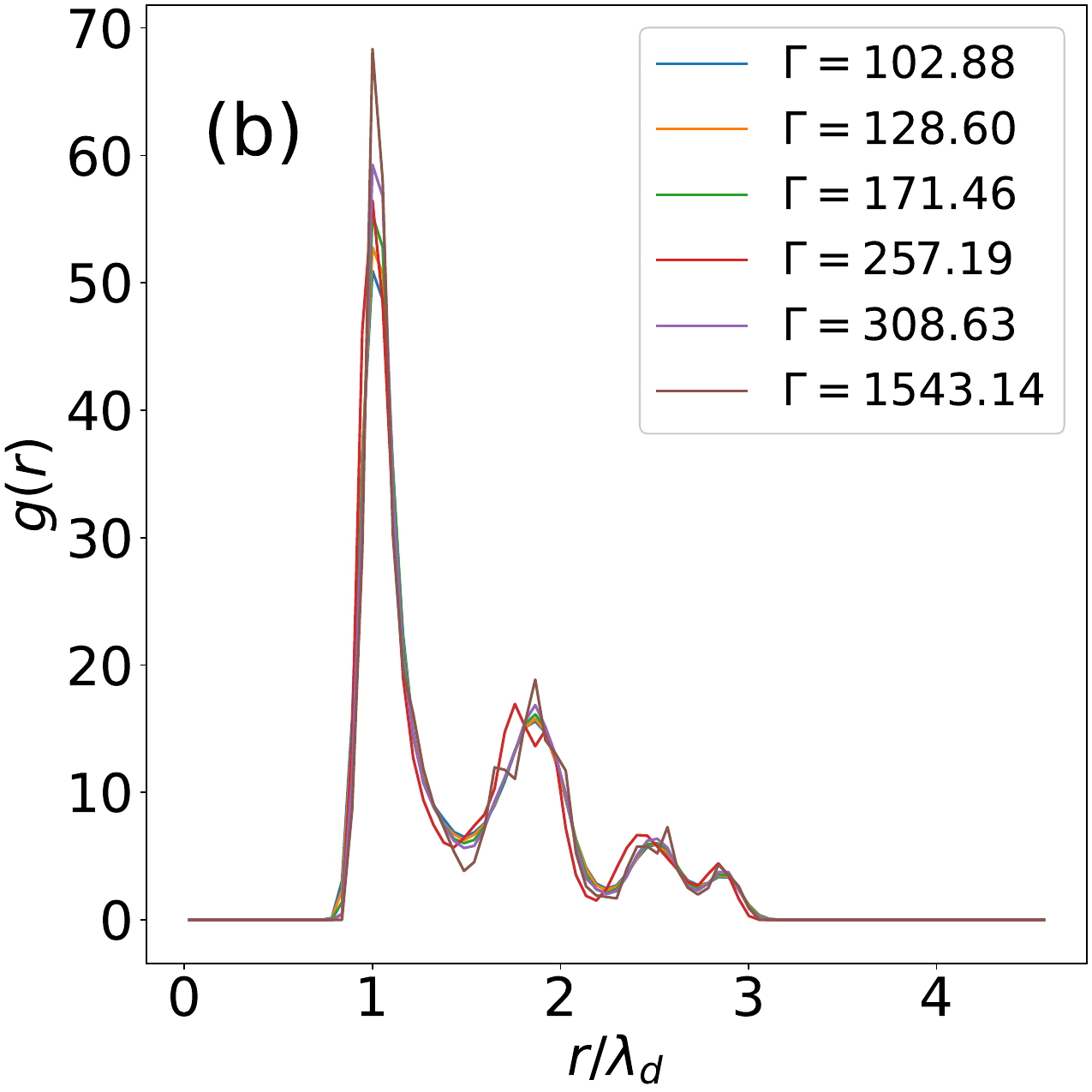}
		
	\end{subfigure}
	\caption{(Color online)RDF of a $N=32$ particle cluster at different values of coupling strengths at two different friction coefficients (a) $\nu=0.3\;Hz$, (b) $\nu=3\;Hz$ at fixed value of screening parameter $\kappa=1.8$.}
	\label{RDF_wFric_wG}
\end{figure}
The RDF of the cluster at different values of coupling strenghts at a fixed value of screening parameter is shown in Fig. \ref{RDF_wFric_wG} for two different values of dust-neutral collision frequency. Similar to the Newtonian Dynamics case at a large enough value of coupling strength ($\Gamma$) the second peak of the RDF starts showing substructure which may be an indication of glassy behaviour. 
\begin{figure}[htbp]
	\includegraphics[width=0.3\linewidth]{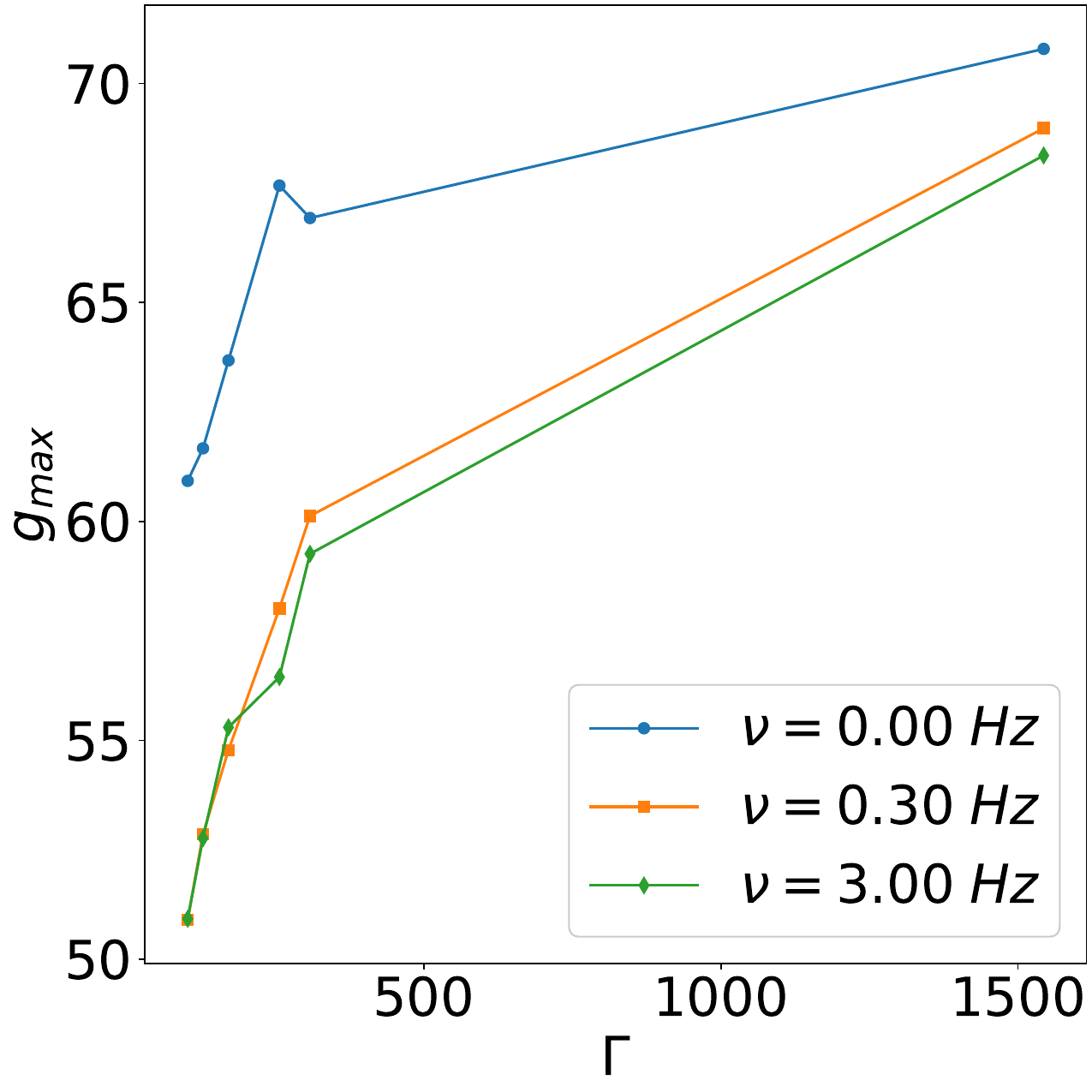}
	\caption{(Color online) First peakheight of the Radial Distribution Function as a function of Coulomb coupling parameter at different values of dust-neutral collision frequencies.}
	\label{MAX_RDF}
\end{figure}
The height of the first peak of the RDF as a function of Coulomb coupling parameter is shown in Fig. \ref{MAX_RDF} for three different dust-neutral collision frequencies. It is seen that for fMD, the first peak height is greater for all the values of coupling parameter suggesting a larger nearest neighbour coordination number than for LD.
\begin{figure}[htbp]
	\begin{subfigure}{0.45\columnwidth}
		\includegraphics[width=0.5\linewidth]{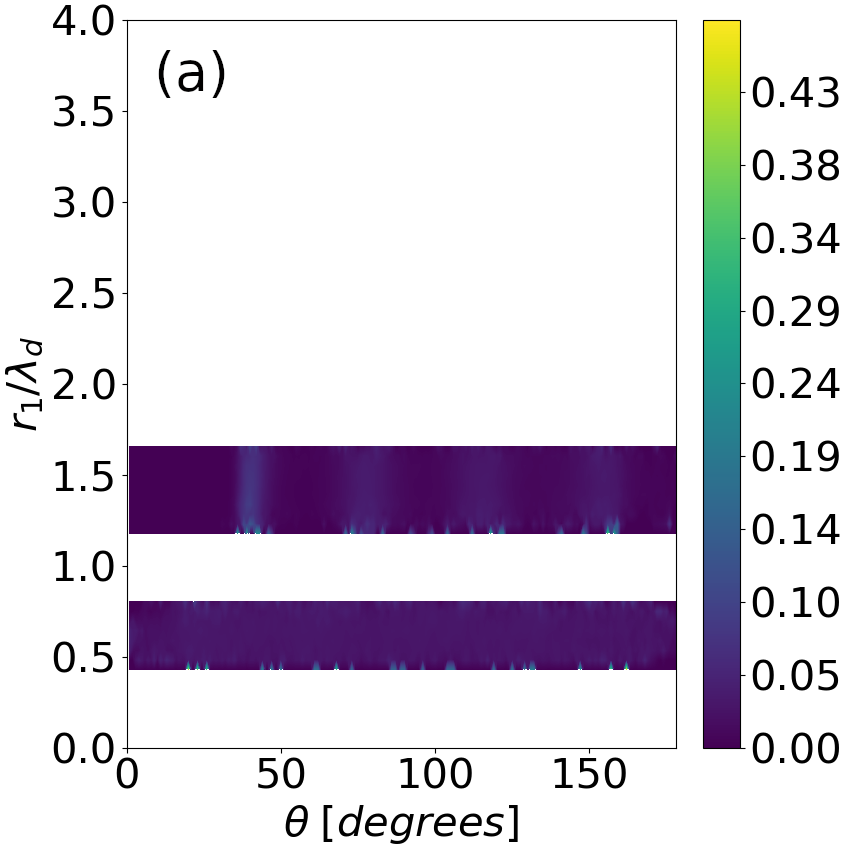}
		
	\end{subfigure}
	\begin{subfigure}{0.45\columnwidth}
		\includegraphics[width=0.5\linewidth]{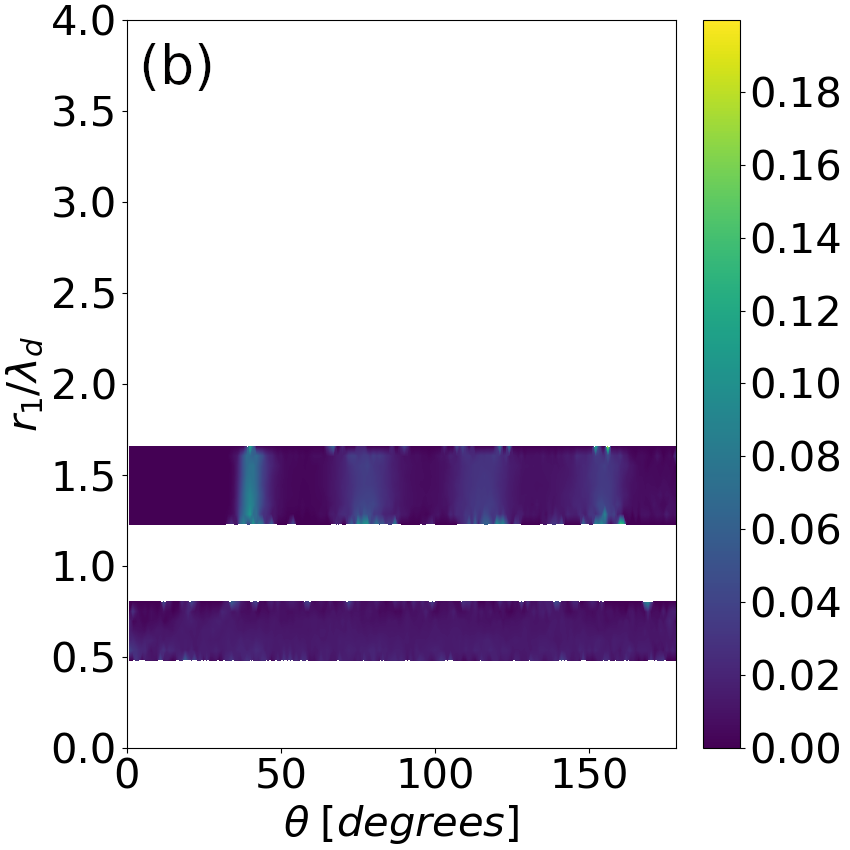}
		
	\end{subfigure}
	\begin{subfigure}{0.45\columnwidth}
		\includegraphics[width=0.5\linewidth]{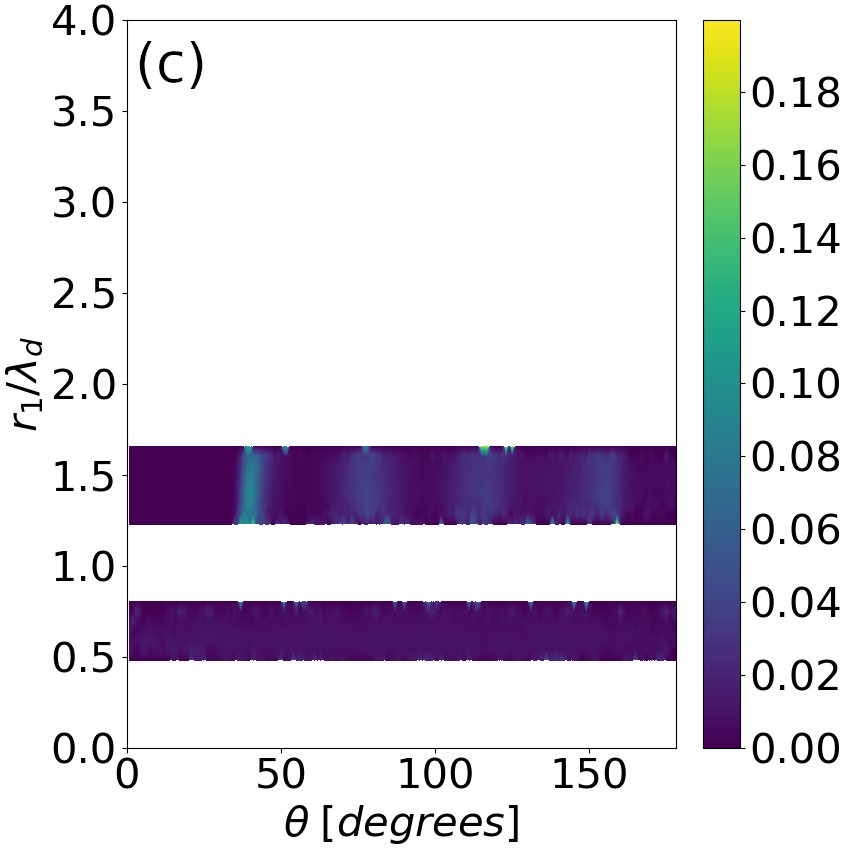}
		
	\end{subfigure} 
	%\hspace{1.6cm}
	\begin{subfigure}{0.45\columnwidth}
		\includegraphics[width=0.5\linewidth]{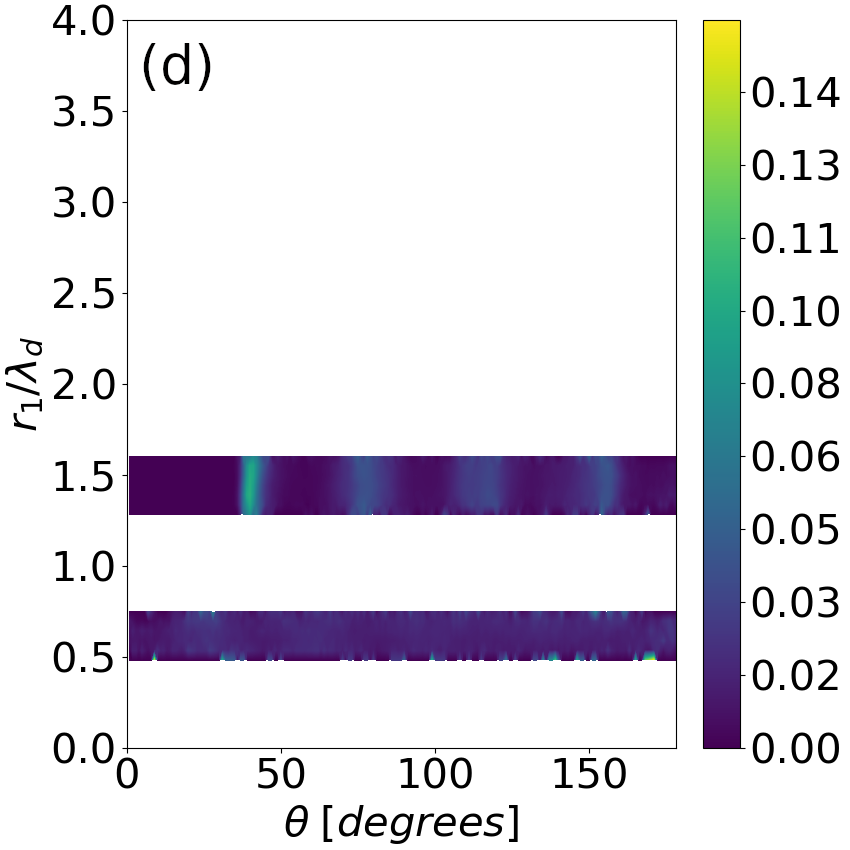}
	\end{subfigure}
	\caption{(Color online)C2P at four different values of coupling strengths (a) $\Gamma=102.88$, (b) $\Gamma=128.60$, (c) $\Gamma=171.46$, (d) $\Gamma=257.19$ for $\kappa=1.8$ and $\nu=0.3\;Hz$}
	\label{c2p_Nu0.3_wG}
\end{figure}

\begin{figure}
	\begin{subfigure}{0.45\columnwidth}
		\includegraphics[width=0.5\linewidth]{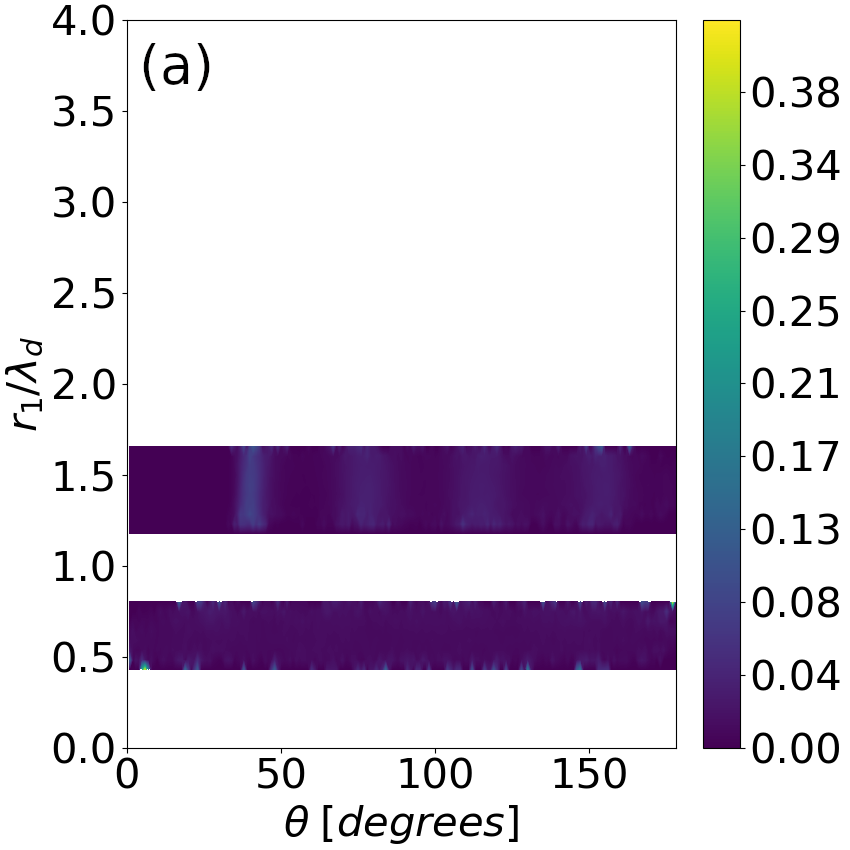}
		
	\end{subfigure}
	\begin{subfigure}{0.45\columnwidth}
		\includegraphics[width=0.5\linewidth]{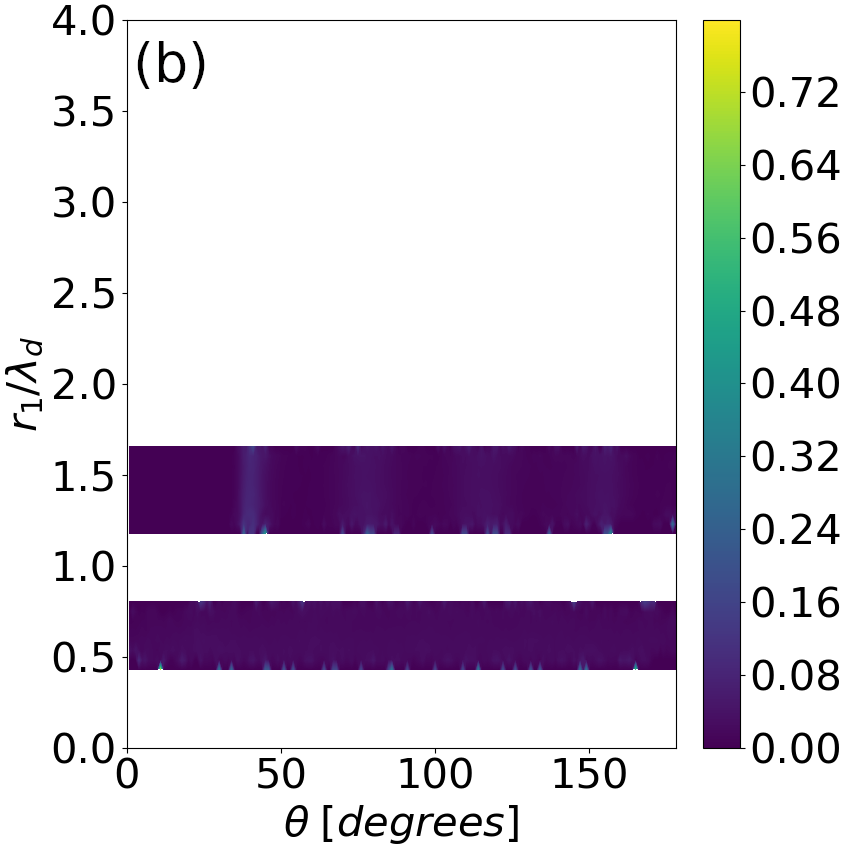}
		
	\end{subfigure}
	\begin{subfigure}{0.45\columnwidth}
		\includegraphics[width=0.5\linewidth]{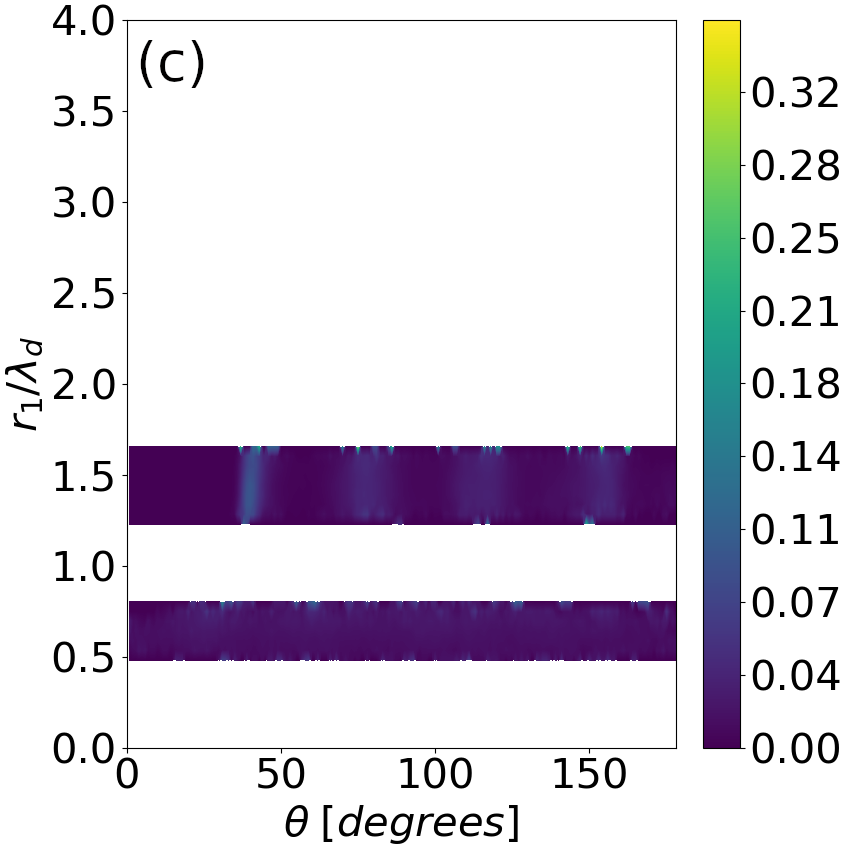}
		
	\end{subfigure} 
	%\hspace{1.6cm}
	\begin{subfigure}{0.45\columnwidth}
		\includegraphics[width=0.5\linewidth]{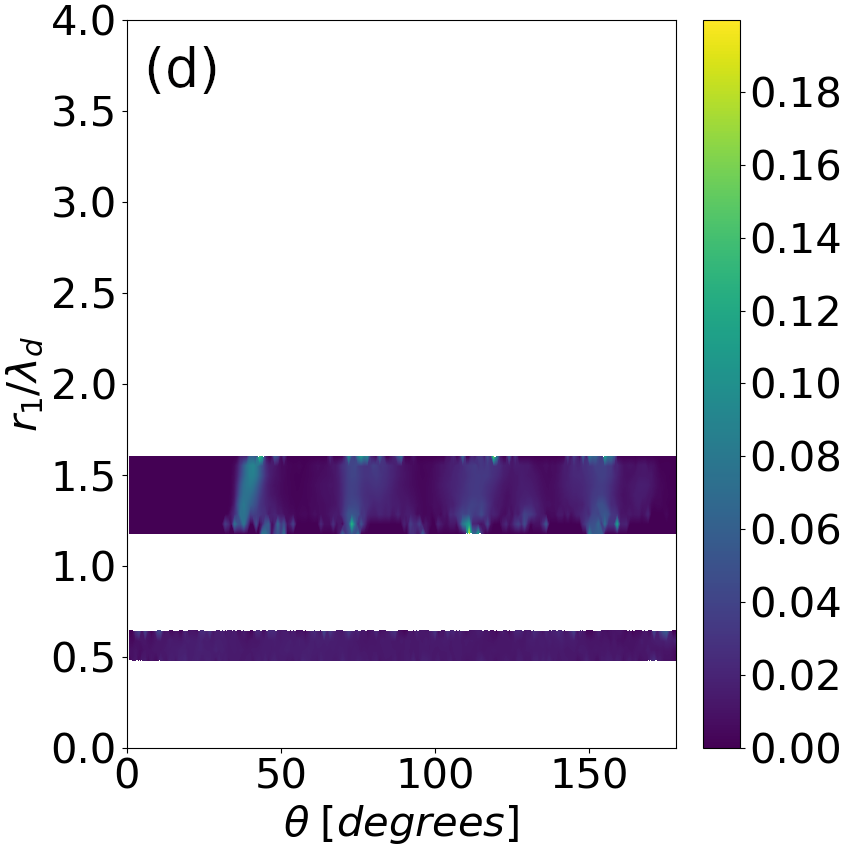}
	\end{subfigure}
	\caption{(Color online)C2P at four different values of coupling strengths (a) $\Gamma=102.88$, (b) $\Gamma=128.60$, (c) $\Gamma=171.46$, (d) $\Gamma=257.19$ for $\kappa=1.8$ and $\nu=3\;Hz$}
	\label{c2p_Nu3_wG}
\end{figure}

The center-two-particle correlation function at four different values of coupling strengths is shown at $\nu = 0.3\;Hz$ and $\nu=3\;Hz$ in Fig. \ref{c2p_Nu0.3_wG} and \ref{c2p_Nu3_wG} respectively by keeping screening parameter fixed as $\kappa=1.8.$ As expected the C2P doesn't show much difference from its frictionless counterpart (see Fig. \ref{c2pWG_withoutFric}).

%\subsubsection{Changes with respect to $N$}
\subsection{Dynamic properties}
To understand the Dynamic properties of the cluster with variation of coupling and screening strength we have evaluated the self part of Van Hove autocorrelation function both using frictionless Molecular Dynamics and Langevin Dynamics simulation. The Van Hove function reveals the single particle dynamics of the particles in the cluster.
\subsubsection{Frictionless Molecualr Dynamics}
\begin{figure}[htbp]
	\begin{subfigure}{0.45\columnwidth}
		\includegraphics[width=0.5\linewidth]{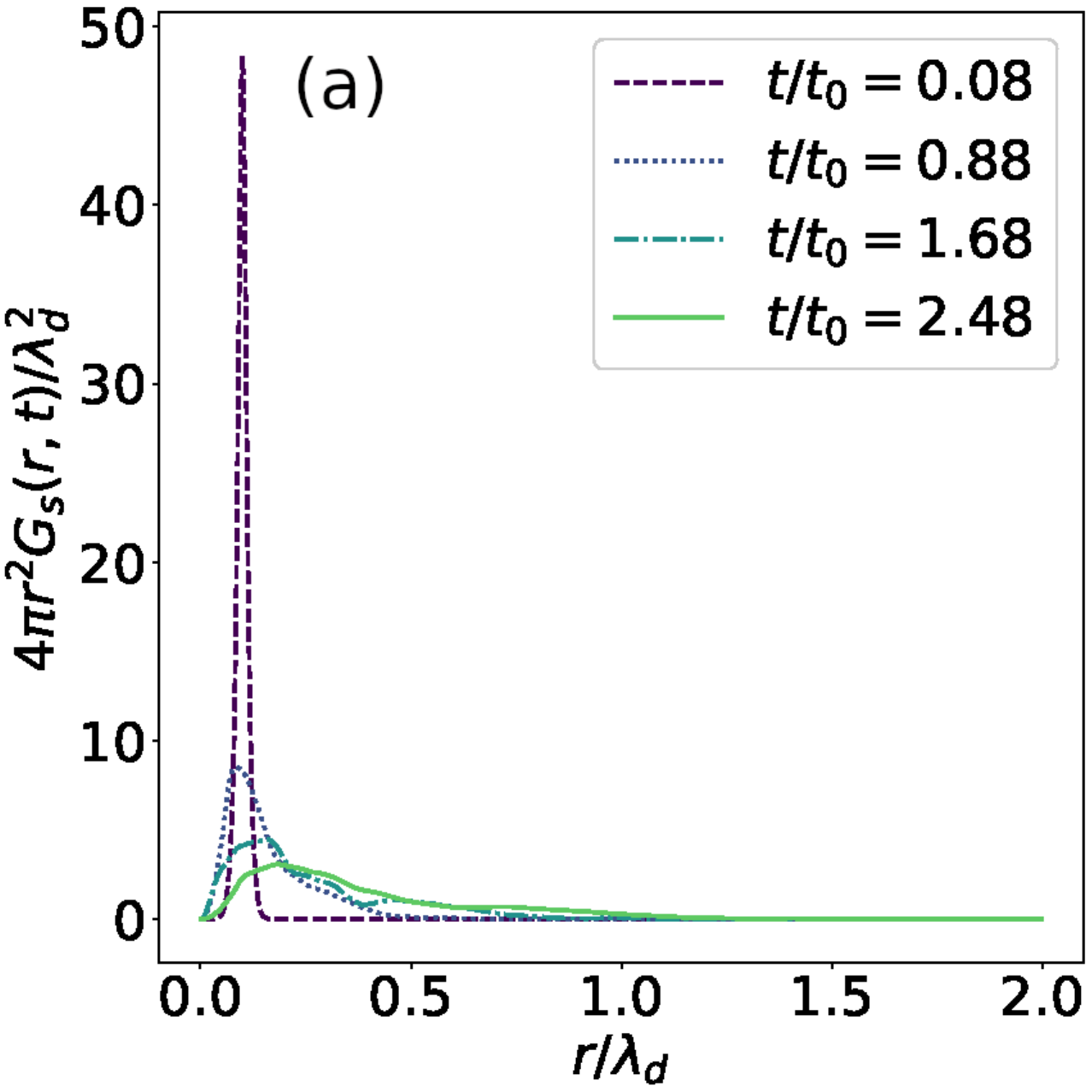}
		
	\end{subfigure}
	\begin{subfigure}{0.45\columnwidth}
		\includegraphics[width=0.5\linewidth]{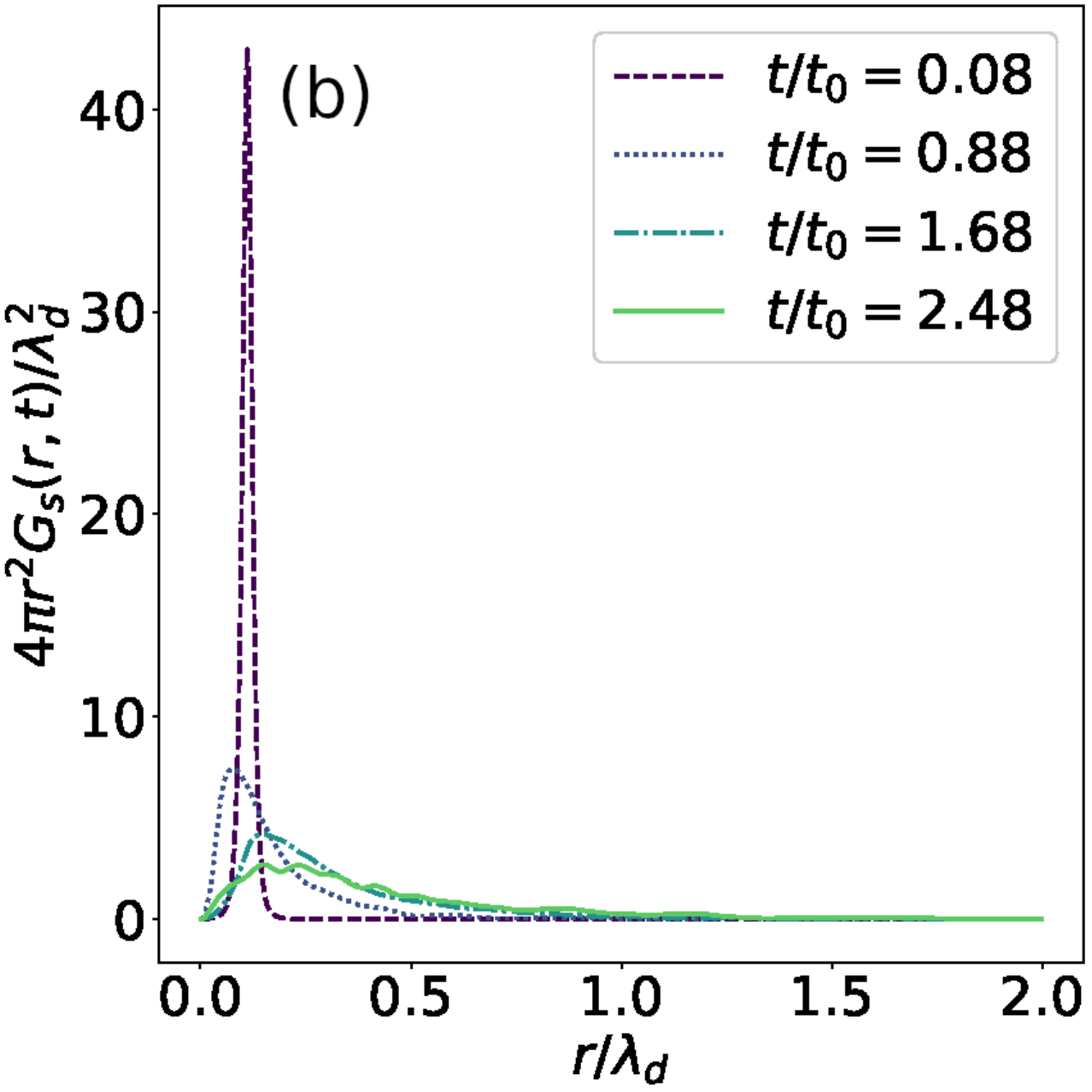}
		
	\end{subfigure}
	\begin{subfigure}{0.45\columnwidth}
		\includegraphics[width=0.5\linewidth]{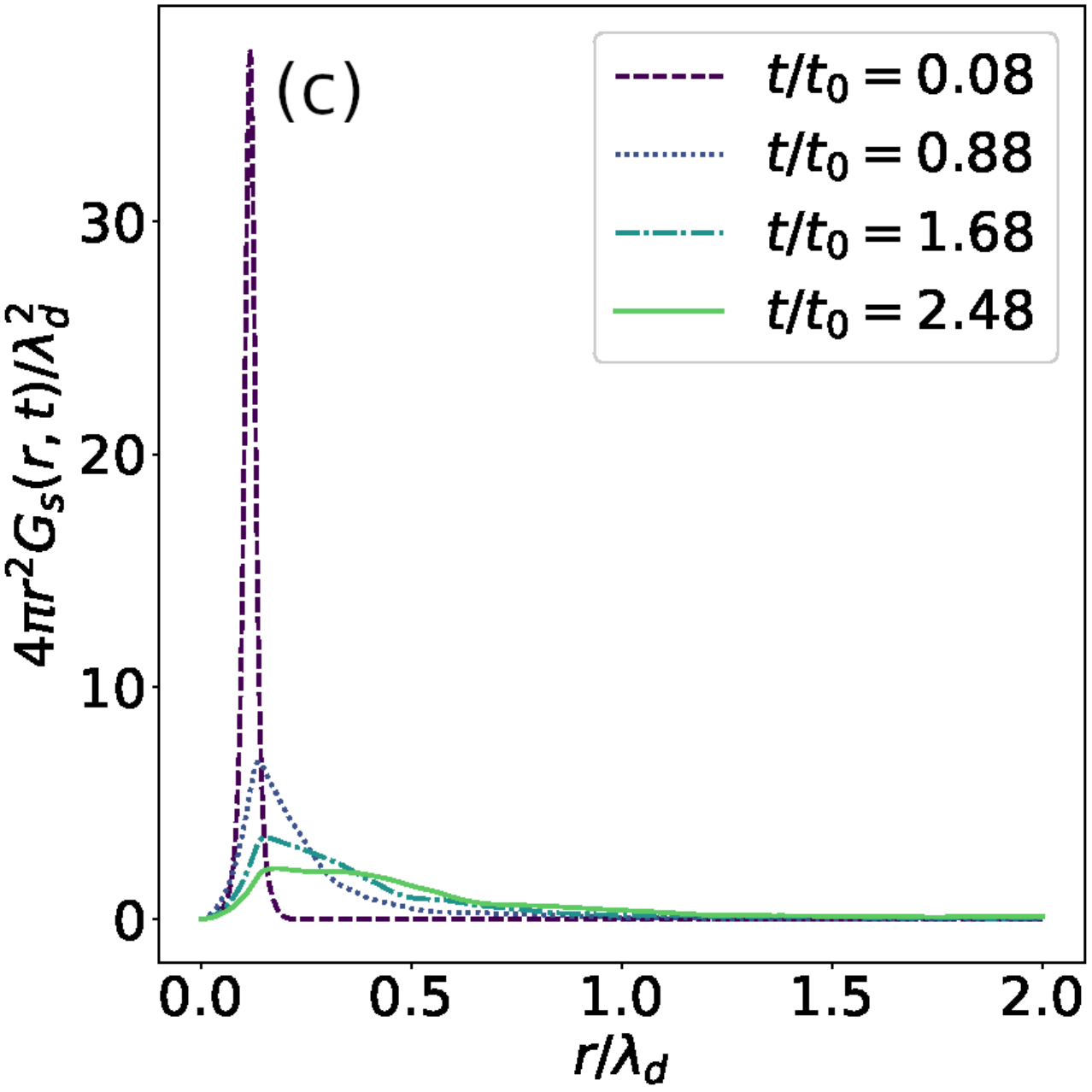}
		
	\end{subfigure} 
	%\hspace{1.6cm}
	\begin{subfigure}{0.45\columnwidth}
		\includegraphics[width=0.5\linewidth]{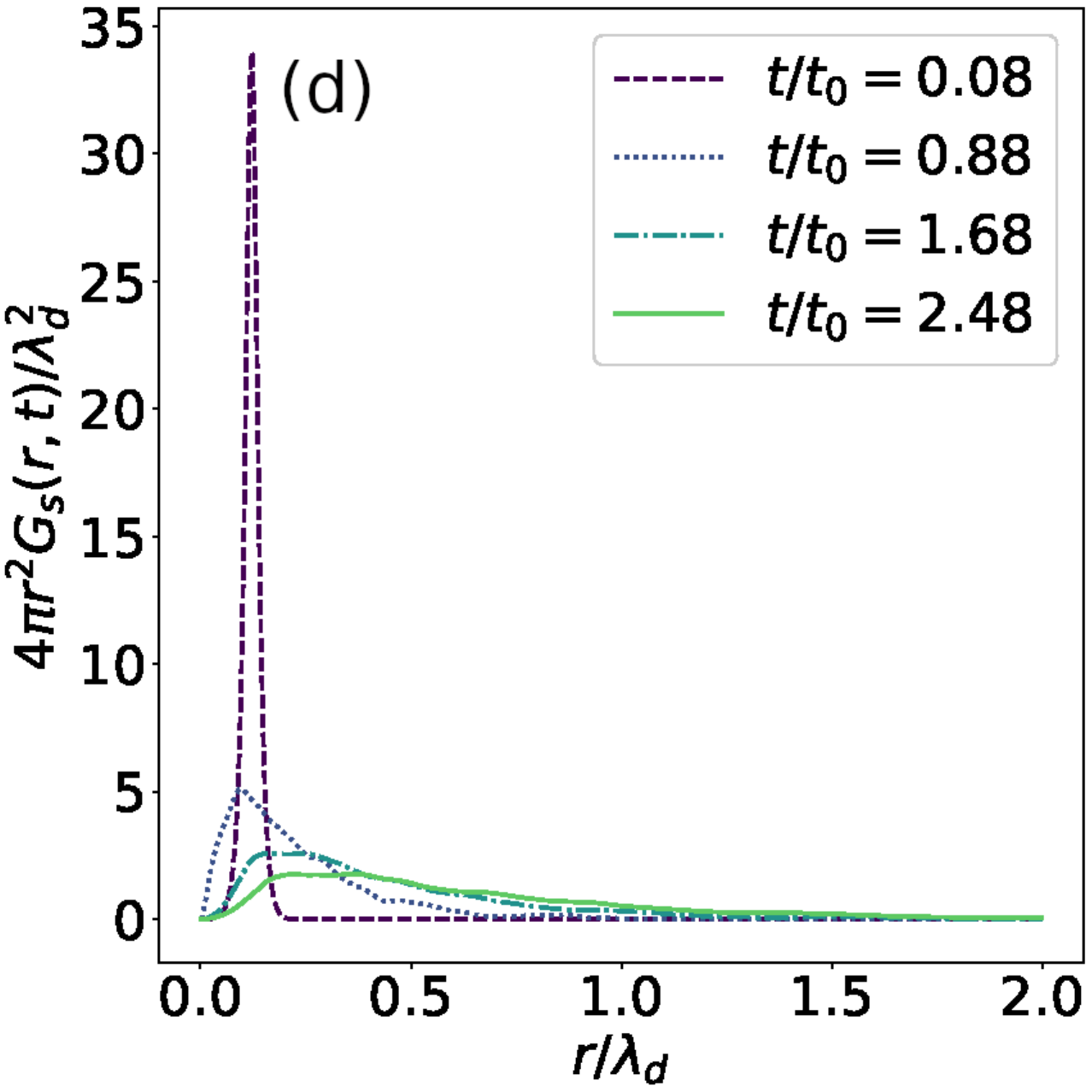}
	\end{subfigure}
	\caption{(Color online)Self Van Hove autocorrelation function for a cluster of $N=32$ particles  at four different values of coupling parameters,(a) $\Gamma=257.19$, (b) $\Gamma=171.46$, (c) $\Gamma=128.60$, (d) $\Gamma=102.88$ for $\kappa=1.8$ in frictionless MD simulation}
	\label{VHcorr_wG_withoutFric}
\end{figure}

To investigate the dynamic structure of the Yukawa ball the Van Hove self autocorrelation function is obtained in fricntionless MD simulation and plotted in Fig. \ref{VHcorr_wG_withoutFric}. The results are shown at four different values of coupling parameter and for each of them $G_s(r,t)$ is plotted at four different delay times.
From Fig. \ref{VHcorr_wG_withoutFric} it is seen that a peak occurs  at $r \sim 0.1\lambda_d$ at all delay times for all the coupling strengths. Initally, a particle is localized at $r\sim 0.1\lambda_d$. At fixed coupling, the peak height at $r \sim 0.1\lambda_d$ decreases with increasing delay times indicating the diffusion of the particles to other locations. However, at a fixed delay time the peak heights in the plots decreases with decreasing $\Gamma$.

\begin{figure}[htbp]

	\includegraphics[width=\textwidth]{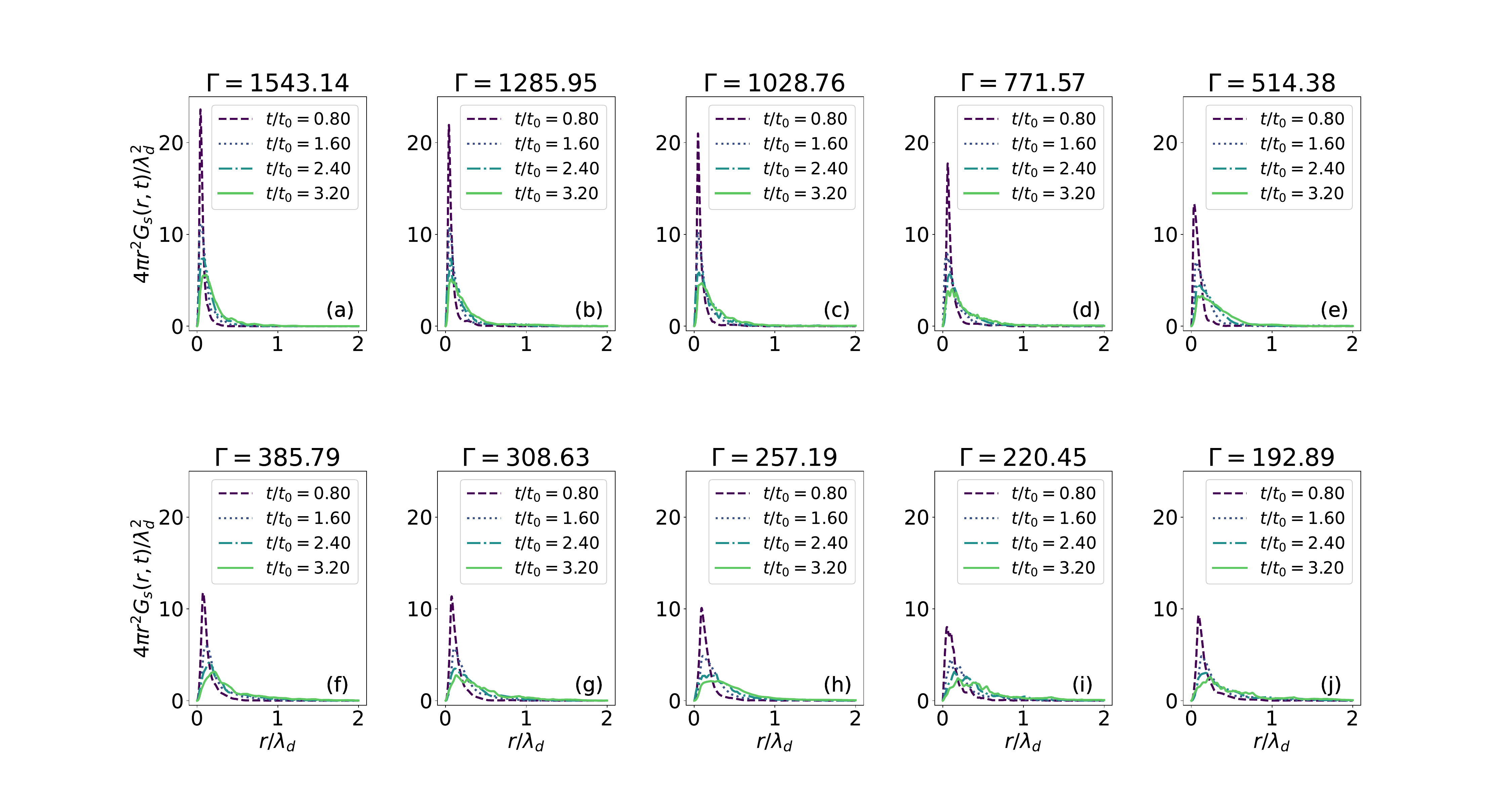}

	\caption{(Color online)Self Van Hove autocorrelation function for a cluster of $N=32$ particles in frictionless MD simulation at $\kappa=1.8$.}
	\label{VH_withoutFric_tenG}
\end{figure}

The second peak of the Radial Distribution Function as shown in Fig. \ref{rdf-wg} and \ref{RDF_wFric_wG}  and that of intra-shell angular correlation function as shown in Fig. \ref{AngCorrWg_withoutFric} splits up at sufficiently large values of Coulomb coupling parameter ($\Gamma$). To see if there is any corresponding change in the Van Hove self correlation function with rise in the value of $\Gamma$, it is shown at ten different coupling parameters and for each of them it is plotted at four different delay times as shown in Fig. \ref{VH_withoutFric_tenG}. It is seen that as the coupling strength is increased, $G_s(r,t)$ at all the delay times tend to decay to zero as $r\to \lambda_d$  indicating the  tendency of the system of particles to freeze.
%\begin{widetext}
%\end{widetext} 
\begin{figure}[htbp]
	\begin{subfigure}{0.45\columnwidth}
		\includegraphics[width=0.5\linewidth]{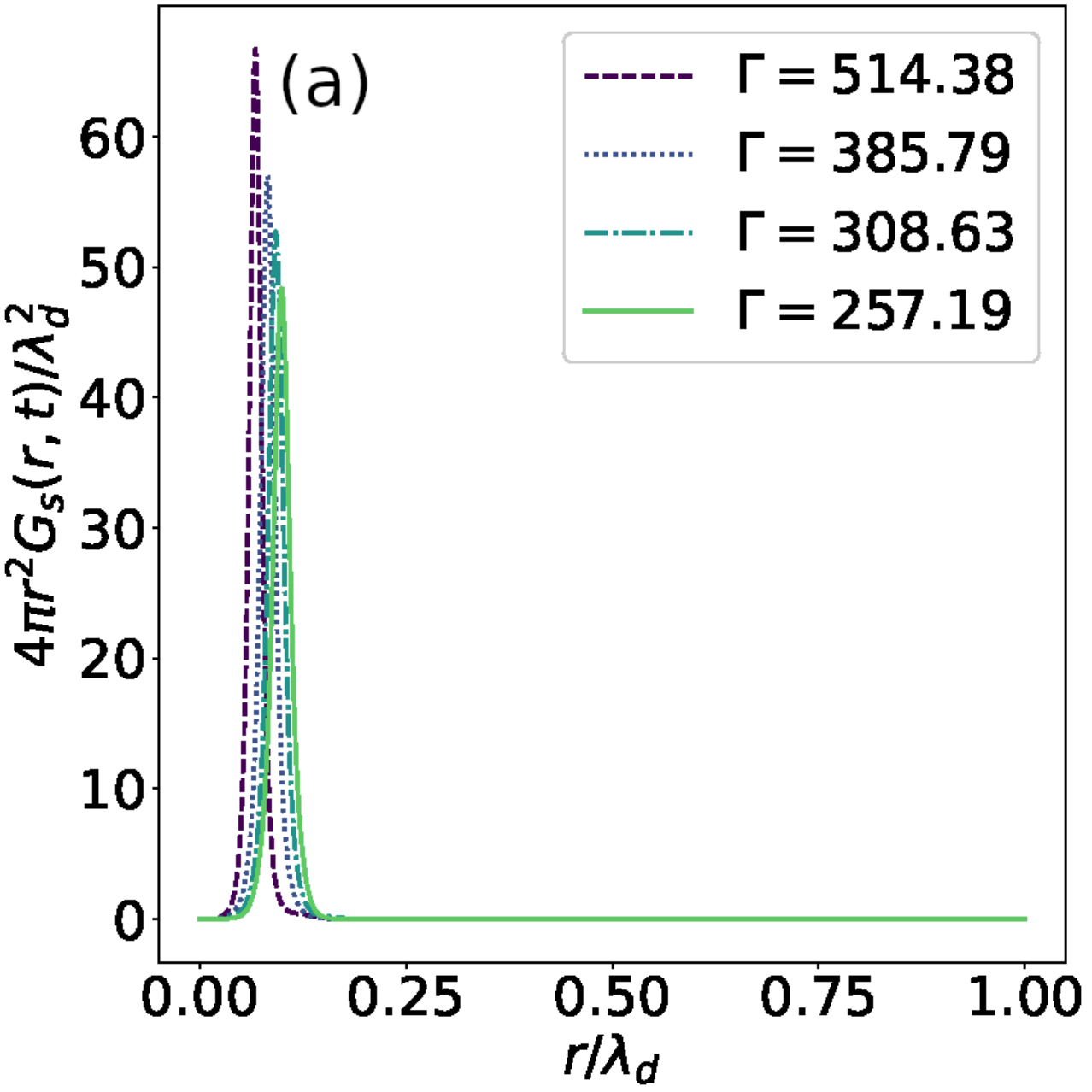}
	\end{subfigure}
	\begin{subfigure}{0.45\columnwidth}
		\includegraphics[width=0.5\linewidth]{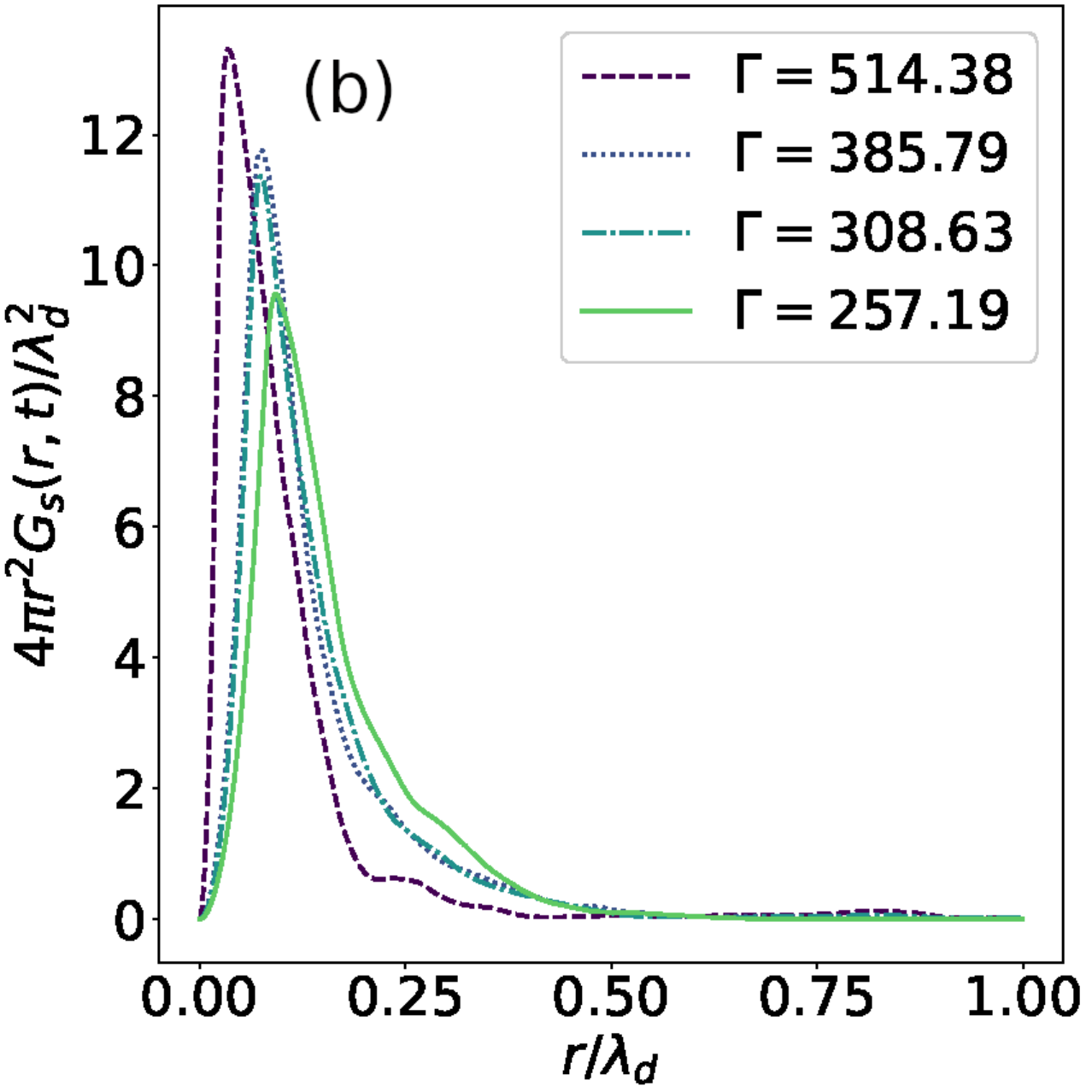}
	\end{subfigure}
	\begin{subfigure}{0.45\columnwidth}
		\includegraphics[width=0.5\linewidth]{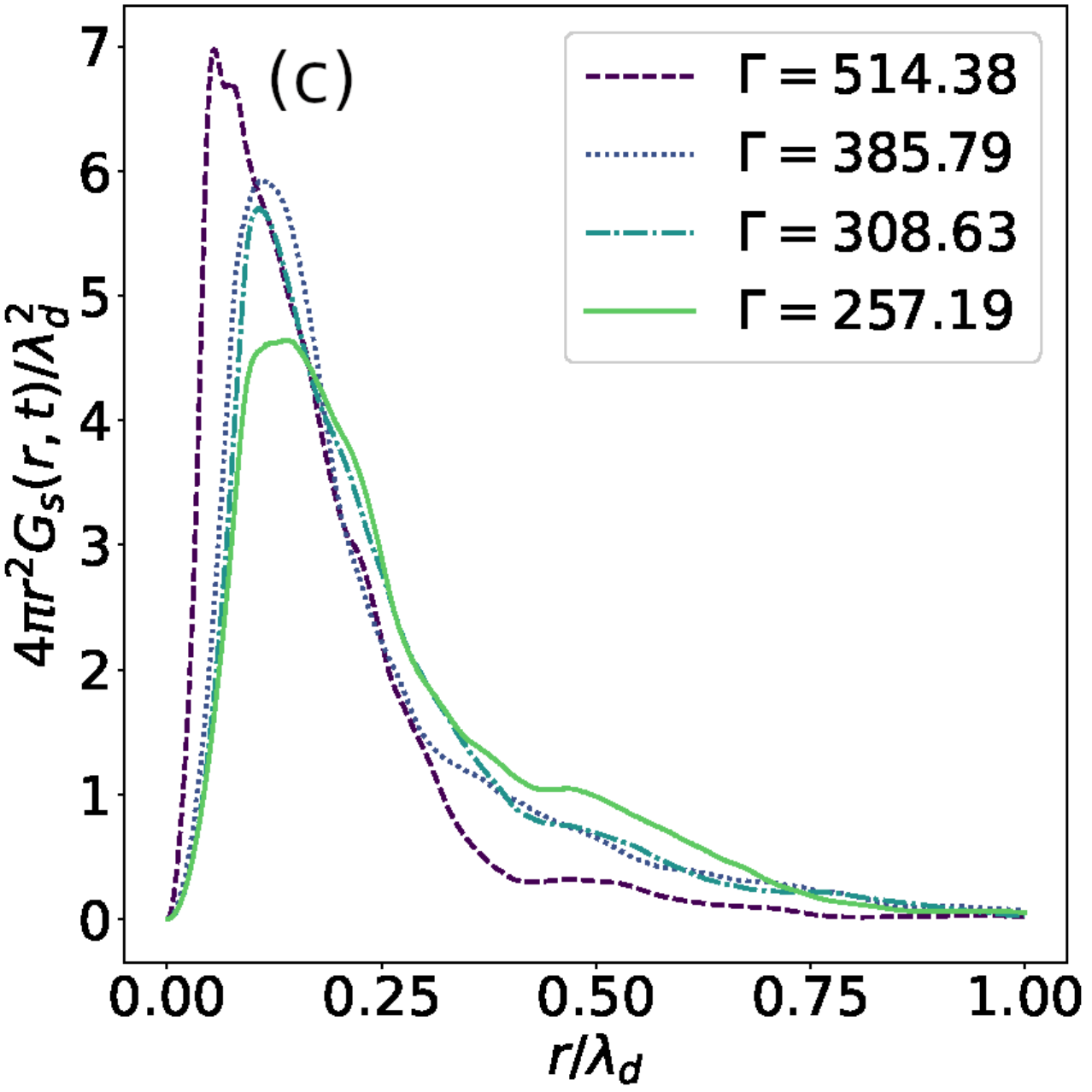}
	\end{subfigure}
	%\hspace{1.4cm}
	\begin{subfigure}{0.45\columnwidth}
		\includegraphics[width=0.5\linewidth]{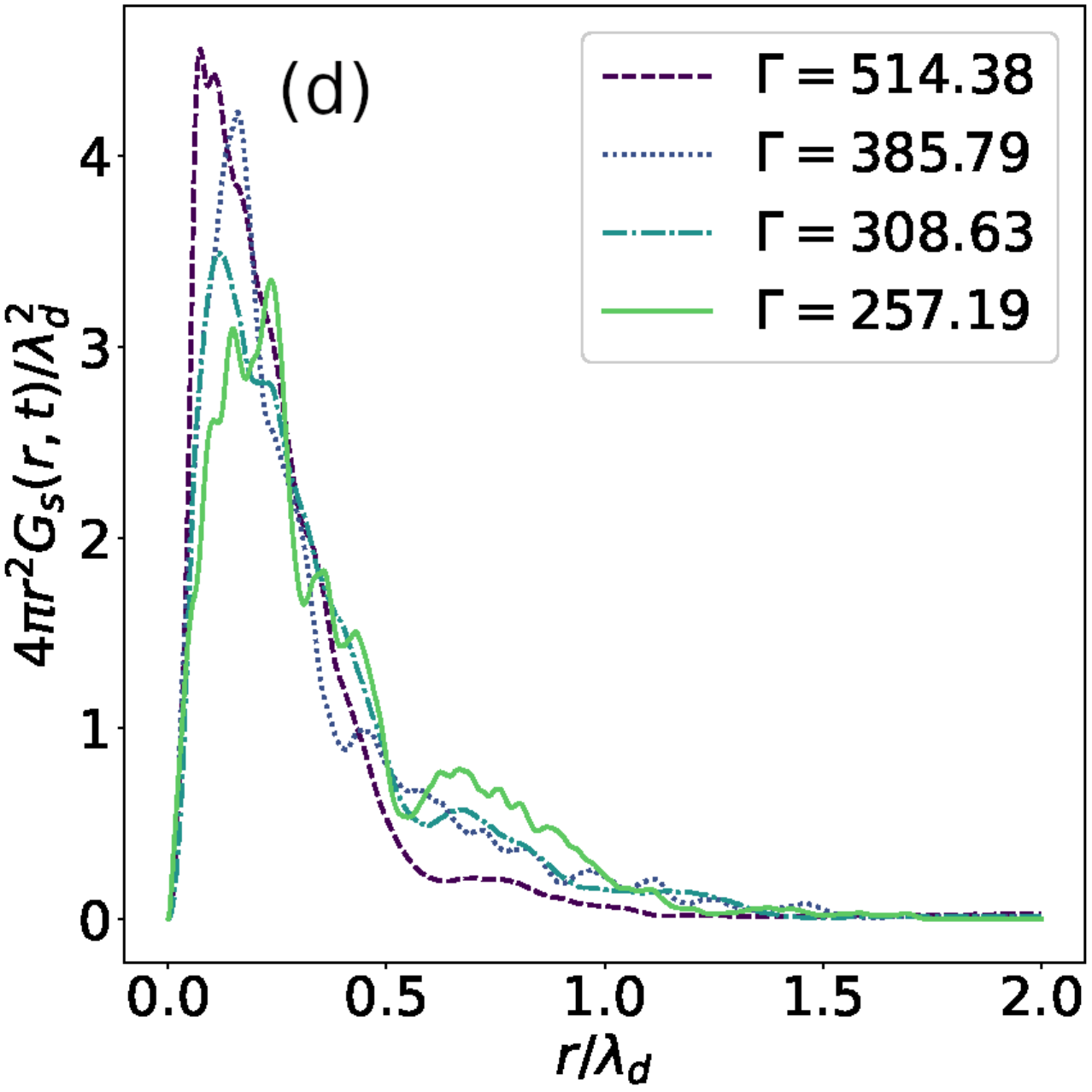}
	\end{subfigure}
	\caption{(Color online)Self Van Hove autocorrelation function at a) $t/t_0=0.08$, b) $t/t_0=0.88$, c) $t/t_0=1.68$ and d) $t/t_0=2.48$ for $\kappa=1.8.$}
	\label{vanHove-diffTimes}
\end{figure}

The self Van-Hove autocorrelation function has been plotted at four different values of delay times as shown in Fig. \ref{vanHove-diffTimes}. For each of these figures, the function is plotted at four different coupling strengths as shown. It is clear that the peak shifts towards right as the coupling parameter decreases which means that with decrease in coupling the particle can diffuse upto a large distance from the origin. Also it is seen that at a lower value of coupling parameter ($\Gamma=257.19$) and at a later time ($t/t_0=2.48$) the smoothness of the curve is lost and substructure appears which is a signature of a structural transition of the cluster.

\begin{figure}[htbp]
	\begin{subfigure}{0.45\columnwidth}
		\includegraphics[width=0.5\linewidth]{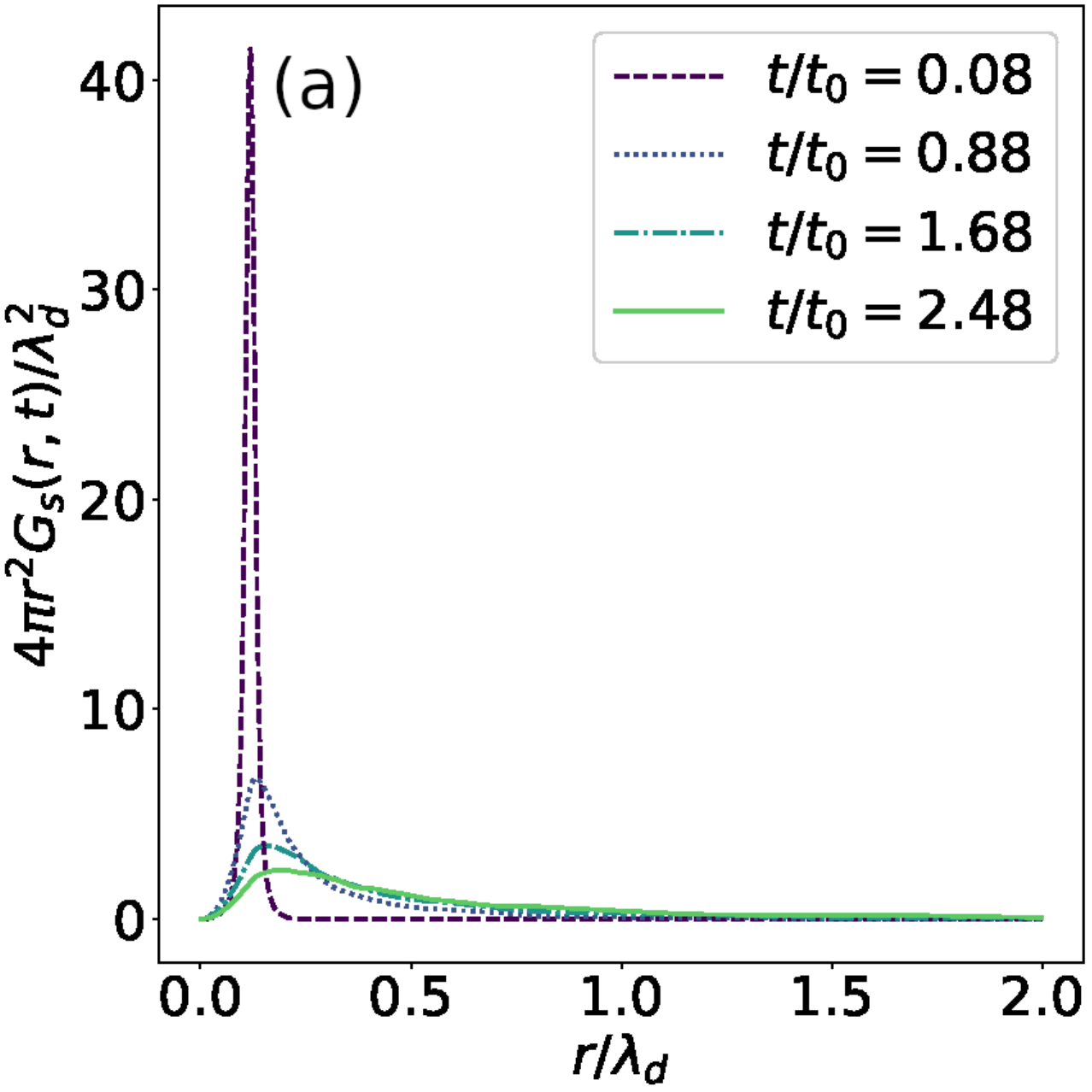}
		
	\end{subfigure}
	\begin{subfigure}{0.45\columnwidth}
		\includegraphics[width=0.5\linewidth]{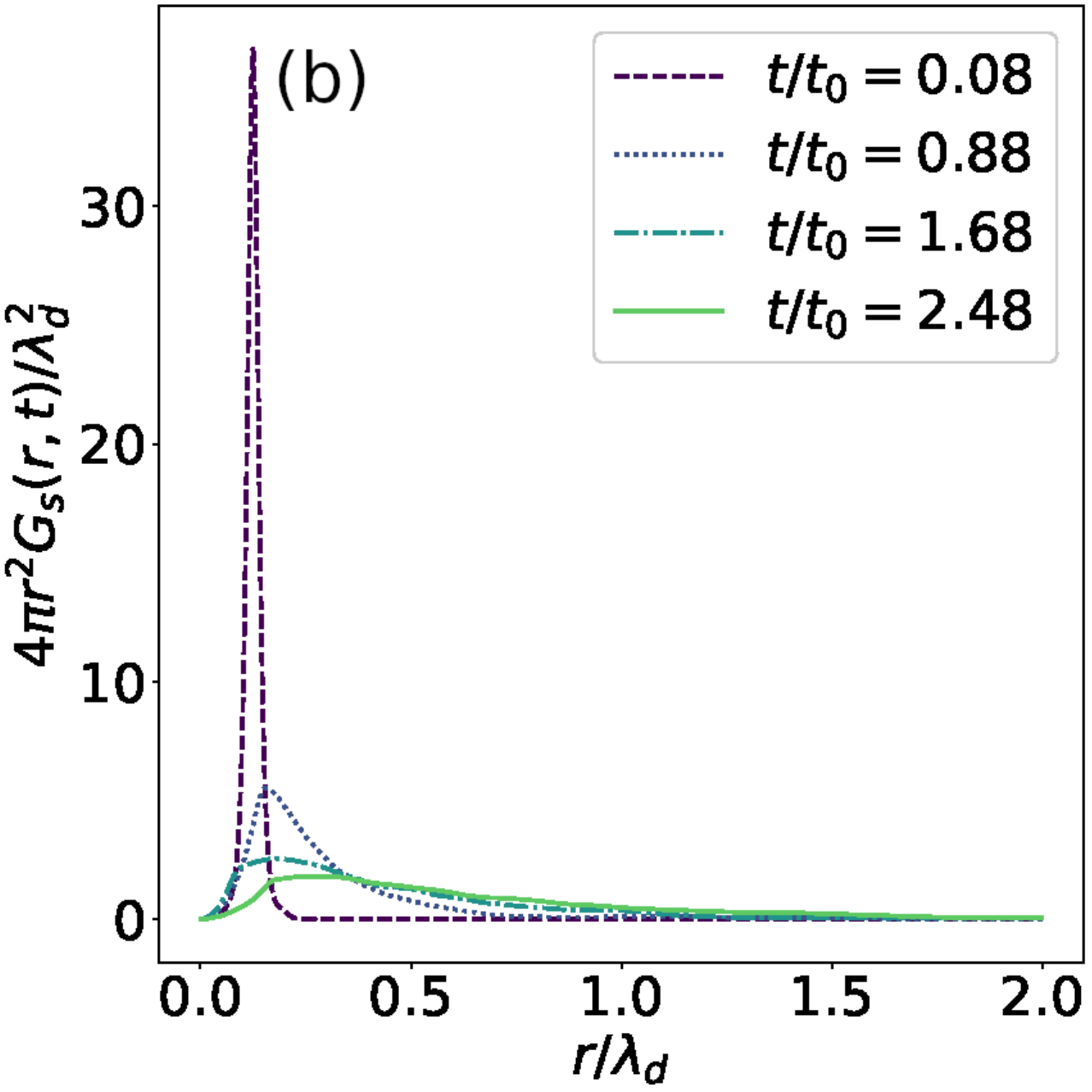}
		
	\end{subfigure}
	\begin{subfigure}{0.45\columnwidth}
		\includegraphics[width=0.5\linewidth]{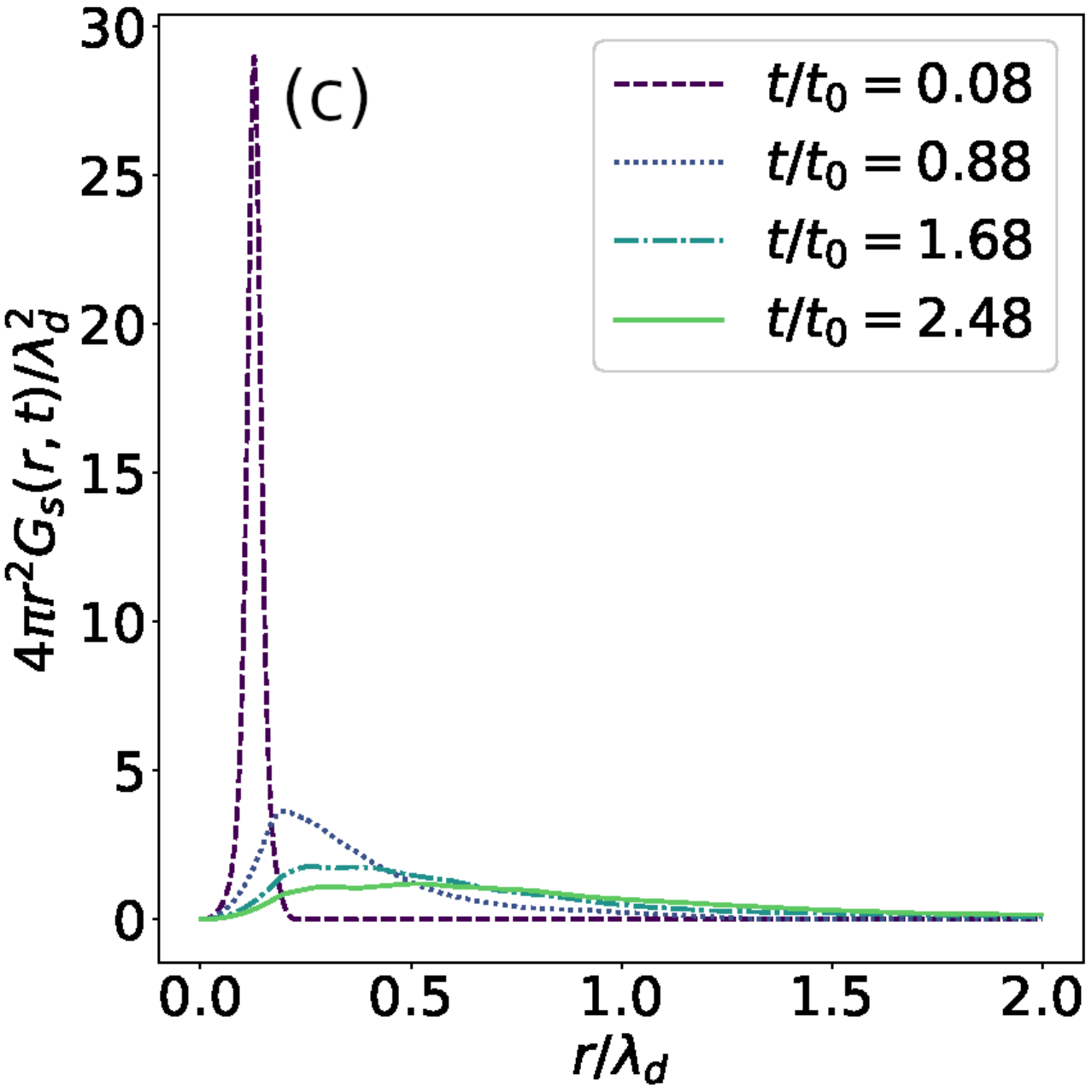}
		
	\end{subfigure} 
	%\hspace{1.6cm}
	\begin{subfigure}{0.45\columnwidth}
		\includegraphics[width=0.5\linewidth]{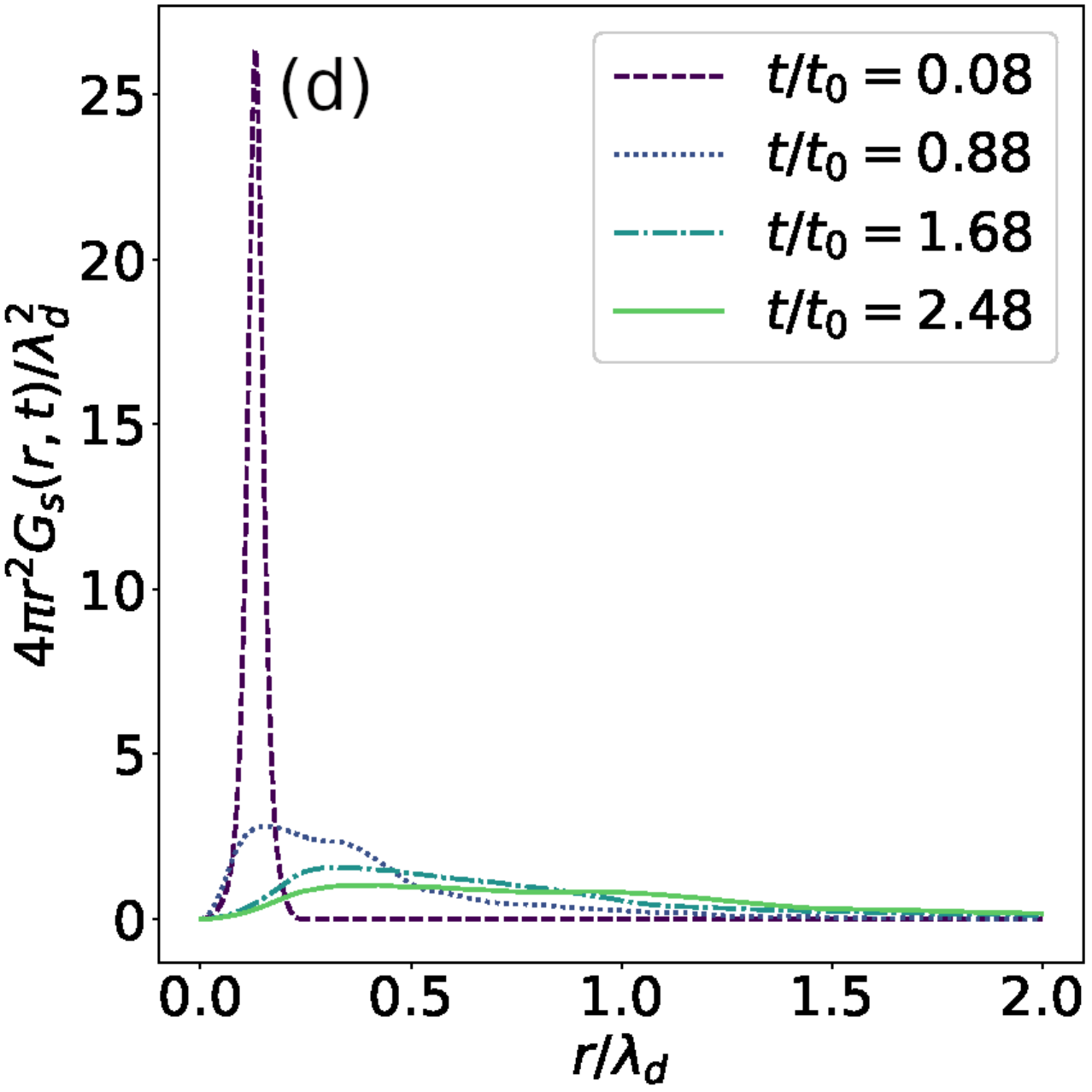}
	\end{subfigure}
	\caption{(Color online)Self Van-Hove autocorrelation function plotted at four different values of screening parameter (a) $\kappa=2.59$, (b) $\kappa=3.19$, (c) $\kappa=4.13$ and (d) $\kappa=4.53$ at  $\Gamma=257.19$ for a cluster with $N=32.$}
	\label{VHcorr_wK_withoutFric}
\end{figure}
The self part of Van Hove autocorrelation function is plotted at four different screening parameters in Fig. \ref{VHcorr_wK_withoutFric} at a fixed value of coupling strength $\Gamma=257.19$ at different delay times. At the shortest delay time $t=0.08\;t_0$ a peak appears in $G_s(r,t)$ which shifts towards left with increase in $\kappa.$ 
\subsubsection{Langevin Dynamics}
\begin{figure}[htbp]
	\begin{subfigure}{0.45\columnwidth}
		\includegraphics[width=0.5\linewidth]{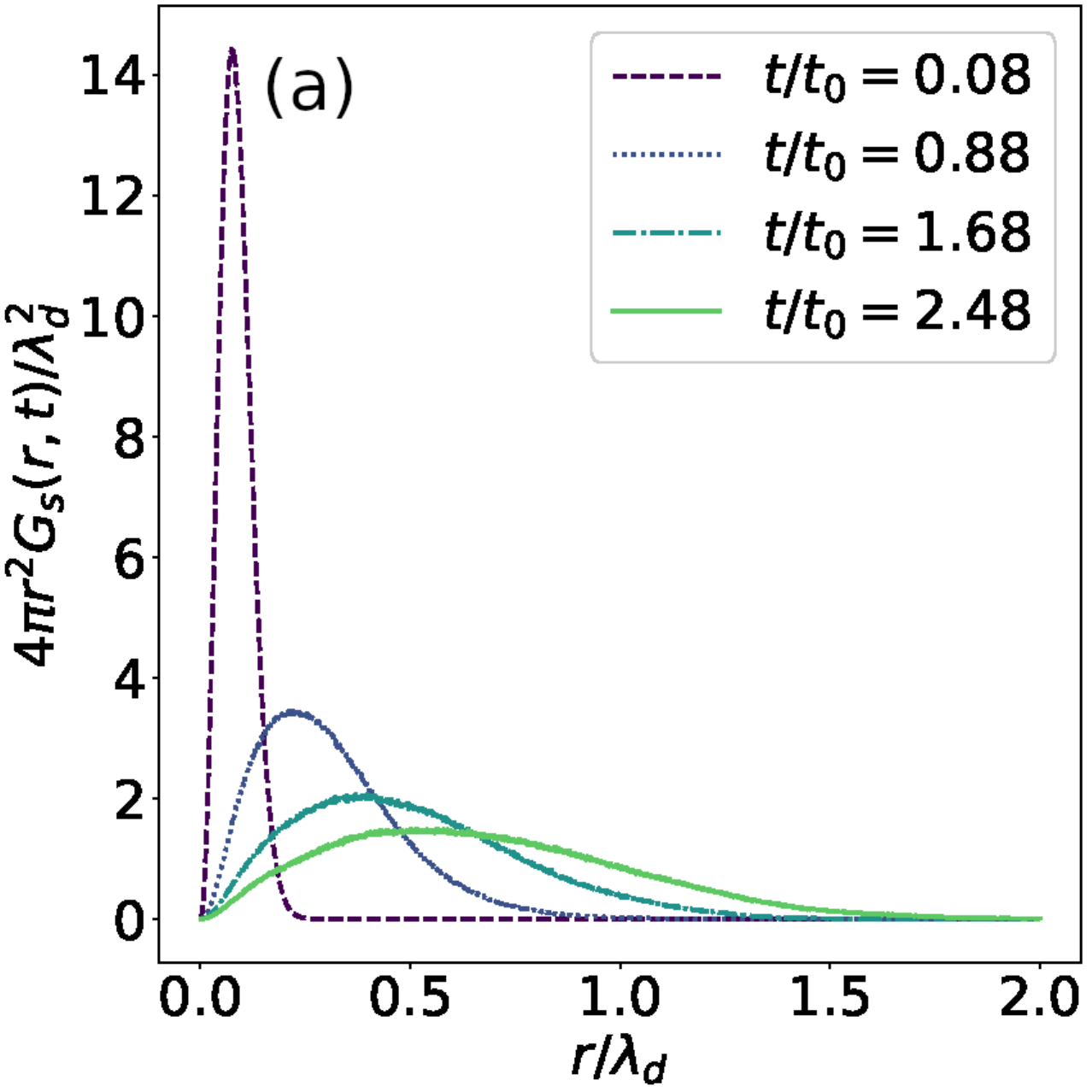}
		
	\end{subfigure}
	\begin{subfigure}{0.45\columnwidth}
		\includegraphics[width=0.5\linewidth]{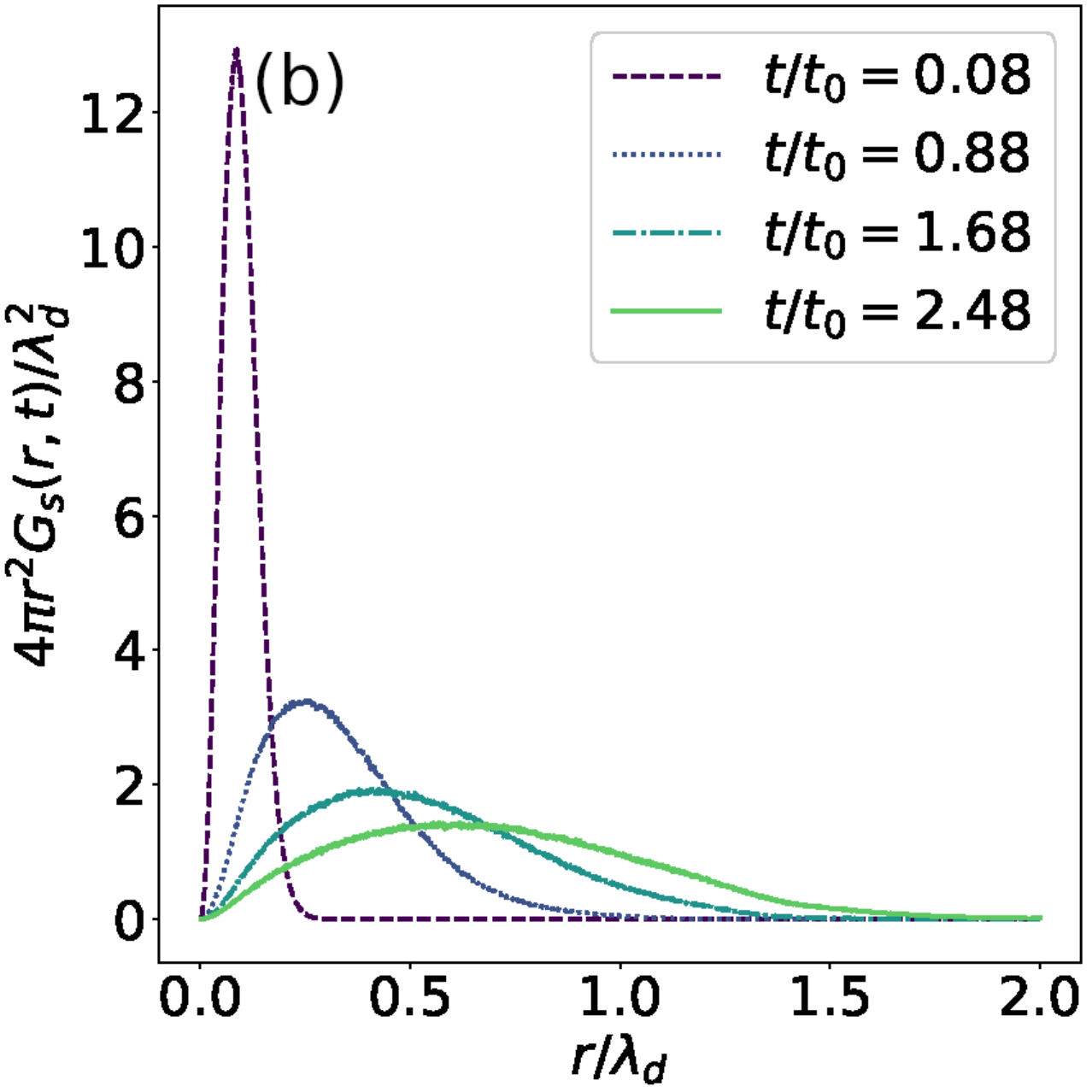}
		
	\end{subfigure}
	\begin{subfigure}{0.45\columnwidth}
		\includegraphics[width=0.5\linewidth]{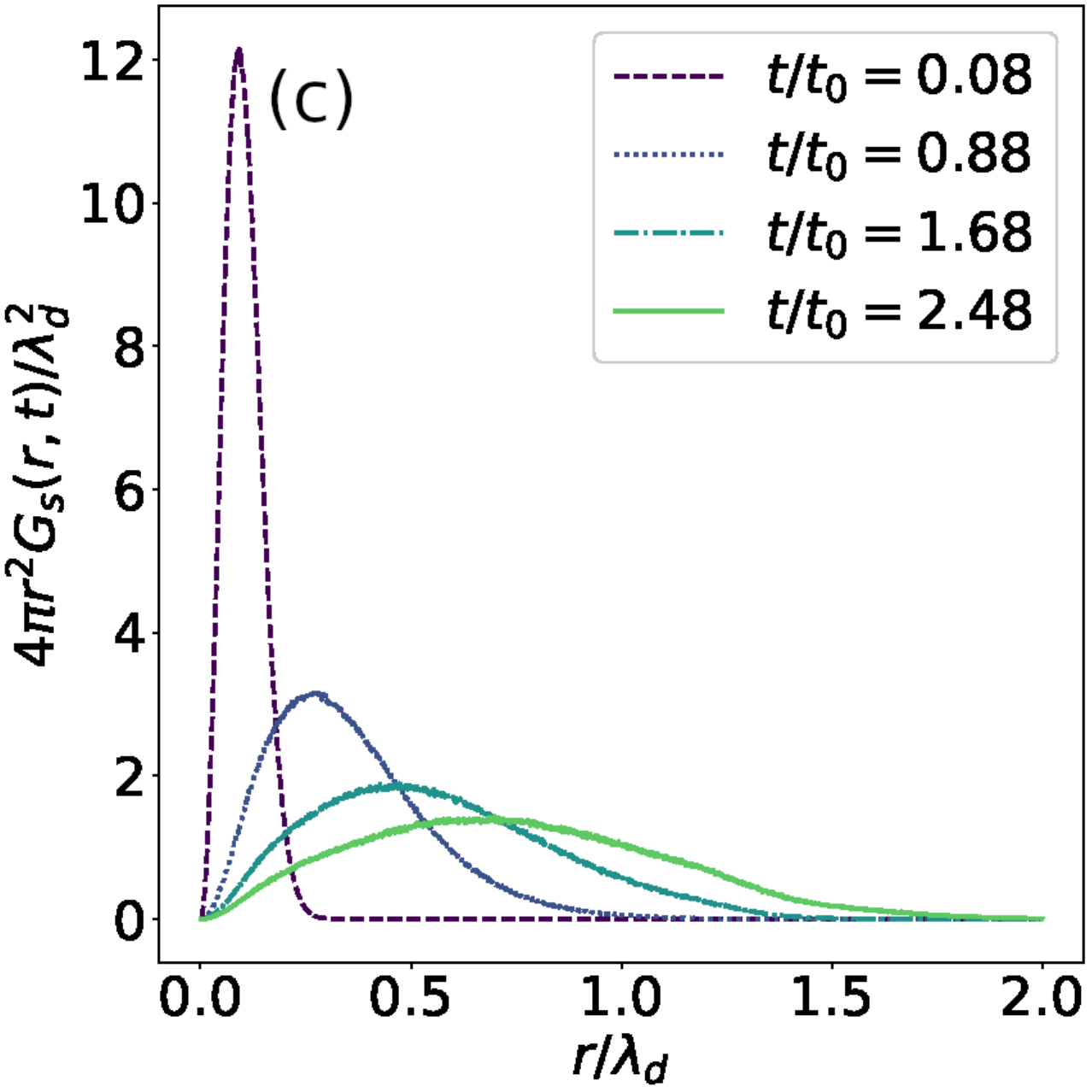}
		
	\end{subfigure} 
	%\hspace{1.6cm}
	\begin{subfigure}{0.45\columnwidth}
		\includegraphics[width=0.5\linewidth]{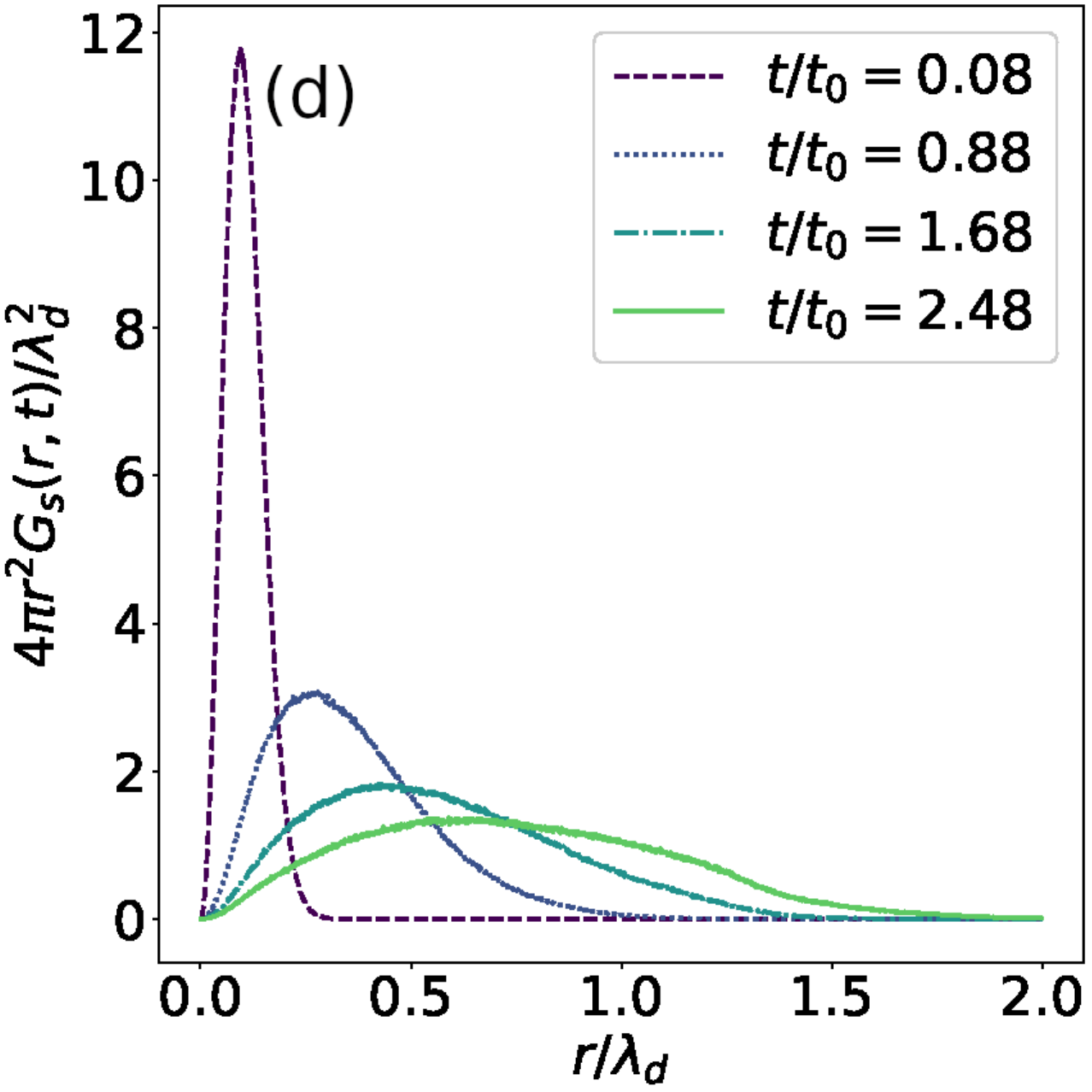}
	\end{subfigure}
	\caption{(Color online)Self Van Hove autocorrelation function at four different values of coupling parameter (a) $\Gamma=257.19$, (b) $\Gamma=171.46$, (c) $\Gamma=128.60$, (d) $\Gamma=102.88$ for $\kappa=1.8.$ and $\nu = 0.3\; Hz$}
	\label{VHcorr_wG_nu0.3}
\end{figure}

\begin{figure}[htbp]
	\begin{subfigure}{0.45\columnwidth}
		\includegraphics[width=0.5\linewidth]{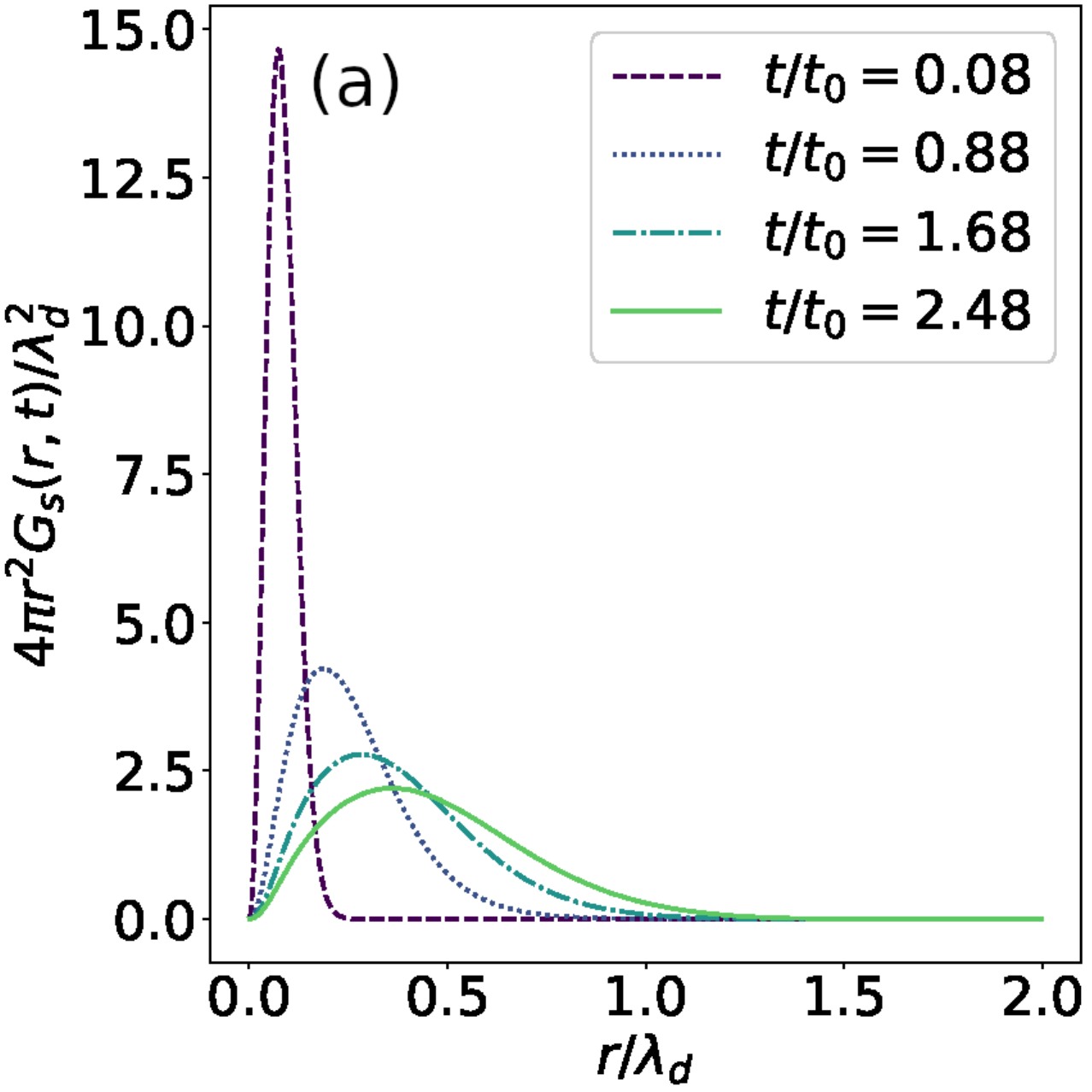}
		
	\end{subfigure}
	\begin{subfigure}{0.45\columnwidth}
		\includegraphics[width=0.5\linewidth]{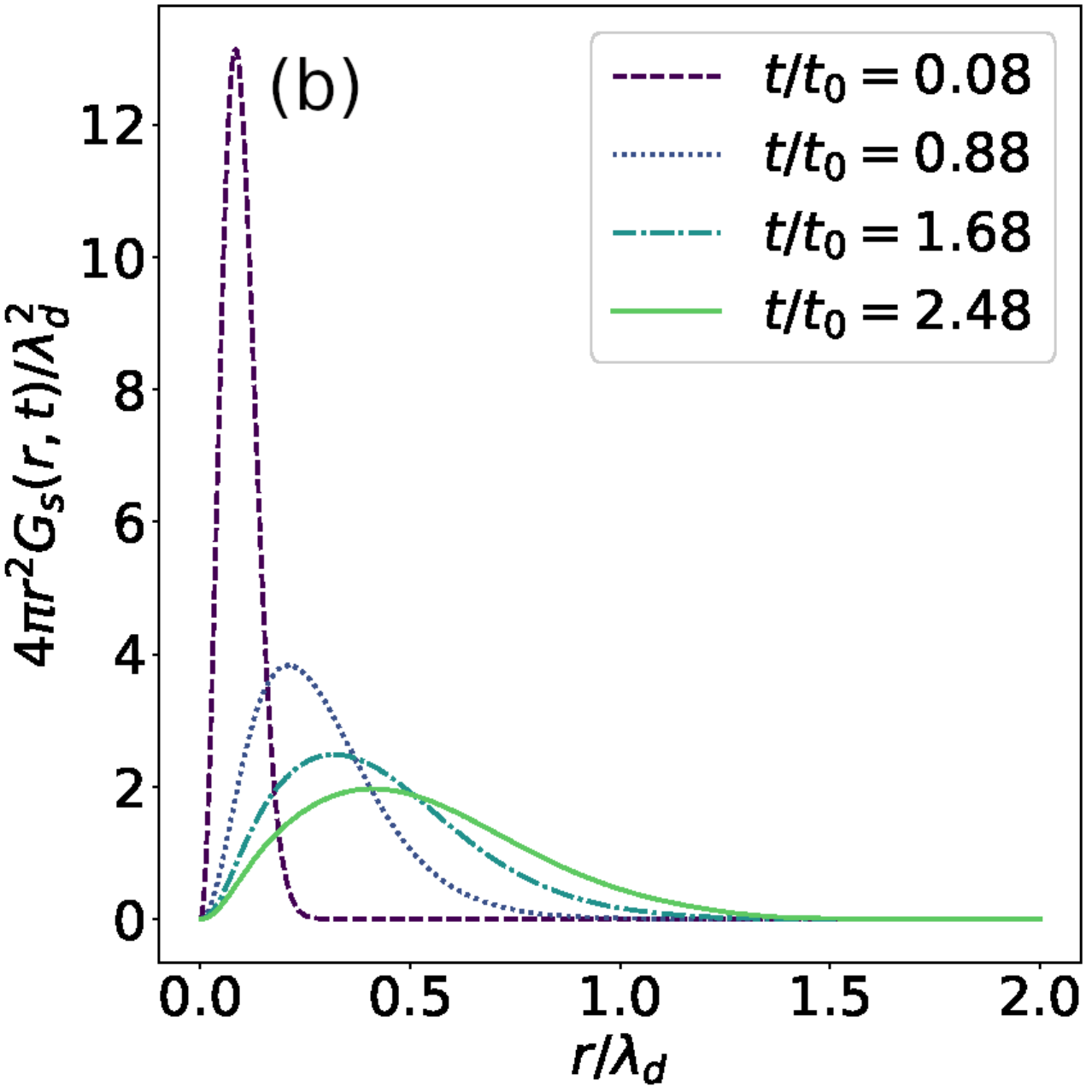}
		
	\end{subfigure}
	\begin{subfigure}{0.45\columnwidth}
		\includegraphics[width=0.5\linewidth]{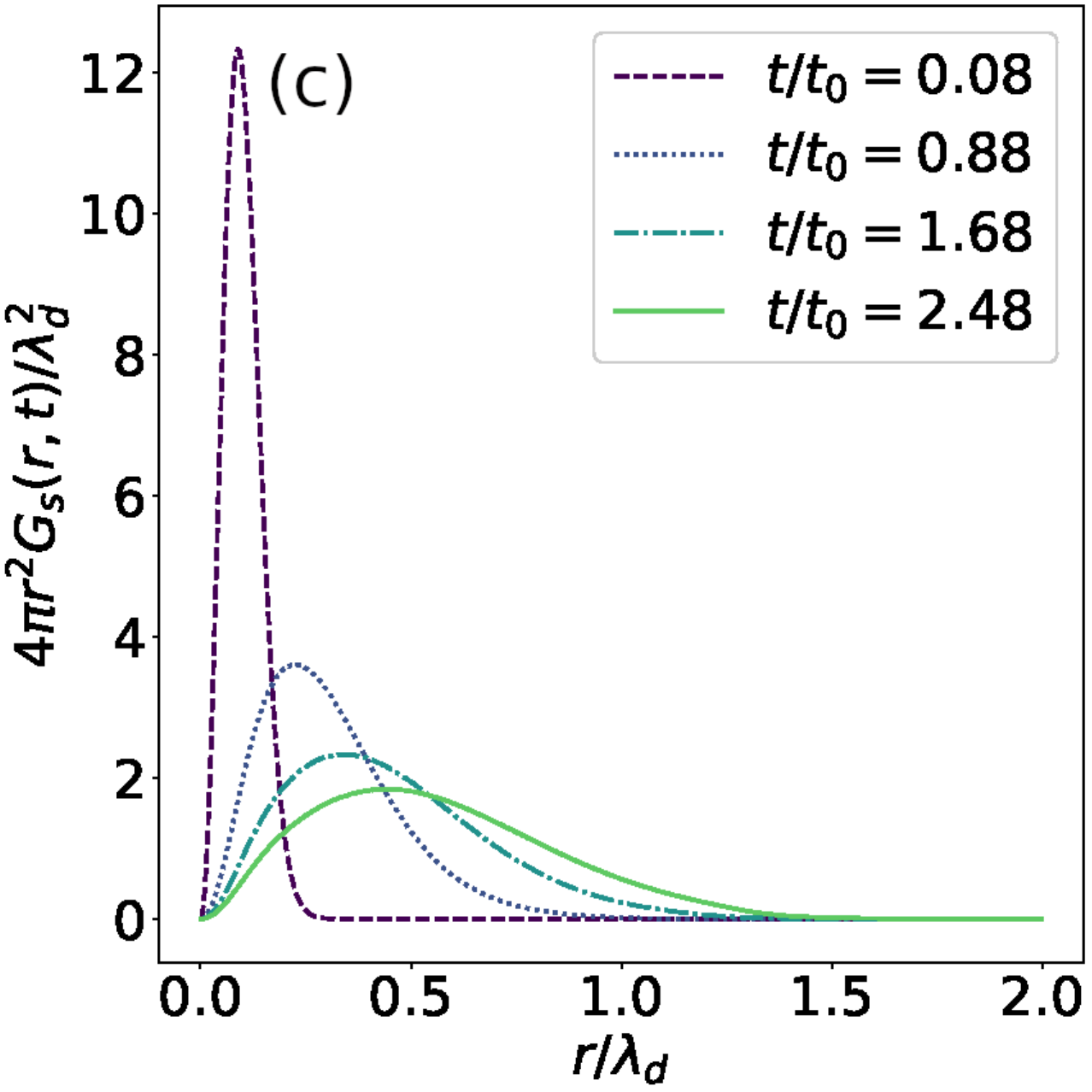}
		
	\end{subfigure} 
	%\hspace{1.6cm}
	\begin{subfigure}{0.45\columnwidth}
		\includegraphics[width=0.5\linewidth]{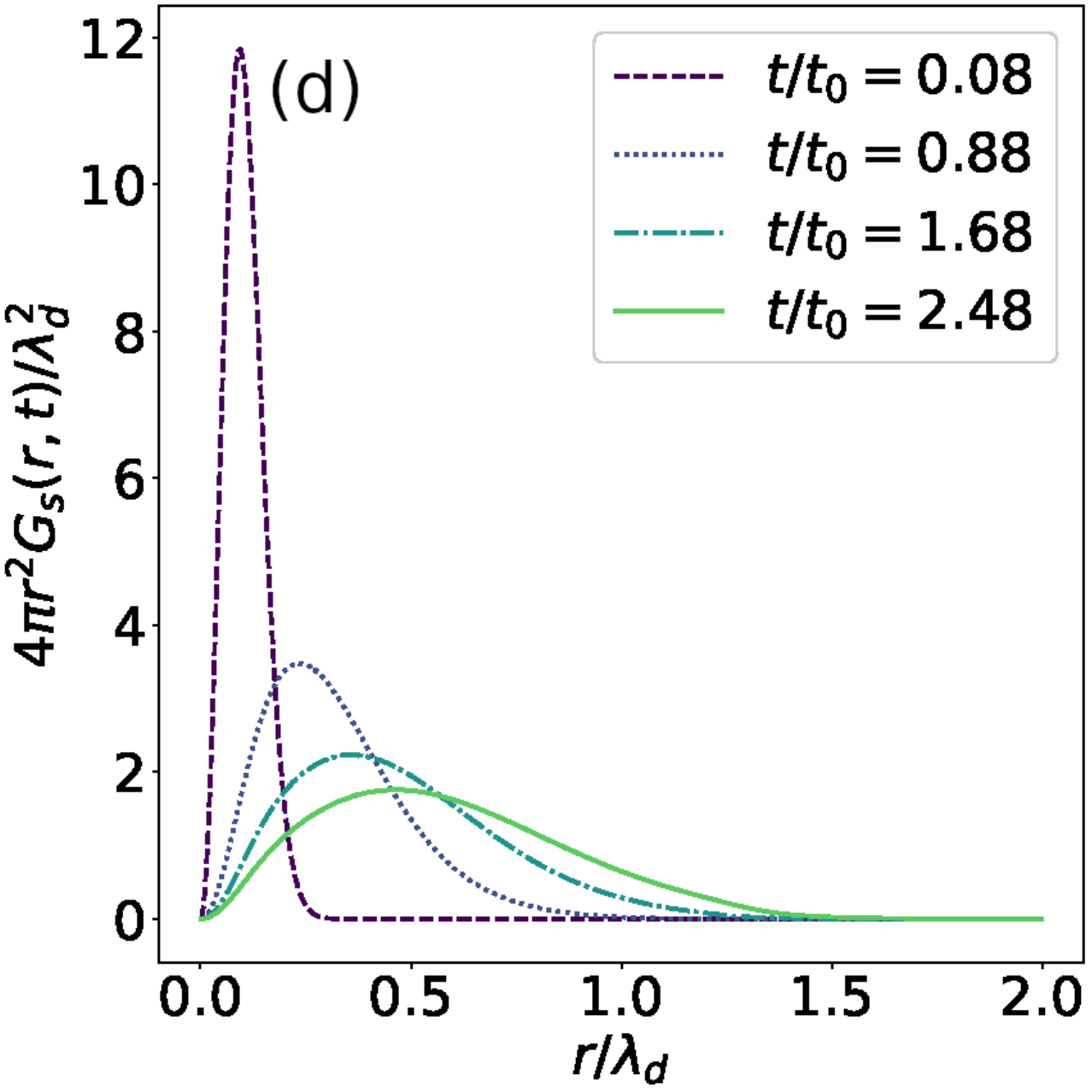}
	\end{subfigure}
	\caption{(Color online)Self Van Hove autocorrelation function at four different coupling strengths (a) $\Gamma=257.19$, (b) $\Gamma=171.46$, (c) $\Gamma=128.60$, (d) $\Gamma=102.88$ for $\kappa=1.8$ and $\nu = 3\; Hz.$}
	\label{VHcorr_wG_nu3}
\end{figure}

To understand the effect of introducing dust-neutral collision in the single particle dynamics of the cluster, we have obtained the Van-Hove self correlation function using Langevin Dynamics simulation. The results are shown in Fig. \ref{VHcorr_wG_nu0.3} and \ref{VHcorr_wG_nu3} at two different values of friction coefficients. 

Fig. \ref{VHcorr_wG_nu0.3} shows the $G_s(r,t)$ plots at four different coupling strengths. The plots clearly show that the probability of a particle lying at a distance $r$ after time $t$ lowers with decreasing value of $\Gamma.$ 

%\begin{widetext}
\begin{figure}[htbp]

	\includegraphics[width=\textwidth]{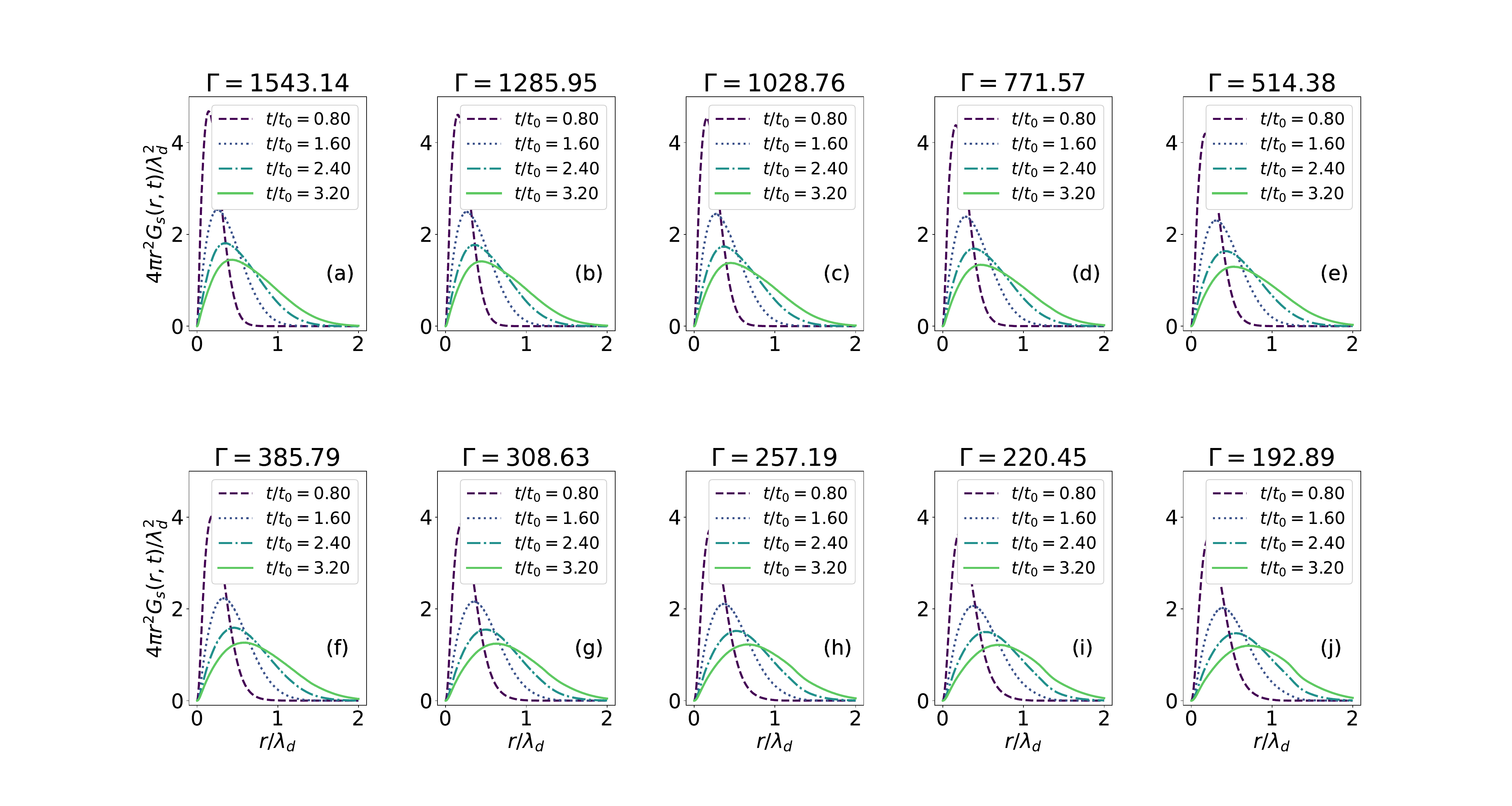}

	\caption{(Color online)Self Van Hove autocorrelation function for a cluster of $N=32$ particles in Langevin Dynamics simulation at $\nu=0.3\;Hz$ and $\kappa=1.8$.}
	\label{VH_withFric_tenG}
\end{figure}
%\end{widetext}
The Van Hove self correlation function is plotted at four different delay times and ten different coupling strengths as shown in Fig. \ref{VH_withFric_tenG}. This should be compared with Fig. \ref{VH_withoutFric_tenG} which was the Newtonian Dynamics case.

It is to be noted that the $G_s(r,t)$ in the Newtonian and Langevin dynamics has some important differences.  The peak heights of $G_s(r,t)$ in the Newtonian Dynamics are much higher than the corresponding peak heights in Langevin dynamics at all coupling strengths and delay times. Moreover, the witdh of the plots are much larger in Langevin Dynamics than the corresponding widths in Newtonian Dynamics. This means that the particles in Langevin Dynamics have a wider range of displacement suggesting greater mobility than in Newtonian Dynamics. 
\subsection{Disappearance of the rotational motion}

\begin{figure}[htbp]
	\begin{subfigure}{0.45\columnwidth}
		\includegraphics[width=0.5\linewidth]{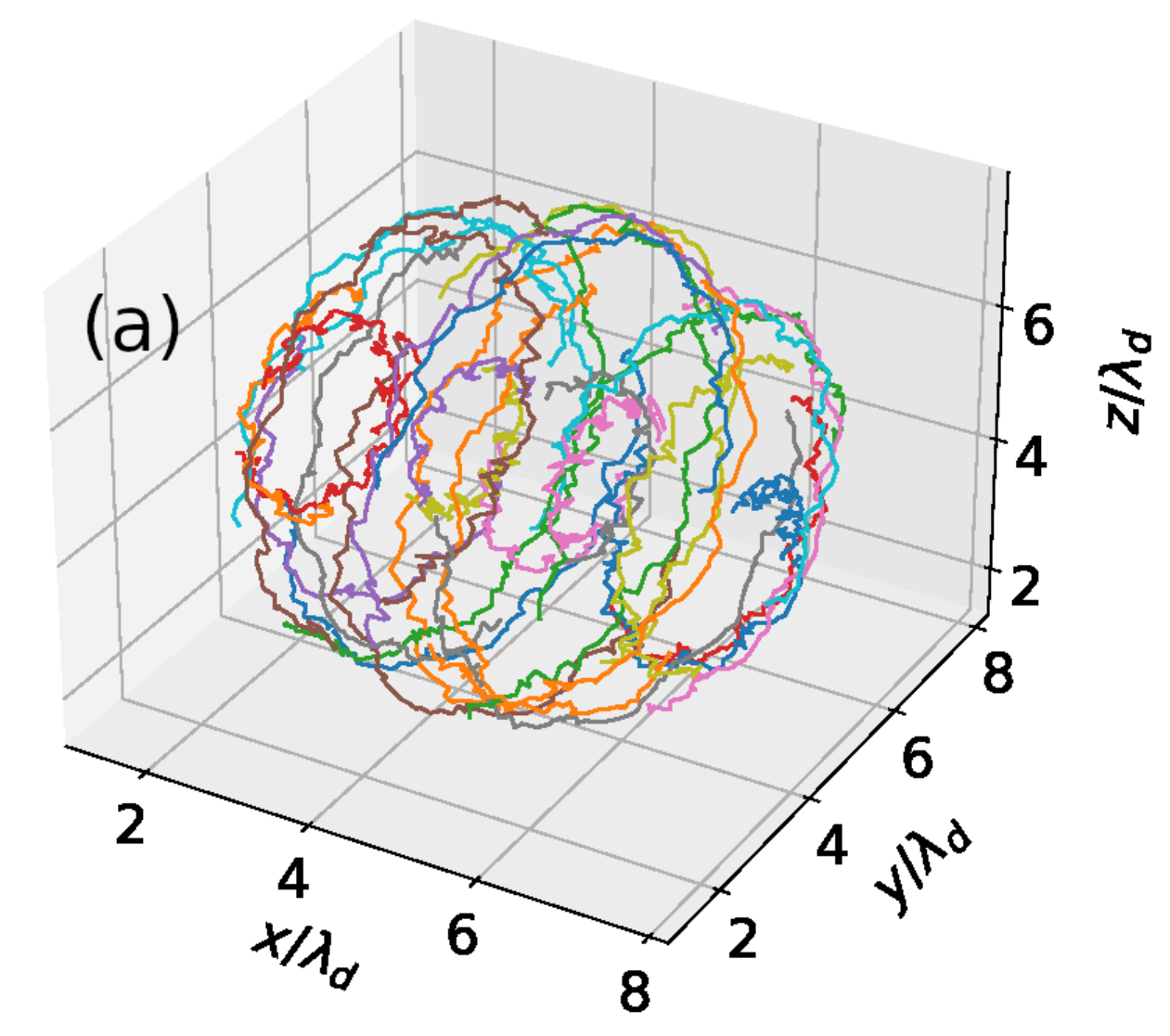}
		
	\end{subfigure} 
	%\hspace{1.6cm}
	\begin{subfigure}{0.45\columnwidth}
		\includegraphics[width=0.5\linewidth]{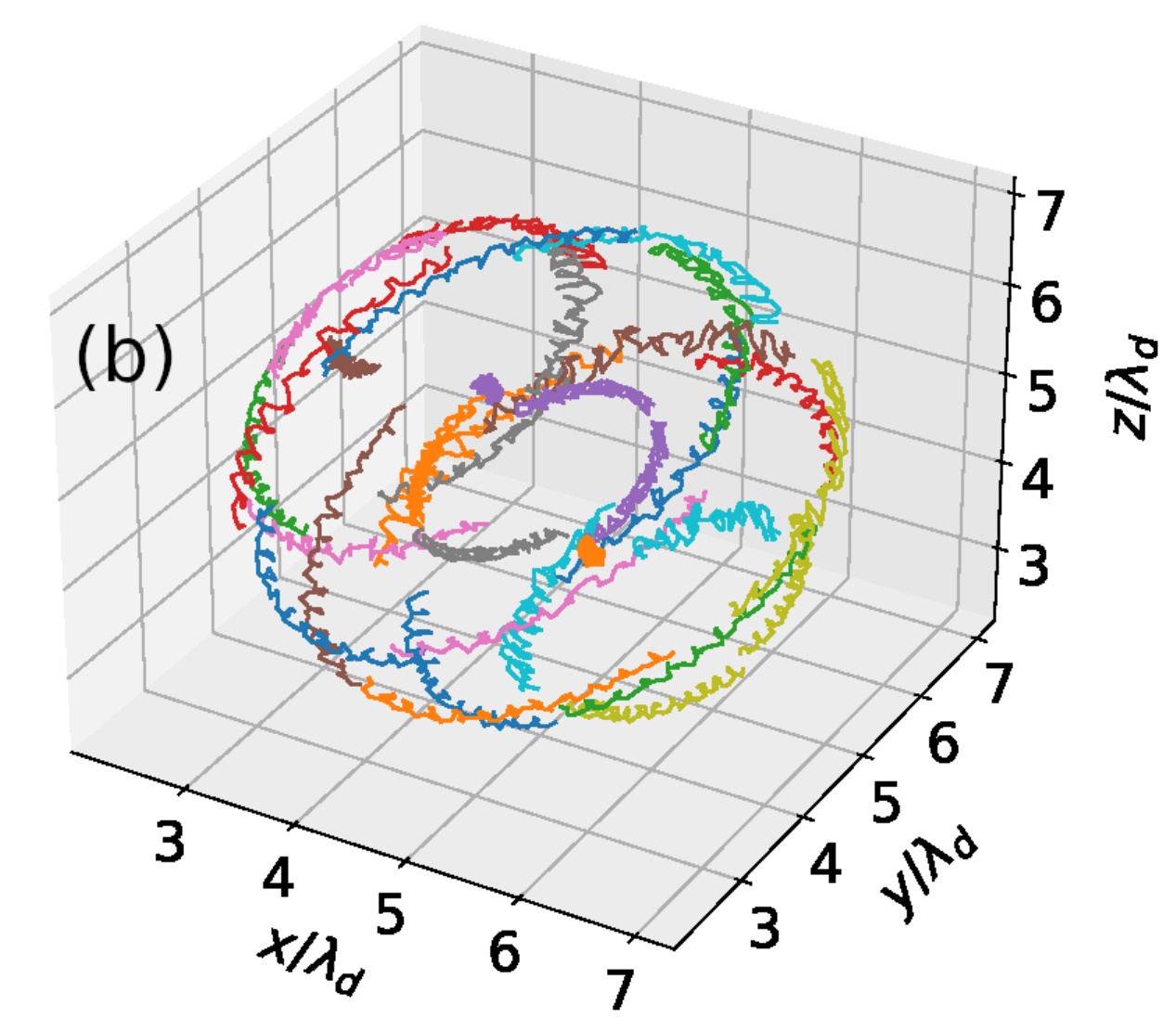}
	\end{subfigure}
	\begin{subfigure}{0.45\columnwidth}
		\includegraphics[width=0.5\linewidth]{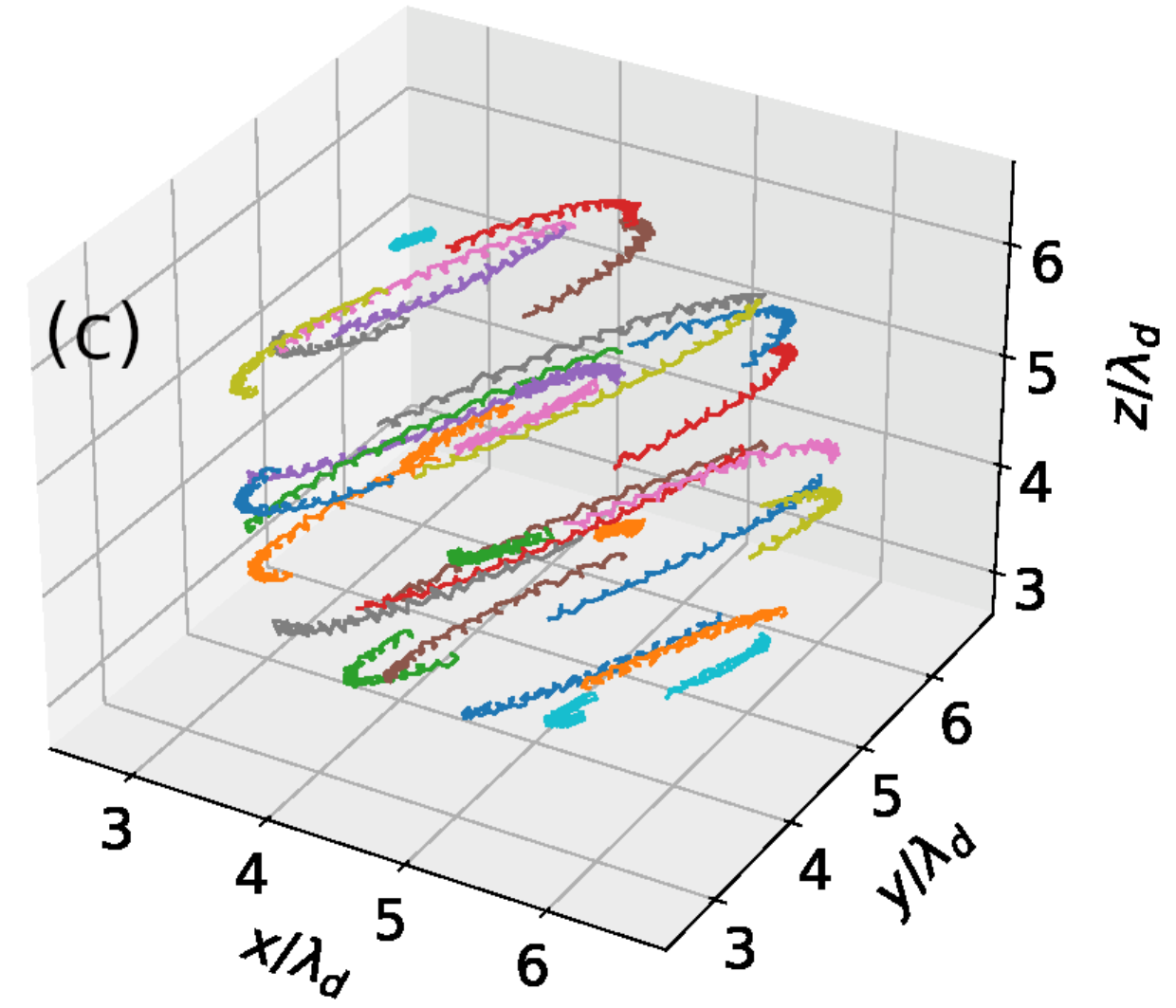}
	\end{subfigure}
	%\hspace{1.6cm}
	\begin{subfigure}{0.45\columnwidth}
		\includegraphics[width=0.5\linewidth]{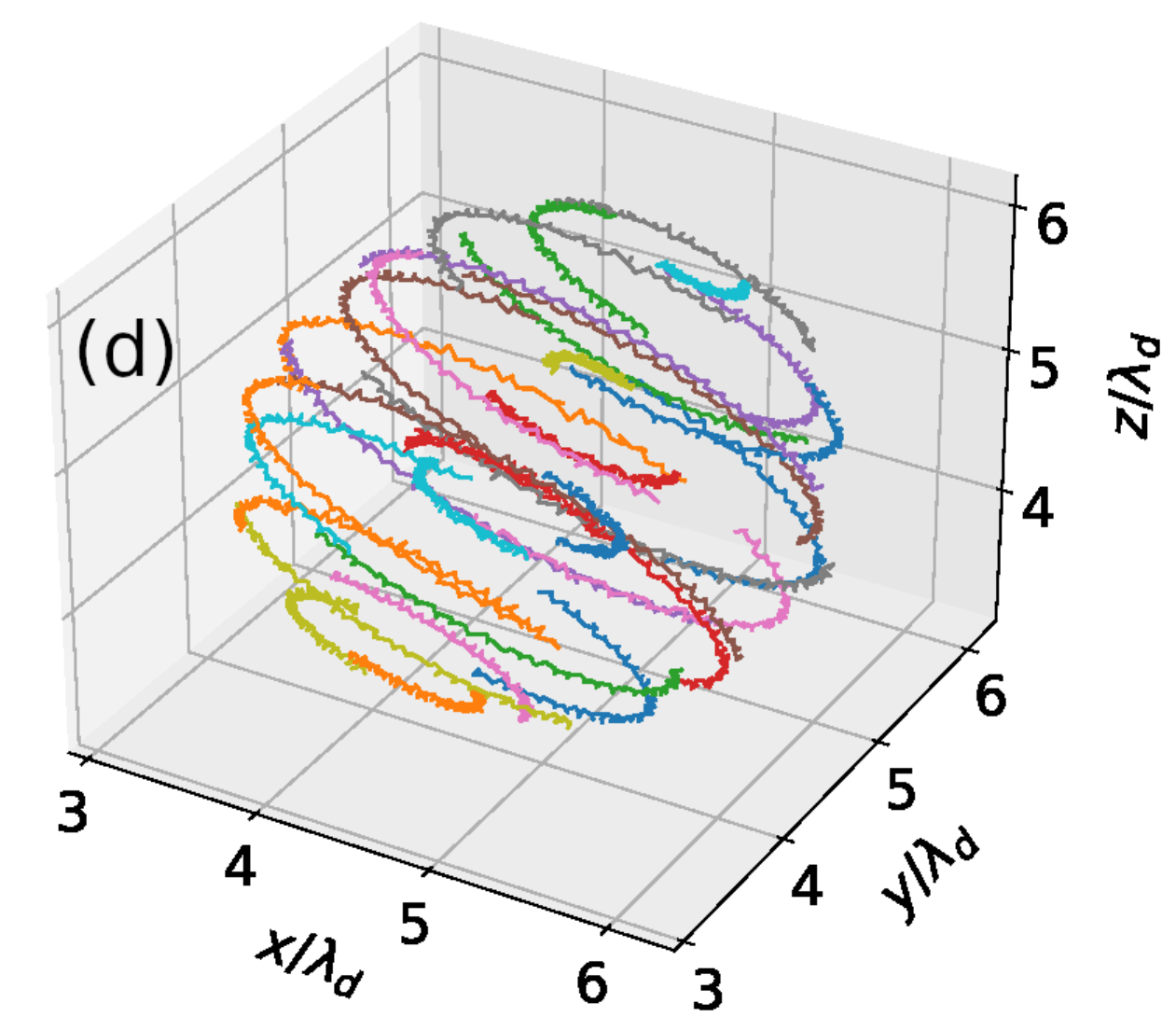}
	\end{subfigure}
	\caption{(Color online)Particle trajectores for a duration of $t/t_0=100$ at different values of harmonic confining strengths (a) $\omega_0 = 10\;Hz$, (b) $\omega_0=20\;Hz$, (c) $\omega_0=30\;Hz$ and (d) $\omega_0=50\;Hz$ obtained from fMD simulation.}
	\label{fMD_traj_withOmega}
\end{figure}

\begin{figure}[htbp]
	\begin{subfigure}{0.45\columnwidth}
		\includegraphics[width=0.5\linewidth]{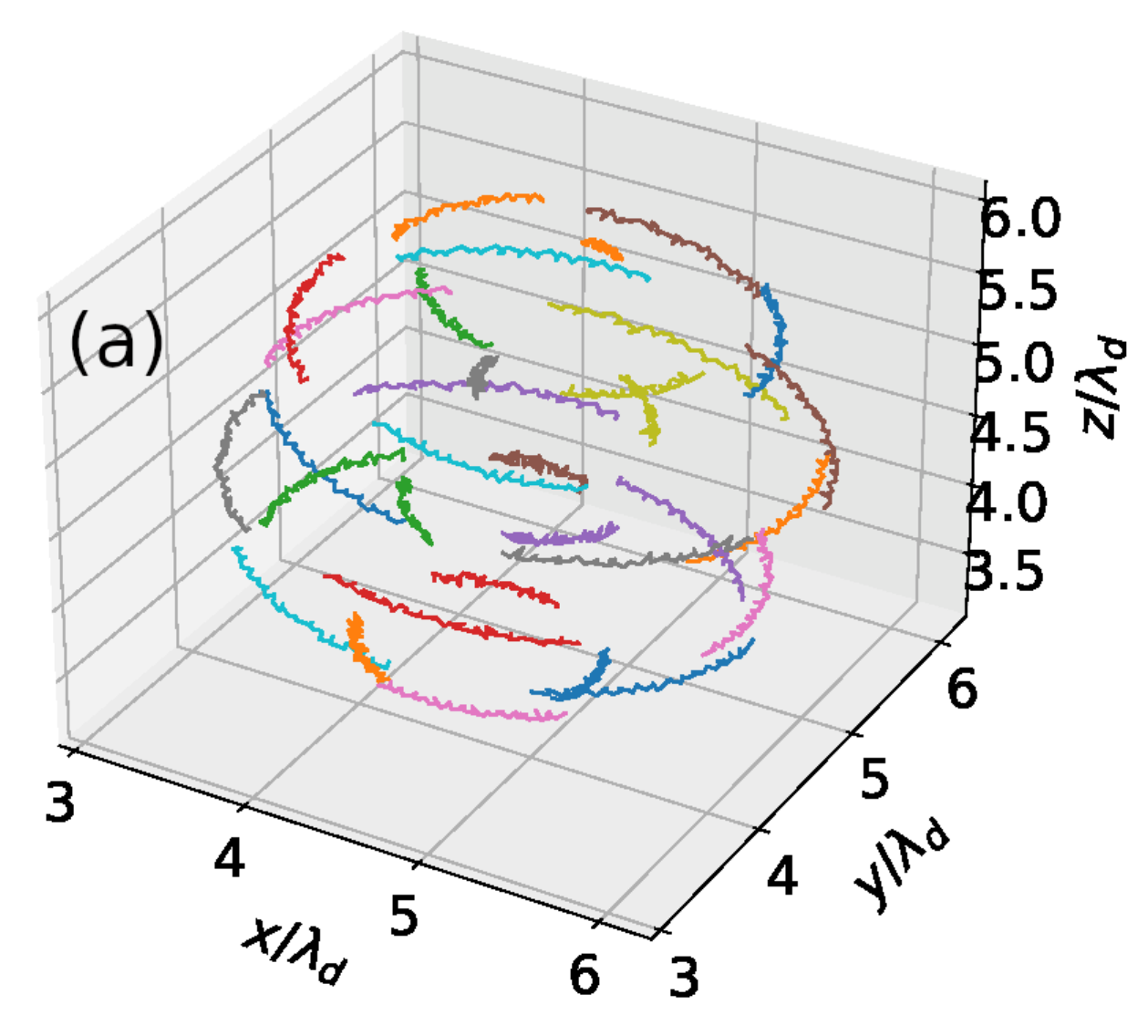}
		
	\end{subfigure} 
	%\hspace{1.6cm}
	\begin{subfigure}{0.45\columnwidth}
		\includegraphics[width=0.5\linewidth]{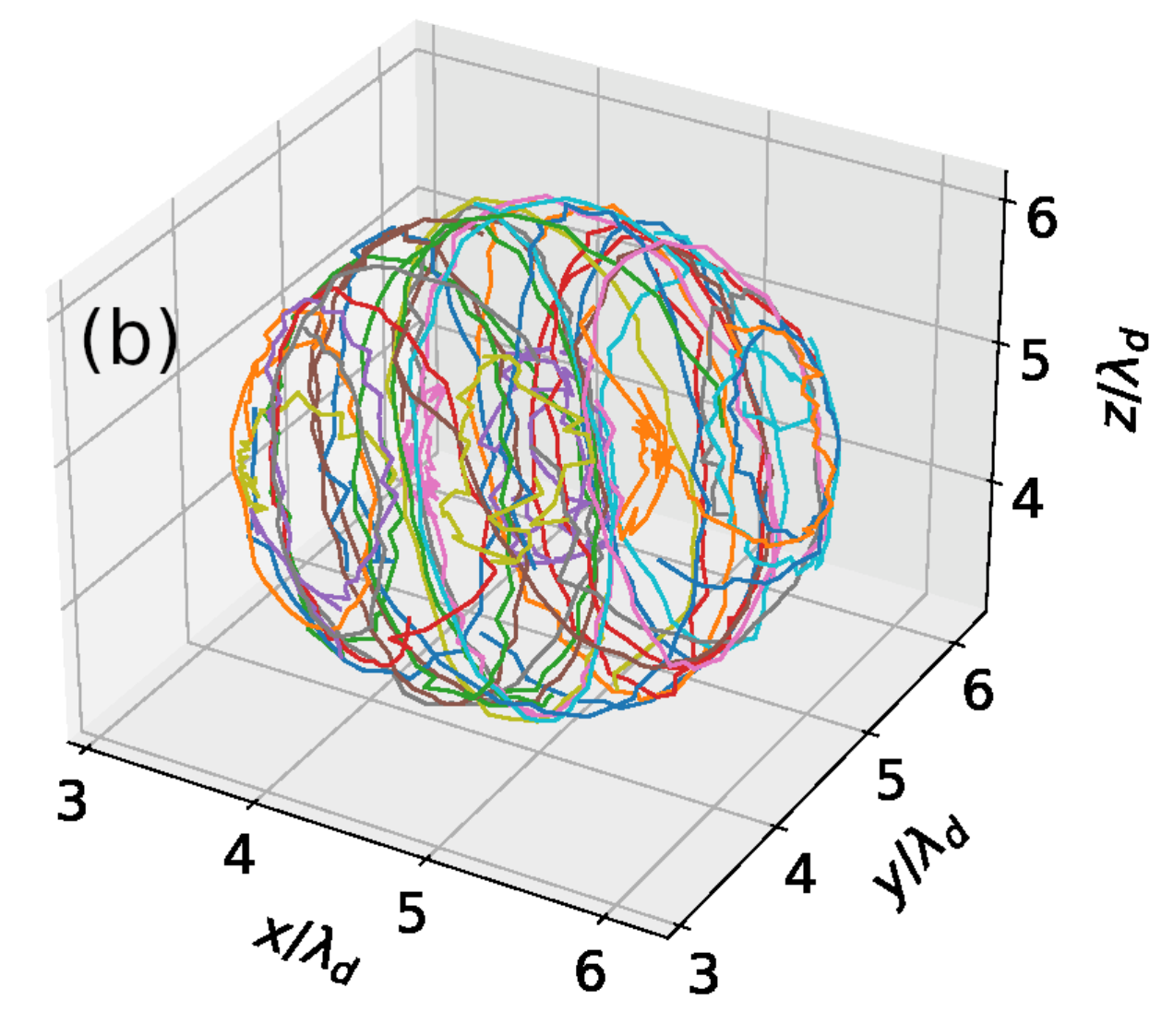}
	\end{subfigure}
	\begin{subfigure}{0.45\columnwidth}
		\includegraphics[width=0.5\linewidth]{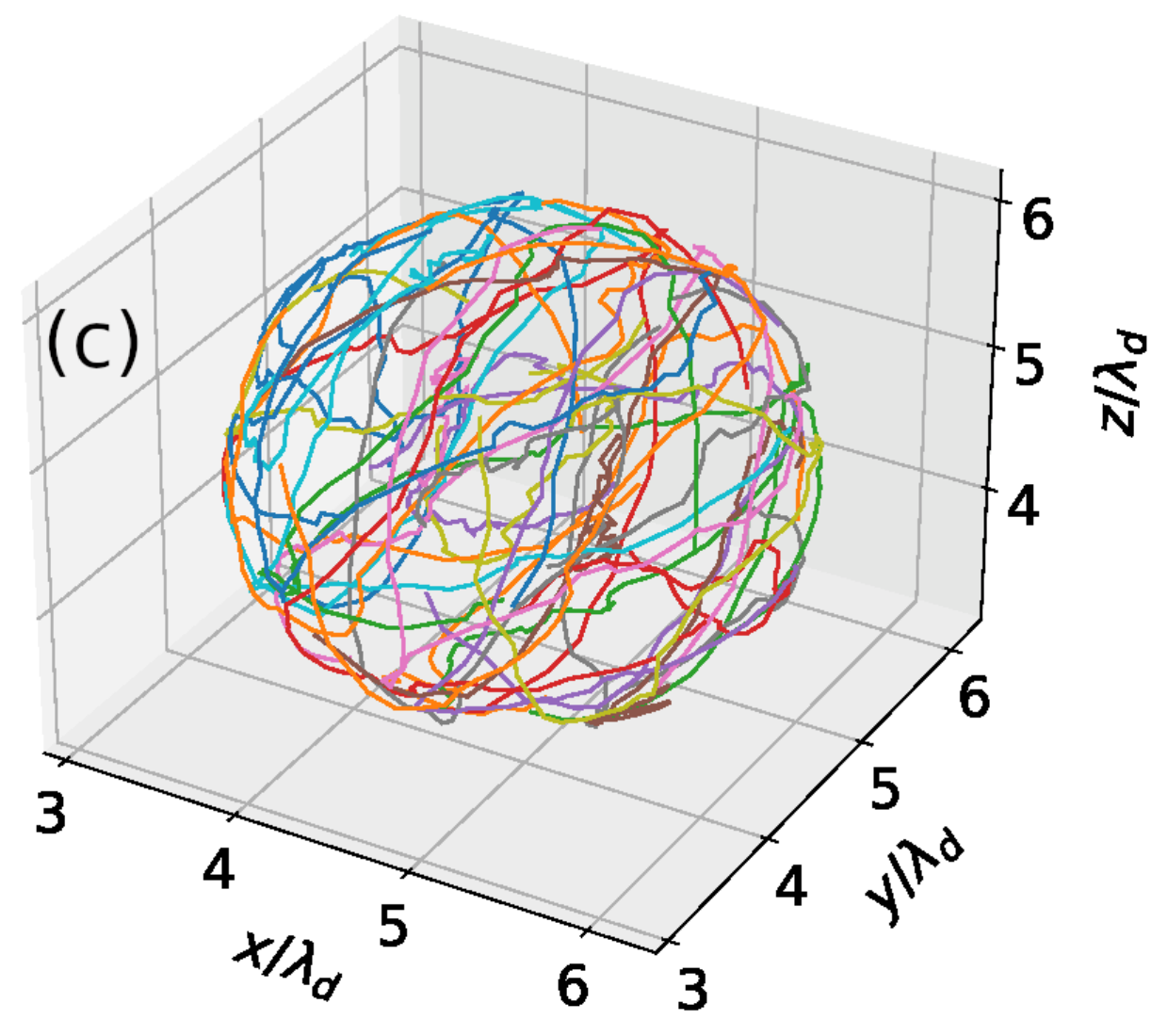}
	\end{subfigure}
	%\hspace{1.6cm}
	\begin{subfigure}{0.45\columnwidth}
		\includegraphics[width=0.5\linewidth]{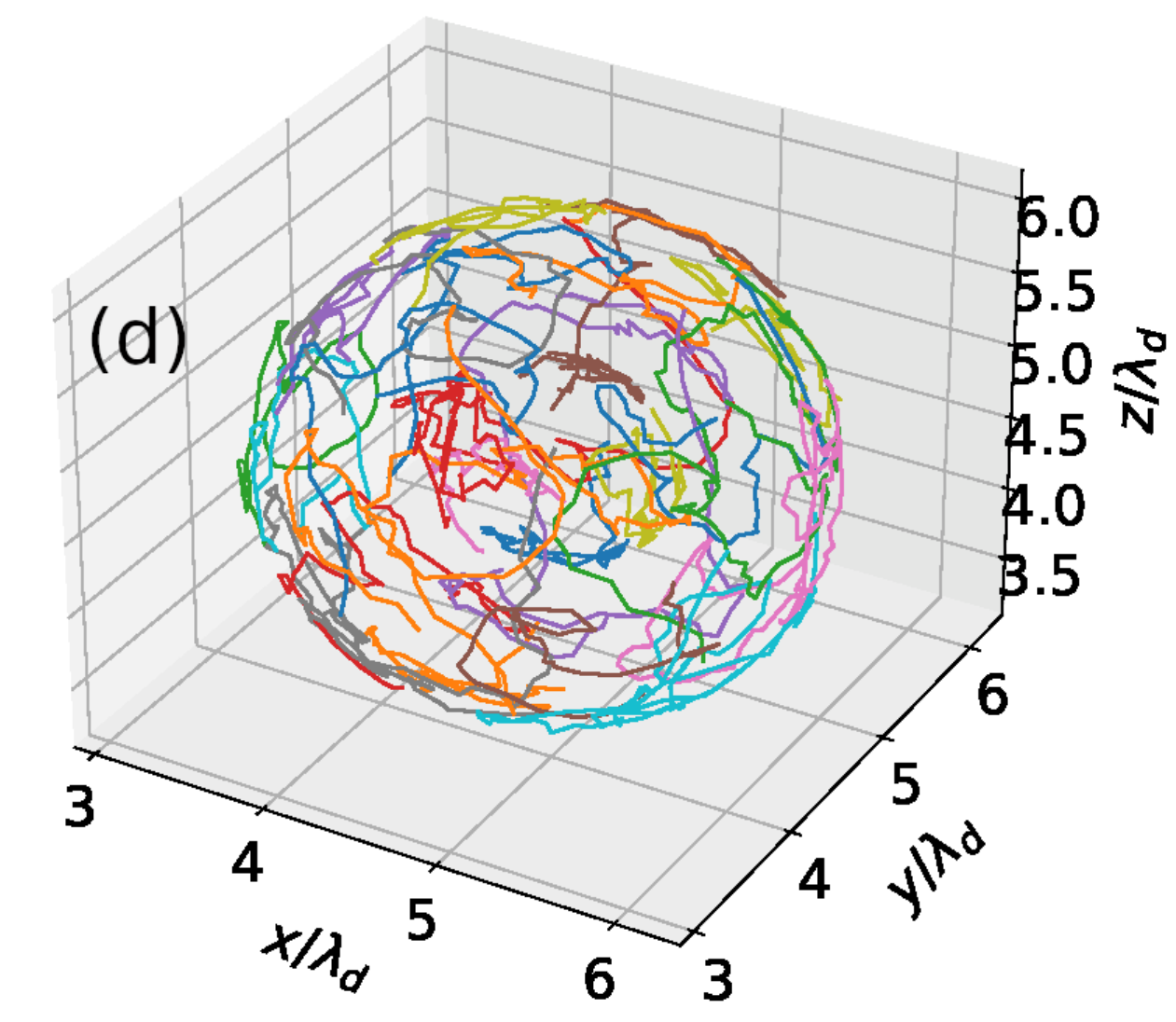}
	\end{subfigure}
	
	\caption{(Color online)Trajectories of dust particles for a time duration of $t/t_0=20$ obtained from (a) frictionless MD simulation and Langevin Dynamics simulation at friction coefficient (b) $\nu = 0.003\;Hz$, (c) $\nu = 0.08\;Hz$ and (d) $\nu = 0.3\;Hz$. The coupling and screening parameter for both the figures is fixed as $\Gamma=1543.14$ and $\kappa=1.8$ respectively.}
	\label{trajectories}
\end{figure}

To get an idea of the mobility of the particles, the plot of trajectories of the dust particles in the cluster are shown in Fig. \ref{trajectories} for both Newtonian and Langevin Dynamics. It is seen that for frictionless MD all the particles in the cluster exhibit predominantly rotational motion about a common axis with oscillatory motion superimposed over the trajectories. This rotational motion of course depends upon the harmonic confinement strength as shown in Fig. \ref{fMD_traj_withOmega}. Qualitatively, it can be seen that as the harmonic confining strength increases the rotation of the particles become much ordered. In LD simulation, for a smaller value of neutral friction it can be seen that the particles still tend to exhibit rotational motion about a common axis but at higher values of neutral friction the rotational motion of the particles about a common axis disappears.

\begin{figure}[htbp]
	\centering
	\includegraphics[width=0.3\linewidth]{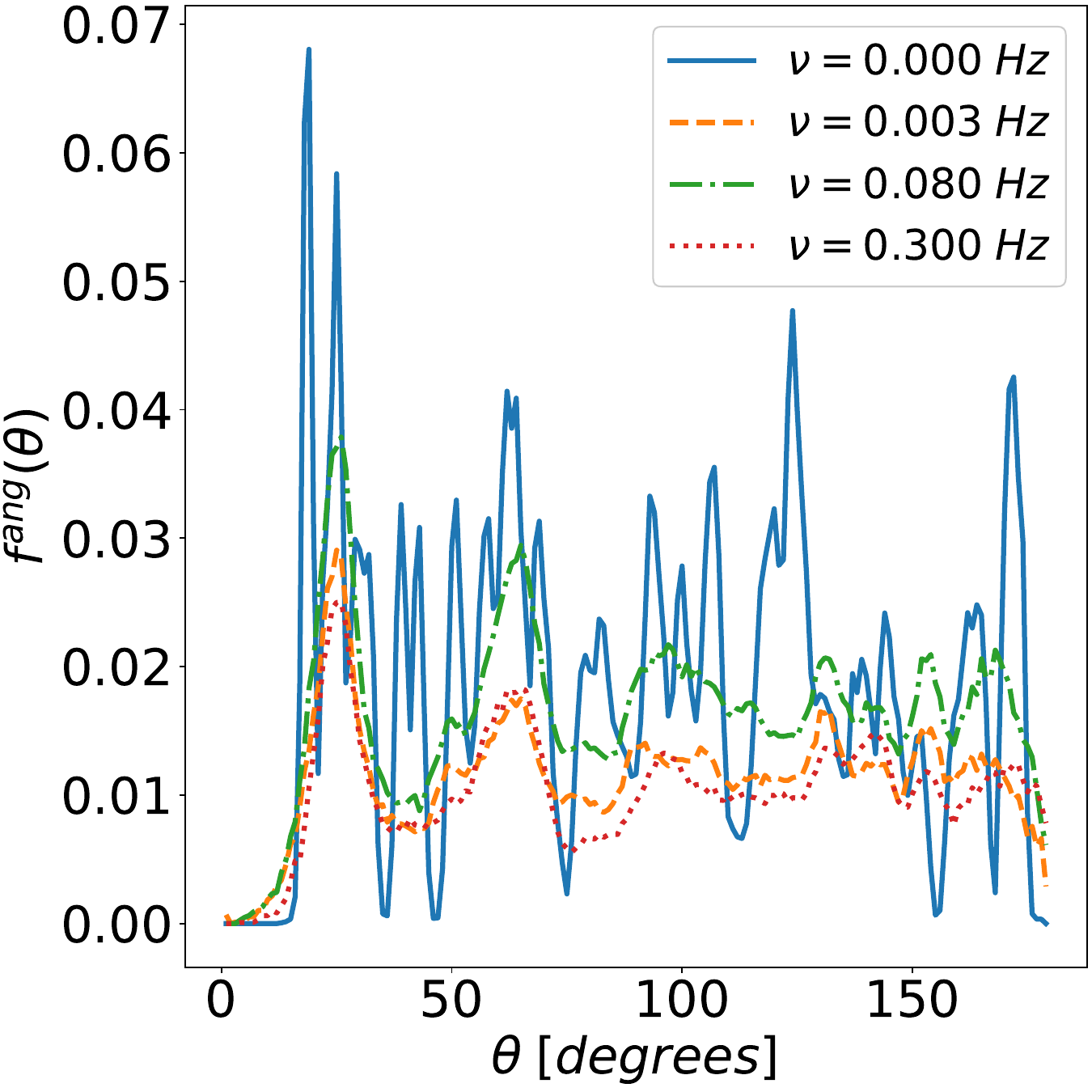}
	\caption{(Color online)Inter-shell angular correlation at different values of dust-neutral collision frequencies at $\Gamma=1543.14$ and $\kappa=1.8$.}
	\label{Inteshell_ang_corr}
\end{figure}

An analysis of the inter-shell angular correlation function indicates that the two nested spherical shells exhibit highly correlated motion in fMD as evidenced by the appearance of sharp peaks in the intershell angular correlation function in Fig.\ref{Inteshell_ang_corr}. However, in LD for all values of friction coefficients there are no sharp peaks in the intershell angular correlation function.

\begin{figure}[htbp]
		\begin{subfigure}{\columnwidth}
			\includegraphics[width=0.4\linewidth]{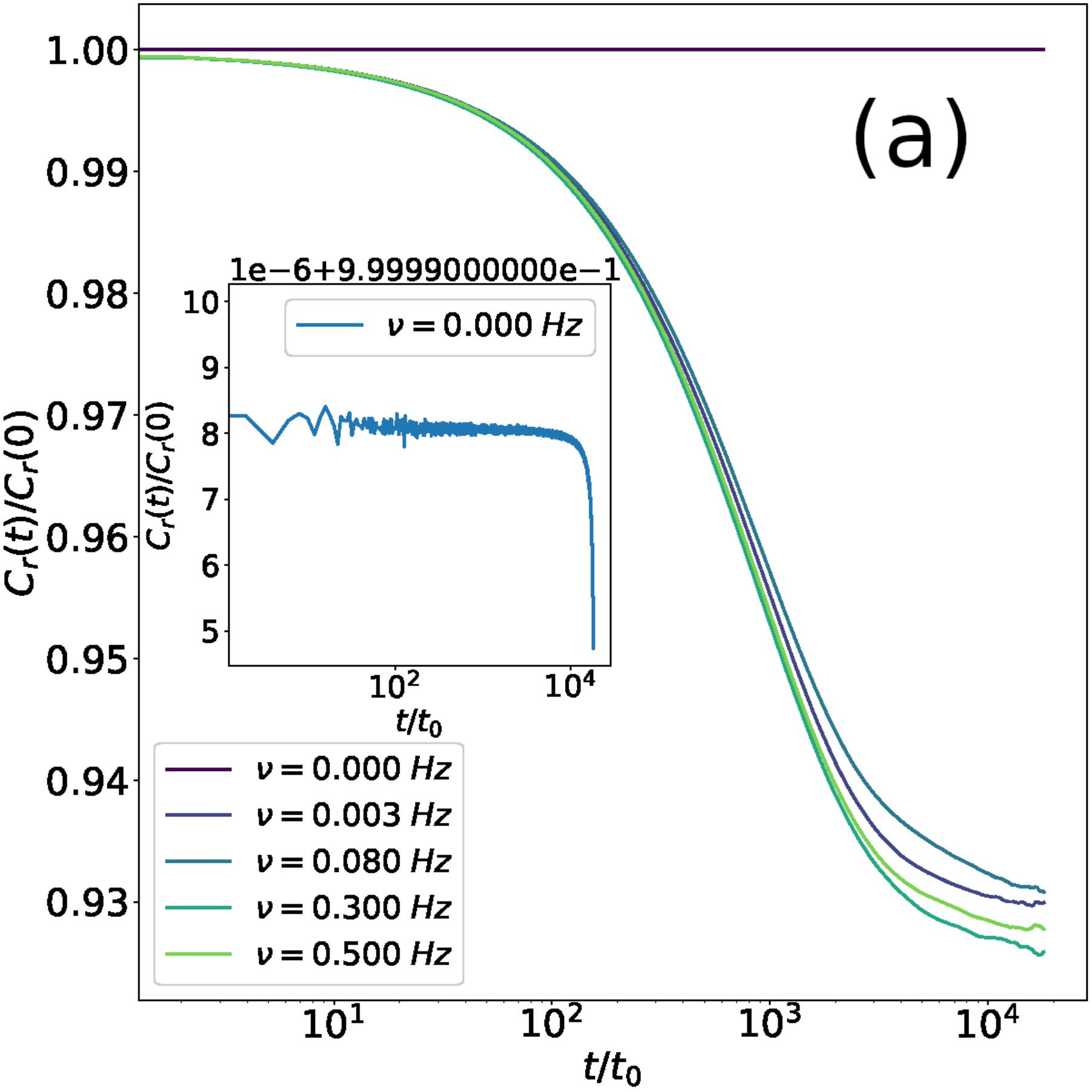}
		\end{subfigure}
		\hspace{1.5cm}
		\begin{subfigure}{\columnwidth}
			\includegraphics[width=0.4\linewidth]{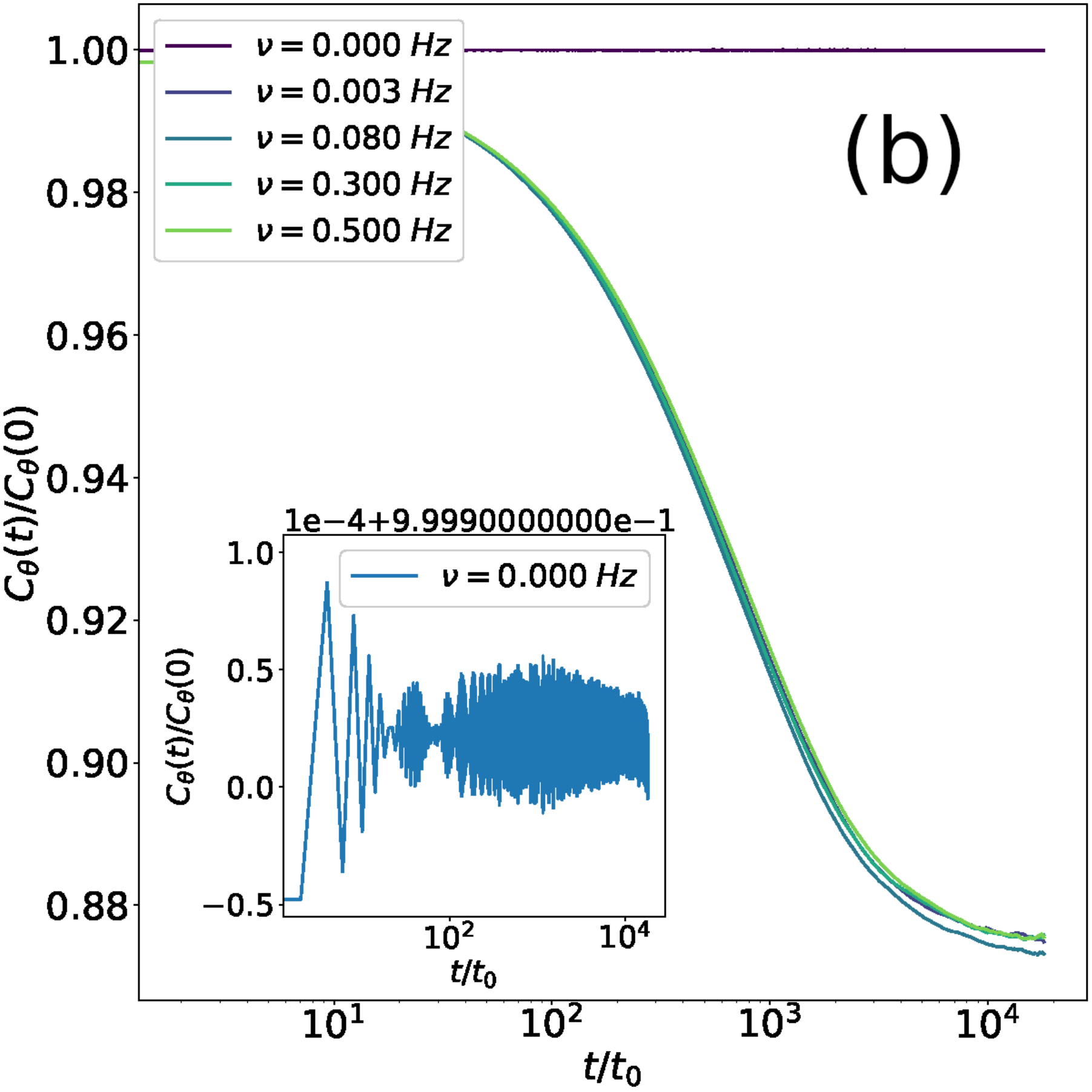}
		\end{subfigure}
	
	\caption{(Color online)Time correlation of  (a) interparticle distance and (b) interparticle angular separation at different values of dust-neutral collision frequency at fixed values of  $\Gamma=1543.14$ and $\kappa=1.8$. The time correlation function for fMD is shown separately in the inset to show the very long time decay of correlations.}
	\label{angular-velocityVsfriction}
\end{figure}

 The fact that the rotational motion of a dust particle in the cluster around the spherical surface of the cluster changes with change in dust-neutral collision frequency motivated us to examine the time correlation function of interparticle distance and interparticle angular separation at different values of dust-neutral collision frequency. The time autocorrelation function of interparticle distance and interparticle angular separation are obtained as $C_r(t)=\Big<r_{ij}(0)r_{ij}(t)\Big>$ and $C_{\theta}(t)=\Big<\theta_{ij}(0)\theta_{ij}(t)\Big>$ respectively, where $\Big<.\Big>$ represents an average over a large number of time origins and all the particle pairs. The angular separation between the $i$th and $j$th particles ($\theta_{ij}$) is measured as the angle between the two vectors drawn from center of the harmonic trap to the two particles. Moreover, we average $C_r(t)$ and $C_{\theta}(t)$ over 80 independent random initial conditions. The result is presented in Fig. \ref{angular-velocityVsfriction}. It can be seen that for fMD there is a very small change of these two quantities as a function of time, whereas for LD the correlation decays much faster with time. The cluster therfore exhibits a persistent rotational motion like a rigid body in fMD which is absent in LD.
\subsection{Mean Squared Displacement}
The Mean Squared Displacement reveals a lot about the single particle dynamics of a many-body system. In glassy systems, the MSD exhibits a plateau region \cite{chong2001mean, deng2019configuration}. 
The observed change in the static structure as shown by the RDF, C2P and intra-shell angular correlation function with increase in coupling strength (i.e, the splitting up of the 2nd peak of RDF and the intra-shell angular correlation function)  motivates us to examine the Mean Squared Displacement (MSD) around this range of $\Gamma$ values.

\begin{figure}[htbp]
	\begin{subfigure}{0.45\columnwidth}
		\includegraphics[width=0.6\linewidth]{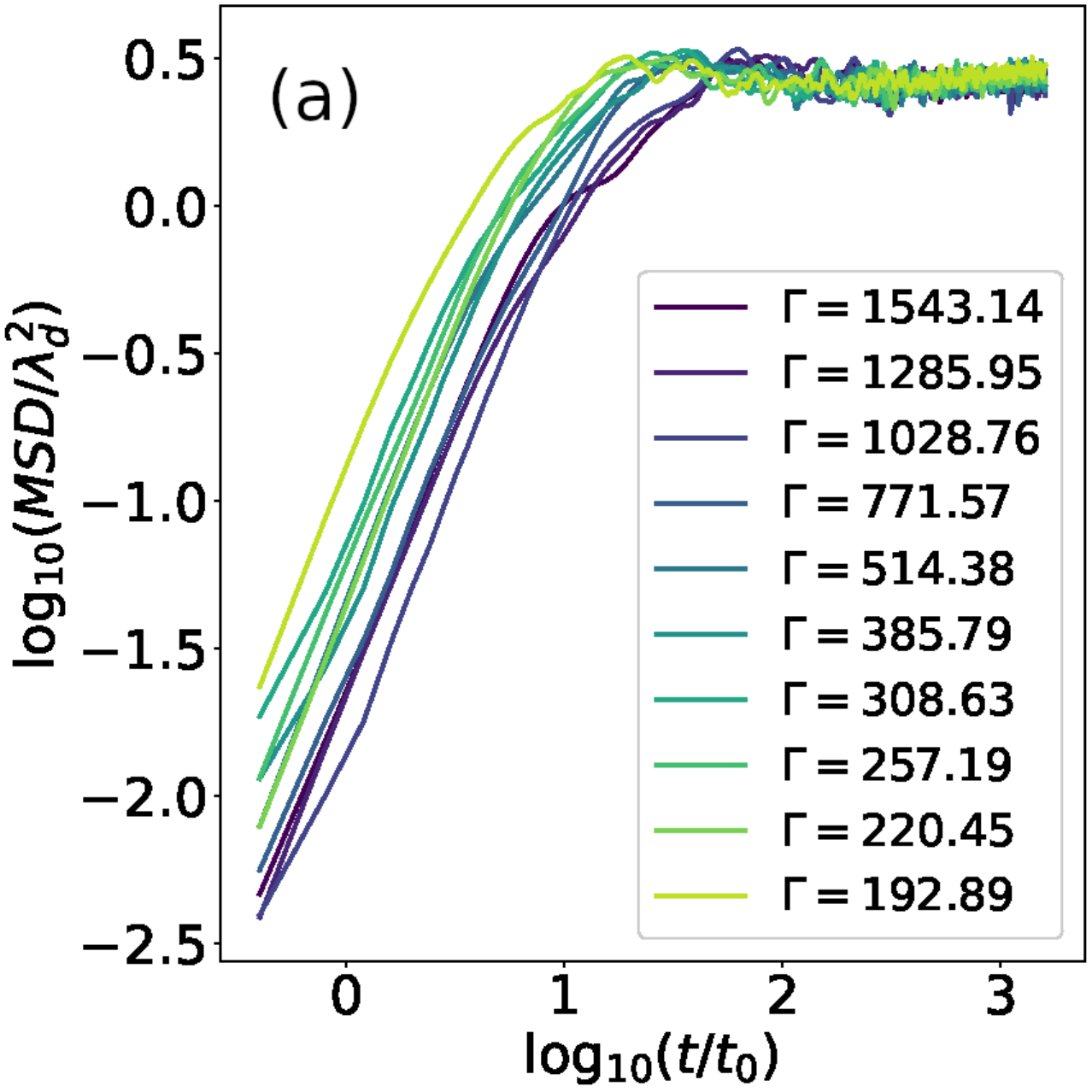}
		
	\end{subfigure}
	%\hspace{1.5cm}
	\begin{subfigure}{0.45\columnwidth}
		\includegraphics[width=0.6\linewidth]{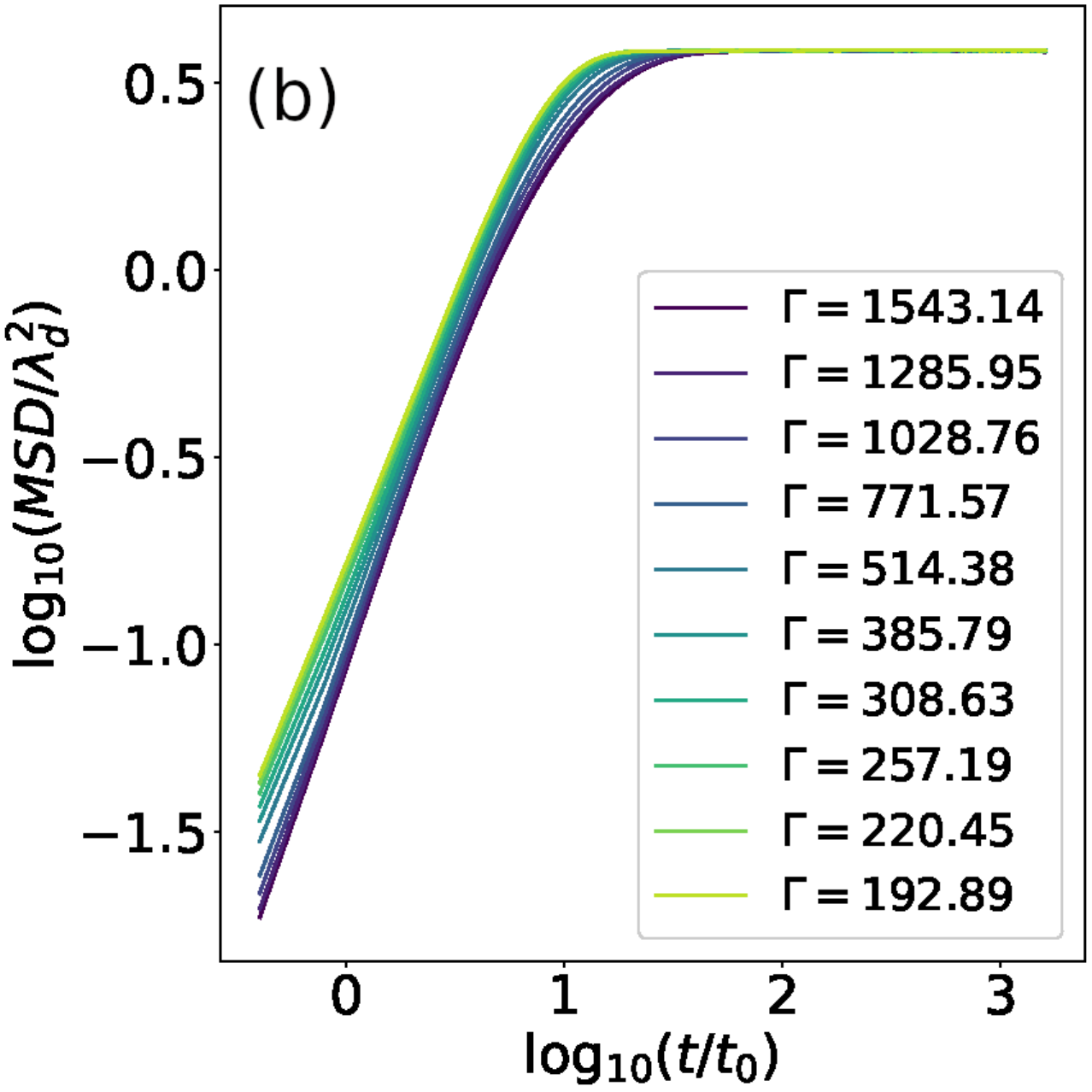}
		
	\end{subfigure}
	%\hspace{1.5cm}
	\begin{subfigure}{0.45\columnwidth}
		\includegraphics[width=0.6\linewidth]{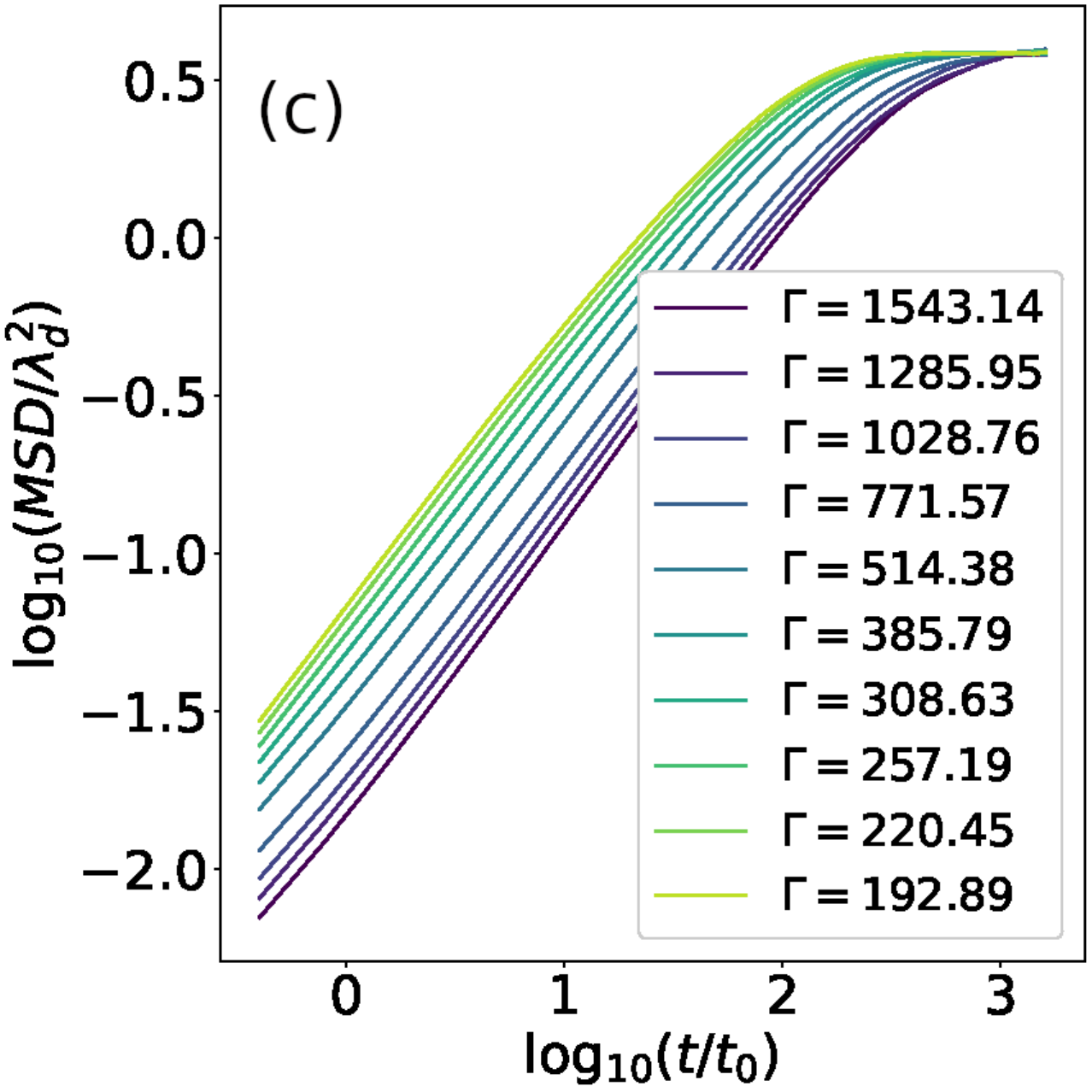}
		
	\end{subfigure}
	\caption{(Color online) Mean squared displacement obtained from (a) frictionless MD simulation and Langevin dynamics simulation at (b) $\nu=0.3\;Hz$ and (c) $\nu=10\;Hz$.}
	\label{MSD}
\end{figure}

The MSDs for both the frictionless MD and Langevin Dynamics cases have been shown in Fig. \ref{MSD} which are obtained according to the Eq. \ref{msd-simulation} by averaging over 80 random independent initial conditions.
\begin{equation}
         MSD(t)=\frac{1}{N\tau_{max}}\sum_{s=1}^{\tau_{max}}\sum_{j=1}^{N}\Big[\textbf{r}_j(t+s)-\textbf{r}_j(s)\Big]^2.
	\label{msd-simulation}
\end{equation}
In the above formula, $\tau_{max}$ is the maximum number of time origins corresponding to the delay time $t$.
As can be seen, for the fMD case, the MSDs lack a regular trend with changing $\Gamma$, whereas for LD the MSD decreases with increasing value of $\Gamma$. Moreover, for the fMD case, the MSDs at all values of $\Gamma$ exhibit oscillatory behaviour at the plateau regions which is absent in case of LD results. For a Brownian particle moving in harmonic potential in 1D the total MSD can be calculated analytically and has the form \cite{li2010measurement} :
\begin{equation}
	\big<\Delta x^2(t)\big>=\frac{2k_BT_d}{m\omega_0^2}\big[1-\exp\big(-\frac{\nu t}{2}\big)\{\cos(\bar\omega_0 t)-\frac{\nu}{2\bar \omega_0}\sin(\bar \omega_0 t)\}\big],
\end{equation} 
where, $\bar\omega_0=\sqrt{\omega_0^2 - (\nu/2)^2}.$
For a harmonically confined Brownian particle moving in three dimensions, the total MSD then becomes $3\big<\Delta x^2(t)\big>$ and at long times such that $t>>1/\nu$ the MSD approaches $\frac{6k_BT_d}{m\omega_{0}^2}$. For a Yukawa cluster like the one considered here, this will happen only when the coupling parameter is so low that the particles' motion becomes almost uncorrelated. For the MSDs shown in Fig. \ref{MSD} this doesn't happen because the coupling strengths are much larger. It is to be noted that although we see development of substructure in the static structural properties namely the RDF and intra-shell angular correlations at sufficiently larger value of Coulomb coupling parameter, no noticeable change is seen in the MSDs at larger values of coupling parameter that corresponds to the observed change in the static properties.

\section{Summary and Conclusions}

In summary, we have investigated the static and dynamic properties of a three dimensional cluster of harmonically confined Yukawa interacting charged dust particles via both frictionless Molecular Dynamics and Langevin Dynamics. Among the static structural properties the RDF and C2P remains largely unaffected by the dynamics employed. However, the inter-shell angular correlation in fMD shows sharp peaks indicating highly correlated motion of the two shells which is absent in LD. The dynamic properties shown by the Van-Hove self correlation function and the self diffusion of the particles in the cluster are affected by the dynamics considered. 

We have investigated the effect of increasing coupling  strength on both the static and dynamic properties of the finite harmonically confined cluster. With increase in coupling, the ordering of particles in the cluster changes which is reflected in the static structural properties of the cluster namely, the RDF, C2P and intra-shell angular correlation function. It is seen that with increase in the coupling, the second peak of RDF and the intra-shell angular correlation function splits up, i.e, substructure appears. 

The change in coupling strength also affects the dynamical properties of the cluster. The Van Hove self autocorrelation function exhibits such change as the coupling strength is varied. We see that for frictionless Molecular Dynamics with decrease in the coupling strength and at a fixed delay time the Van Hove function broadens. Also the main peak height at all delay times can be seen to be reducing with decreasing coupling strength. With the introduction of neutral friction, i.e, when the particle dynamics is governed by the Langevin equation of motion the basic nature of the Van-Hove self correlation function remains the same although the peak height remains smaller and the widths broader at all delay times than the corresponding plots of Newtonian Dynamics. This indicates greater particle mobility in Langevin Dynamics than in frictionless Molecular Dynamics. The trajectory plot of fMD reveals a predominantly rotational motion of the particles about a common axis in the cluster. This rotational motion of the cluster disappears with the introduction of dust-neutral collision. An analysis of the time autocorrelation function of the interparticle distance and interparticle angular separation reveals that persistent rotation of the cluster is not possible in the presence of dust-neutral collision as the time autocorrelation of interparticle distance ($r_{ij}$) and interparticle angular separation ($\theta_{ij}$) decays much faster in case of Langevin Dynamics as compared to Newtonian Dynamics. 
%%%%%%%%%%%%%%%%%%%%%%%%%%%%%%%%%%%%%%%%%%%%%%%%%
\section*{Acknowledgement}
H.S. is thankful to Tezpur University for providing financial support under the Research and Innovation Grant ( DoRD/RIG/10-73/ 1592-A).  

\bibliography{newpaper}

\end{document}